\documentclass[aps,prd,twocolumn,superscriptaddress, nofootinbib,floats,floatfix]{revtex4-2}
\usepackage{bm}
\usepackage{amsmath}
\usepackage{amsfonts}
\usepackage{aas_macros}
\usepackage{graphicx}
\usepackage[hidelinks]{hyperref}
\usepackage{booktabs}
\usepackage{color}
\usepackage[dvipsnames]{xcolor}
\usepackage[capitalise]{cleveref}
\usepackage[caption=false]{subfig}
\usepackage{paralist}
\usepackage{acro}
\usepackage{tabularx}

\newcommand\numberthis{\addtocounter{equation}{1}\tag{\theequation}}

\newcommand{\sigv}{\ensuremath{\left< \sigma v \right>}}
\newcommand{\Msun}{\ensuremath{\mathrm{M}_\odot}}
\newcommand{\Mpch}{\ensuremath{h^{-1}\;\text{Mpc}}}

\DeclareMathOperator{\Li}{Li}
\newcommand{\dd}{\mathrm{d}}

\newcommand\csiborg{\texttt{CSiBORG}}
\newcommand\borg{\texttt{BORG} }
\newcommand\nside{\texttt{nside}}
\newcommand\healpix{\textsc{HEALPix} }
\newcommand\healpy{\textsc{healpy} }
\newcommand\clumpy{\textsc{clumpy} }
\newcommand\numpyro{\textsc{numpyro} }
\newcommand\fermipy{\textsc{FermiPy} }

\DeclareAcronym{GCE}{short = GCE, long  = Galactic Centre Excess}
\DeclareAcronym{MCMC}{short = MCMC, long = Markov Chain Monte Carlo}

\begin{document}

\title{No evidence for \texorpdfstring{$p$}{p}- or \texorpdfstring{$d$}{d}-wave dark matter annihilation from local large-scale structure}

\author{A. Kosti\'{c}}
\email{andrii.kostic@gmail.com}
\affiliation{Max Planck Institute for Astrophysics, Karl-Schwarzschild-Stra{\ss}e 1, 85748 Garching, Germany}
\author{D. J. Bartlett}
\email{deaglan.bartlett@physics.ox.ac.uk}
\affiliation{CNRS \& Sorbonne Universit\'{e}, Institut d’Astrophysique de Paris (IAP), UMR 7095, 98 bis bd Arago, F-75014 Paris, France}\affiliation{Astrophysics, University of Oxford, Denys Wilkinson Building, Keble Road, Oxford, OX1 3RH, UK}
\author{H. Desmond}
\affiliation{Institute of Cosmology \& Gravitation, University of Portsmouth, Dennis Sciama Building, Portsmouth, PO1 3FX, UK}

\begin{abstract}
    If dark matter annihilates into standard model particles with a cross-section which is velocity dependent, then Local Group dwarf galaxies will not be the best place to search for the resulting gamma ray emission. A greater flux would be produced by more distant and massive halos, with larger velocity dispersions. We construct full-sky predictions for the gamma-ray emission from galaxy- and cluster-mass halos within $\sim 200 {\rm \, Mpc}$ using a suite of constrained $N$-body simulations (\texttt{CSiBORG}) based on the Bayesian Origin Reconstruction from Galaxies algorithm. Comparing to observations from the \textit{Fermi} Large Area Telescope and marginalising over reconstruction uncertainties and other astrophysical contributions to the flux, we obtain constraints on the cross-section which are two (seven) orders of magnitude tighter than those obtained from dwarf spheroidals for $p$-wave ($d$-wave) annihilation.
    We find no evidence for either type of annihilation from dark matter particles with masses in the range $m_\chi = 2-500{\rm \, GeV}/c^2$, for any channel. As an example, for annihilations producing bottom quarks with $m_\chi = 10 {\rm \, GeV}/c^2$, we find $a_{1} < 2.4 \times 10^{-21} {\rm \, cm^3 s^{-1}}$ and $a_{2} < 3.0 \times 10^{-18} {\rm \, cm^3 s^{-1}}$ at 95\% confidence, where the product of the cross-section, $\sigma$, and relative particle velocity, $v$, is given by $\sigma v = a_\ell (v/c)^{2\ell}$ and $\ell=1, 2$ for $p$-, $d$-wave annihilation, respectively.
    Our bounds, although failing to exclude the thermal relic cross-section for velocity-dependent annihilation channels, are among the tightest to date.
\end{abstract}

\maketitle

\section{Introduction}

An unambiguous detection of non-gravitational interactions of dark matter continues to evade us. Astrophysical objects are ideal targets in the search for novel dark matter physics, since low interaction rates can be compensated by the large quantities of dark matter, leading to potentially detectable signals. For over a decade, the origin of the observed excess of gamma rays at the centre of our galaxy -- known as the \ac{GCE} \citep{Goodenough_2009,Ajello_2016,Linden_2016} -- has been debated, with explanations such as unresolved point sources or dark matter annihilation being proposed 
\citep[e.g.][]{Hooper_2011,Hooper_2011PRD,Hooper_2013PDU,Zhou_2015,Daylan_2016,Cholis_2021,Grand_2022,Abazajian_2011,OLeary_2015,Petrovic_2015,Lee_2016,Buschmann_2020,Gautam_2021,Hooper_2013,Cholis_2015_pulsars,Leane_2019,Abazajian_2012,Gordon_2013,Abazajian_2014,Calore_2015,Horiuchi_2016,Petrovic_2014,Carlson_2014,Cholis_2015_rays,Gaggero_2015,Macias_2018,Bartels_2018}.
The majority of work on this excess has focused on $s$-wave annihilation, where the product of the self-annihilation cross-section, $\sigma$, and relative velocity, $v$, is independent of $v$. Such models can be strongly constrained by considering the gamma ray flux from nearby dwarf galaxies \citep{dSph_0, dSph, dIrr}, which are able to rule out the thermal relic cross-section for dark matter particle masses relevant to explain the \ac{GCE}. 

Not all dark matter annihilation models require $\sigma v$ to be independent of velocity. For example, for fermionic dark matter annihilation to spin-0 particles, even parity final states cannot have an $s$-wave contribution in parity conserving theories \citep{Kim_2007,Lee_2008}. $p$-wave annihilation -- where $\sigma v$ is proportional to the square of the relative velocity -- dominates when Standard Model fermion-antifermion pairs are produced from Majorana fermions in models with minimal flavour violation due to chirality suppression of $s$-wave annihilation \citep{Kumar_2013}. If dark matter is instead a real scalar, then $d$-wave annihilation -- where the fourth power of the relative velocity is relevant -- is dominant \citep{Giacchino_2013,Toma_2013}.

If the \ac{GCE} were due to velocity-dependent dark matter annihilation, then, despite their close proximity, dwarf spheroidals cannot provide tight constraints on the annihilation cross-section, since the low velocity dispersions in these objects result in heavily suppressed annihilation rates. One should therefore test whether these velocity-dependent models are compatible with other observations of gamma ray fluxes, since these can evade bounds from these objects. In these scenarios, one would expect a large signal from extra-galactic halos, where the velocity dispersion should be larger. It has therefore been recently suggested to use extra-galactic halos to search for velocity-dependent dark matter annihilation signals in gamma ray data \citep{Baxter_2022}.

In this paper we perform this test for $p$-wave and $d$-wave annihilation, where the cross-section is proportional to the square and fourth power of the relative velocity between dark matter particles, respectively. As in our previous work constraining $s$-wave annihilation \citep{FLAT_2022}, we utilise the \csiborg{} \citep{Bartlett_2021_VS,antihalos,Hutt_2022} suite of constrained $N$-body simulations to produce full-sky templates for the dark matter annihilation flux (see \cref{fig:jd_ensemble}), which are then compared to the observations of the \textit{Fermi} Large Area Telescope via a \ac{MCMC} algorithm. The initial conditions for these simulations are inferred using the \borg (Bayesian Origin Reconstruction from Galaxies) algorithm and are designed to produce three-dimensional dark matter density fields which match the observed positions of galaxies in the 2M++ galaxy catalogue \citep{Lavaux_2016, Jasche_Lavaux}. We marginalise over uncertainties in this reconstruction as well as over non-dark-matter contributions to the gamma ray sky. Our constraints on the self-annihilation cross-section are two (seven) orders of magnitude tighter than those obtained using dwarf spheroidal galaxies for $p$-wave ($d$-wave) annihilation.

In \cref{sec:Methods} we describe how we forward model the dark matter annihilation flux, discuss the models used for competing astrophysical contributions to the signal, and outline our inference procedure. These predictions are compared to observations, presented and discussed in \cref{sec:Results,sec:Discussion}, and we conclude in \cref{sec:Conclusions}.

\section{Methods}
\label{sec:Methods}

\subsection{Bayesian large-scale structure inference}

In this work we study the expected gamma ray flux from dark matter annihilation in extragalactic halos. To determine the masses and locations of these halos, we use the \csiborg{} suite of 101 dark matter-only constrained simulations \citep{Bartlett_2021_VS,antihalos,Hutt_2022}. The initial conditions of these simulations produce present-day density fields which match the observed number densities of galaxies in the 2M++ galaxy compilation \citep{Lavaux_2016, Jasche_Lavaux} and are inferred using the \borg algorithm \citep[see e.g.][]{BORG_1,BORG_2,BORG_3,BORG_4,Lavaux_2016}. This algorithm produces a Markov Chain of plausible initial conditions via the application of a Bayesian forward model and marginalises over galaxy bias parameters. The initial conditions are constrained in a box of length $677.7\Mpch$ with $256^3$ voxels. For each \csiborg{} simulation these initial conditions are augmented with white noise to a resolution of $2048^3$ within $155\Mpch$ of the Milky Way (giving a mass resolution of $4.4 \times 10^9\;\Msun$) and run to $z=0$ using \texttt{RAMSES} \citep{ramses}. The watershed halofinder \texttt{PHEW} \citep{PHEW} is applied on-the-fly to the dark matter particles with the standard threshold of $200\rho_c$.
We do not consider sub-halos and the minimum halo mass used ($M_{200\text{m}}$) is $4.4 \times 10^{11} \Msun$. The resulting halo catalogues are publicly available at \citep{max_zenodo}.

\subsection{Velocity-dependent \texorpdfstring{$J$}{J} factor}

In this work we consider dark matter annihilation of particles of mass $m_\chi$, with a cross-section $\sigma$ which depends on the relative velocity, $v$,
between annihilating dark matter particles as
\begin{equation}
    \sigma v = a_\ell S_\ell 
    \left( \frac{v}{c} \right),
\end{equation}
where $a_\ell$ is a constant and $S_\ell$ can be an arbitrary function of $v/c$. At observed energy $E_\gamma$, the differential photon flux, $\dd \Phi_\gamma / \dd E_\gamma$, is
\begin{equation}
    \label{eq:fractional_flux}
    \frac{\dd\Phi_{\gamma}}{\dd E_\gamma} 
    = \frac{a_\ell}{8 \pi m_\chi^2} J^{\ell} \sum_i {\rm Br}_i  \left. \frac{\dd N_i}{\dd E_\gamma^\prime} \right\vert_{E_\gamma^\prime = E_\gamma \left( 1 + z \right)},
\end{equation}
where we sum over annihilation channels with branching ratios ${\rm Br}_i$. This equation is valid if dark matter is entirely comprised of a single particle whose antiparticle is itself. We introduced the $J$ factor 
\begin{equation}
    \label{eq:J_factor_Sv}
    \begin{split}
        J^{\ell}_{\mu,p} = \int_{\mu,p} \dd\Omega \dd \mu \, \dd^3v_1 \, \dd^3v_2 \,
            &S_\ell \left( \frac{\left| \bm{v}_1 - \bm{v}_2 \right|}{c} \right) \times \\ 
            &f \left( \bm{r}(\mu), \bm{v_1} \right)
            f \left( \bm{r}(\mu), \bm{v_2} \right),
   \end{split}
\end{equation}
where $f\left( \bm{r}(\mu),\bm{v} \right)$ is the distribution function of DM particles, which we integrate over the velocities of the two particles, $\bm{v}_1$ and $\bm{v}_2$, and along the line of sight coordinate, $\mu$. 
Since we define these quantities on a discretised map,
we also integrate across the \healpix \footnote{\url{http://healpix.sf.net}} pixel area $p$. 
For simplicity, we assume that the decays occur via one channel, so the branching ratio, ${\rm Br}_i$ is either 0 or 1. Note that the photon distribution function, $\dd N_i / \dd E_\gamma^\prime$ should be evaluated at the redshift of the halo. Since we will be using low-redshift halos in this work ($z \lesssim 0.05$) we will neglect this redshift effect and thus assume that $E_\gamma^\prime \approx E_\gamma$. Introducing redshift dependence would require inserting a redshift dependent optical depth into \cref{eq:fractional_flux}, which, in our case, will only marginally affect the results (see for example the discussion in Section IV.B.5 in \citep{FLAT_2022}).

In this work we will consider $p$-wave and $d$-wave annihilation, i.e. $S_\ell (x) = x^{2 \ell}$, where $\ell=1$ ($p$-wave) or $\ell=2$ ($d$-wave). In the following two sections we describe how one can obtain the $J$ factor for these models from the density profile of a dark matter halo.

\subsubsection{\texorpdfstring{$J$}{J} factor for \texorpdfstring{$p$}{p}-wave annihilation}

\begin{figure*}
\centering
\begin{minipage}{.45\textwidth}
  \centering
  {$p$-wave}
  \includegraphics[width=\linewidth]{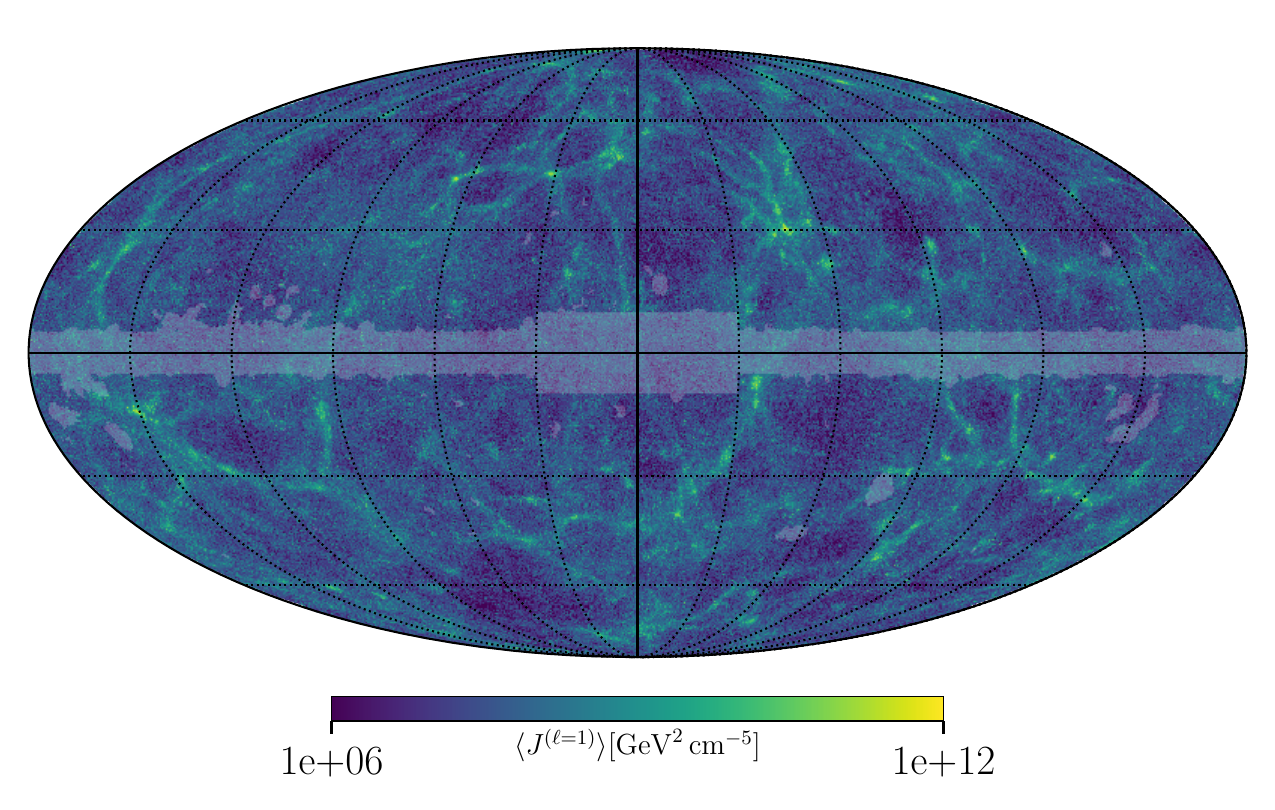}
\end{minipage}
\begin{minipage}{.45\textwidth}
  \centering
  {$d$-wave}
  \includegraphics[width=\linewidth]{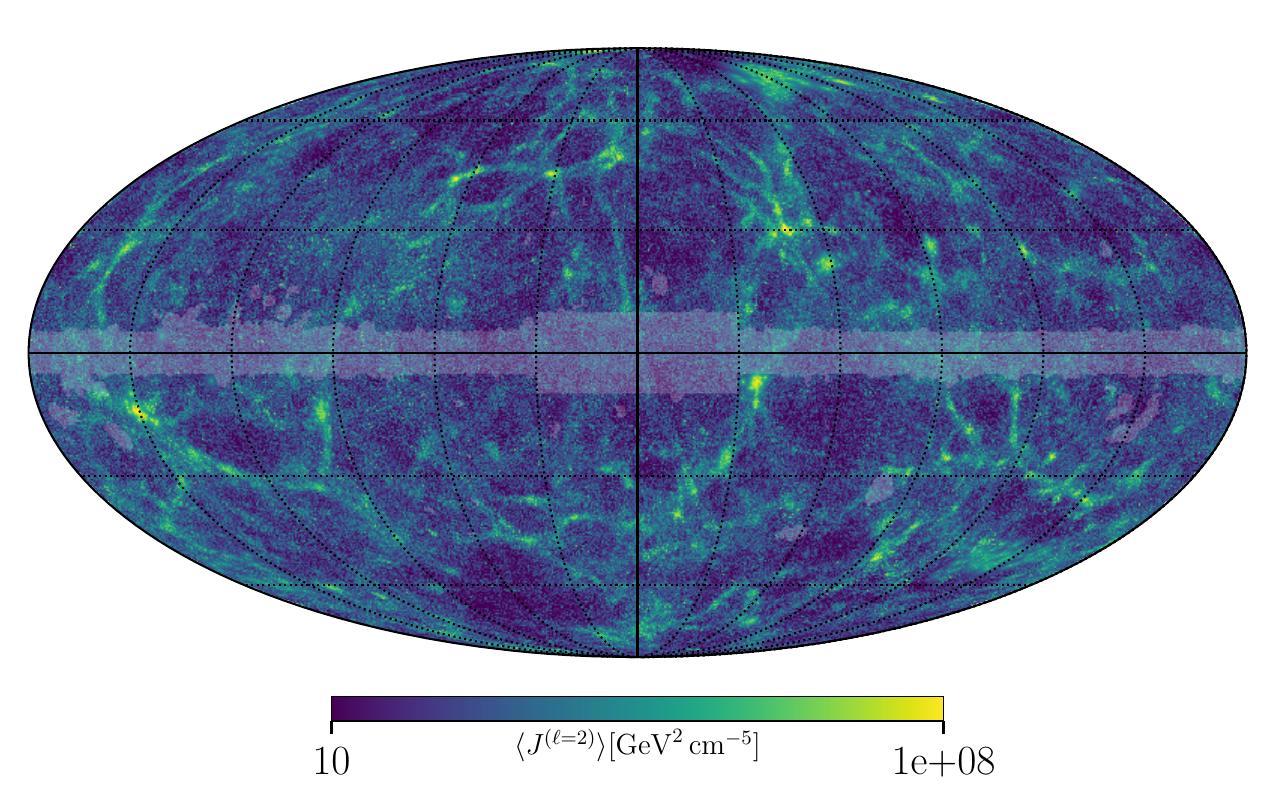}
\end{minipage}
\caption{
The ensemble mean over \csiborg{} realizations of the all-sky $J$-factor maps for our fiducial case of $p$- ($\ell=1$) and $d$-wave ($\ell=2$), assuming isotropic haloes. It is interesting to note that the $d$-wave all-sky map shows a more prominent peaked annihilation signal coming from the central halo regions. As in Figure 1 of \citet{FLAT_2022}, we plot also the completeness mask used in the \borg inference of the initial conditions \citep{Lavaux_2016, Jasche_Lavaux} (transparent gray). These maps were used in our likelihood model as described in \cref{subsec:likelihood_model}.
}
\label{fig:jd_ensemble}
\end{figure*}

\begin{figure}[h!]
    \centering
    \includegraphics[width=\columnwidth]{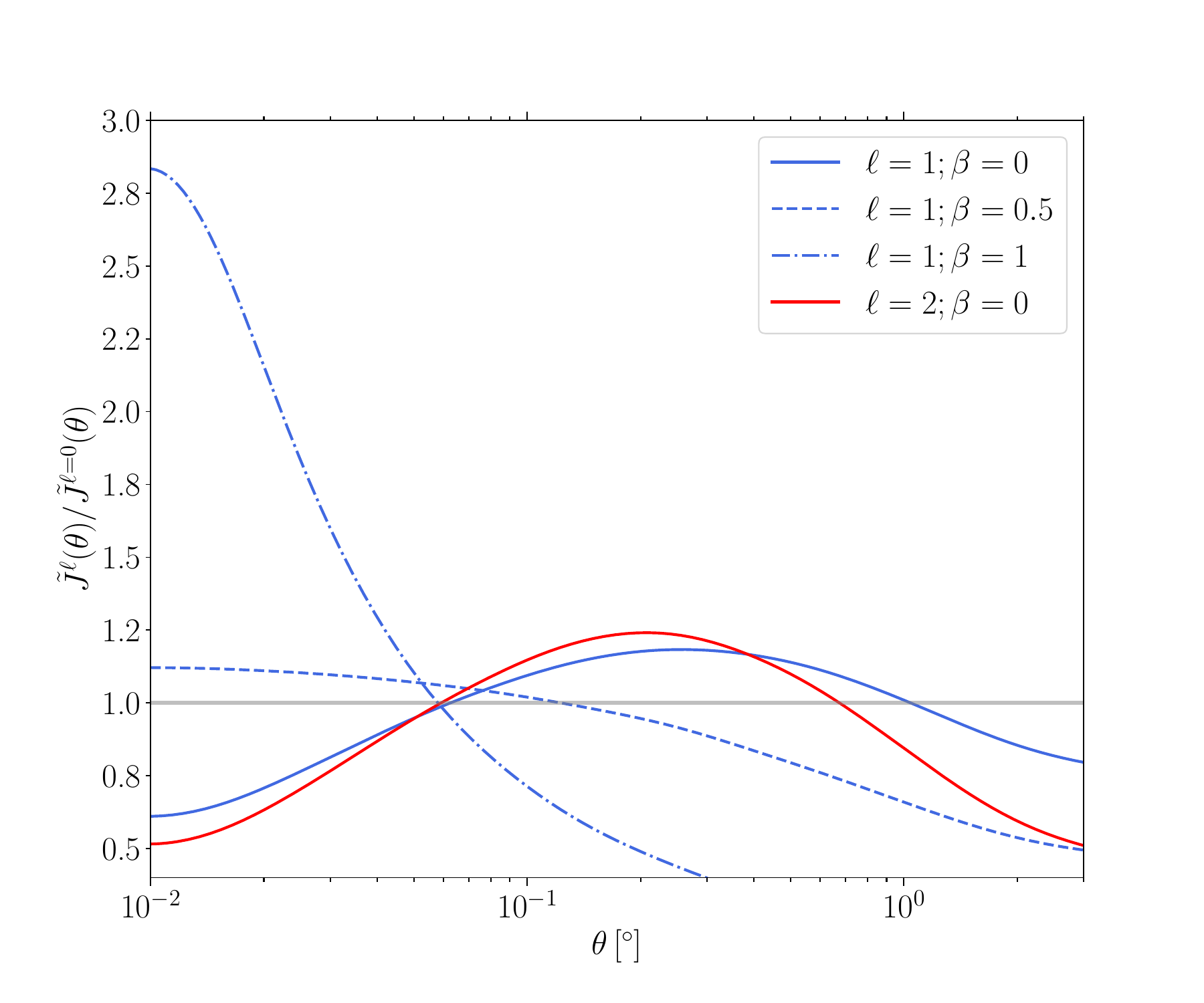}
    \caption{The angular distribution of the normalised $J$ factor, $\tilde{J}$ (Eq.~\ref{eq:Jnorm}), as a function of angle on the sky from the halo centre, $\theta$, for $p$-wave ($\ell=1$) and $d$-wave ($\ell=2$) annihilation channels. The results are shown relative to the $s$-wave ($\ell=0$) result and for different choices of anisotropy parameter, $\beta$. For this example, we used one of the most massive halos from the \csiborg{} 9844 realisation with $M = 1.2 \times 10^{15}M_{\odot}$ at a distance of $96 \, \mathrm{Mpc}$ from the observer. 
    }
    \label{fig:pwave_vs_dwave_Jtheta}
\end{figure}

For $p$-wave annihilation, the $J$ factor can be re-written as
\begin{equation}
    J^{(\ell=1)}_{\mu,p} 
    = 
    \int \dd \Omega \dd \mu
    \left( 
    \frac{
    \lvert
    \bm{v}_1 - \bm{v}_2
    \rvert}
    {c}
    \right)^2
    f(\bm{r}(\mu), \bm{v}_1)
    f(\bm{r}(\mu), \bm{v}_2)
\end{equation}        
which upon expanding the square and remembering that we are assuming velocities are defined with respect to the center of mass, i.e. the velocity distribution is zero-centered, the whole expression can then be simplified to (see \cite{boucher2022j})
\begin{equation}
    \label{eq:Jfactor pwave}
    \begin{split}
        J^{(\ell=1)}_{\mu,p}
        &=
        2 \int_{\mu,p}  \dd\Omega \dd \mu \rho^2 \left( \bm{r}(\mu) \right) \left< \frac{\bm{v}^2}{c^2} \right>(\bm{r}(\mu))\\ 
        &= \frac{2}{c^2} 
        \int_{\mu,p}  \dd\Omega \dd \mu \rho^2 \left( \bm{r}(\mu) \right) \left( \sigma_r^2(\bm{r}(\mu)) + 2 \sigma_\theta^2(\bm{r}(\mu)) \right),
    \end{split}
\end{equation}
where $\rho\left( \bm{r}(\mu) \right) $ is the dark matter density at the position $\mu$ along the line of sight and $\sigma_r$ and $\sigma_\theta$ are the corresponding radial and tangential velocity dispersion, respectively. The factor of 2 follows from the symmetry of the problem, i.e. that the two annihilating particles have the same phase-space distribution function. We keep the explicit dependence on the line of sight position in order to emphasize that the $J$ factor needs to be evaluated with respect to the observer. However, since the total $J$ factor within a pixel is a scalar quantity, one can also first evaluate it within the corresponding volume with respect to the halo centre and then make the appropriate projection. This is precisely how the \clumpy package -- which we utilise here --  works (see \cite{FLAT_2022, Charbonnier_2012, Bonnivard_2016, Hutten_2019}). Now, to obtain the dispersion for a given halo, we assume spherical symmetry, and therefore must solve the Jeans equation
\begin{equation}
    \frac{1}{\rho} \frac{\dd}{\dd r} \left( \rho \sigma_r^2 \right) + 2 \beta \frac{\sigma_{r}^2}{r} = - \frac{\dd \Phi}{\dd r},
\end{equation}
where we introduce an anisotropy parameter
\begin{equation}
    \beta \left( r \right) = 1 -  \frac{\sigma_\theta^2 \left( r \right)}{\sigma_r^2 \left( r \right)},
\end{equation}
and $\Phi$ is the gravitational potential of the halo.

We assume that all our halos have Navarro-Frenk-White (NFW) \citep{NFW_1997} density profiles
\begin{equation}
    \rho \left( r \right) = \frac{\rho_0}{(r/r_{\rm s}) \left( 1 + r / r_{\rm s} \right)^2},
    \label{eq:NFW_profile}
\end{equation}
with scale length $r_{\rm s}$, defined such that the concentration of the halo is $c = r_{\rm vir} / r_{\rm s}$ for virial radius $r_{\rm vir}$. In this work we use the mean mass-concentration relation of \citet{Prada12,Sanchez_Conde_2014} and
assume zero scatter in this relation. Note that the use of an external mass--concentration relation is needed because the halo finder we used \cite{PHEW} does not calculate concentration on-the-fly. However, given the consistency of mass--concentration relations derived from $N$-body simulations \citep[see e.g.][]{Dutton_2014, diemerkravtsov_2015}, the method for adding the concentration~\citep{Prada12,Sanchez_Conde_2014} is not an important source of systematic error. Since the $J$ factor traces the square of the density, one may expect that subhalos or clumps may provide a significant contribution to the $J$ factor and should also be included. Although this is true for $s$-wave annihilation, the small velocity dispersion of these clumps means that the overall signal is dominated by the smooth component and we therefore ignore unresolved substructure. This approximation has been validated by explicitly calculating the $J$ factor for halos from hydrodynamical simulations \citep{Piccirillo_2022}, although we note that this conclusion does not hold for the case of Sommerfeld enhanced annihilation (see for example \citep{Facchinetti2022, Lacroix_2022}). We do not consider this case in this paper and leave it for future work.

Defining the dimensionless radial coordinate $\tilde{r} \equiv r / r_{\rm vir}$, one finds that the radial velocity dispersion for circular orbits is (see Eq. (14) in \citep{Lokas_2001})
\begin{equation}
    \label{eq:Dispersion beta=0}
    \begin{split}
        \frac{\sigma_r^2}{V_{\rm vir}^2}(\tilde{r}) &\stackrel{\beta=0}{=} \frac{c^2 \tilde{r} \left( 1 + c\tilde{r} \right)^2 }{2 \left( \ln \left( 1 + c \right) - c / \left( 1 + c \right) \right)} \left[ \pi^2 + 6 \Li_2 \left( - c\tilde{r} \right)  \right. \\
        & \left. - \ln \left( c\tilde{r} \right) + \left( 1 + \frac{1}{c^2\tilde{r}}^2 - \frac{4}{c\tilde{r}} - \frac{2}{1+c\tilde{r}} \right) \times \ln \left( 1 + c\tilde{r} \right) \right. \\
        & \left. + 3 \ln^2 \left( 1 + c\tilde{r} \right) - \frac{1}{c\tilde{r}} -  \frac{6}{1 + c\tilde{r}}  - \frac{1}{\left( 1 + c\tilde{r} \right)^2} \right],
    \end{split}
\end{equation}
where $V_{\rm vir}$ is the virial velocity,
and $\Li_2$ is the dilogarithm function. Similarly, for the extreme cases of a constant anisotropy parameter $\beta=0.5$ and $\beta=1.0$ one finds (see Eq. (15) and (16) from \citep{Lokas_2001})
\begin{equation}
    \label{eq:Dispersion beta=0.5 and beta=1.0}
    \begin{split}
        \frac{\sigma_r^2}{V_{\rm vir}^2}(\tilde{r}) \stackrel{\beta=0.5}{=} 
        &\frac{c \left( 1 + c\tilde{r} \right)^2 }{\left( \ln \left( 1 + c \right) - c / \left( 1 + c \right) \right)} 
        \left[
        -\frac{\pi^2}{3} 
        -
        2\Li_2(-c\tilde{r})\right. \\
        & +
        \frac{2}{1+c\tilde{r}}
        +
        \frac{\ln(1+c\tilde{r})}{c\tilde{r}}
        +
        \frac{\ln(1+c\tilde{r})}{1+c\tilde{r}} \\
        &\left. -
        \ln^2(1+c\tilde{r})
        -
        \frac{1}{2(1 + c\tilde{r})^2} 
        \right],\\
        \frac{\sigma_r^2}{V_{\rm vir}^2}(\tilde{r}) \stackrel{\beta=1}{=}
        &\frac{\left( 1 + c\tilde{r} \right)^2 }{\tilde{r}\left( \ln \left( 1 + c \right) - c / \left( 1 + c \right) \right)} 
        \left[
        \frac{\pi^2}{6} + \Li_2(-c\tilde{r}) \right.\\
        &\left.
        -\frac{1}{1+c\tilde{r}} - \frac{\ln(1+c\tilde{r})}{1+c\tilde{r}}
        +\frac{\ln^2(1+c\tilde{r})}{2}
        \right].
    \end{split}
\end{equation}

In \cref{fig:pwave_vs_dwave_Jtheta} we show how these different velocity dispersion profiles affect the angular dependence of the normalised $J$ factor (labelled as $\tilde{J}$)
for a halo of mass $M = 1.2 \times 10^{15}M_{\odot}$ (one of the most massive halos in \csiborg{} simulation 9844) at a distance of $\sim 100 \, \mathrm{Mpc}$ from the observer. The quantity $\tilde{J}$ we calculate as
\begin{equation}\label{eq:Jnorm}
    \tilde{J}^{\ell}(\theta) = \frac{J^\ell(\theta)}{\int \dd \Omega J^{\ell}(\theta)},
\end{equation}
where $\theta$ is the angular distance from the halo centre as seen on the sky. We perform this calculation from close to the halo centre, $\theta \sim {0.01}^{\circ}$, all the way up to the two times the virial radius of the halo. Similar results for the case of the Milky Way halo (modelled as a NFW profile) were obtained through the Eddington inversion method in \citep{Boddy_2018}. We show the results for $p$-wave with different values of the anisotropy parameter $\beta$ (\cref{eq:Dispersion beta=0,eq:Dispersion beta=0.5 and beta=1.0}) and $d$-wave calculations. The $\beta=0.5$ (dashed) and $\beta=1.0$ (dotted-dashed) cases are more strongly peaked at the centre than the corresponding $\beta=0$ case, reflecting the behaviour of the velocity dispersion profiles (see also Fig.~1 in \citep{Lokas_2001}). Furthermore, for $\beta=1.0$, $\tilde{J}$ diverges at the centre and hence needs to be regularised (see \cref{eq:core_radius}). The $d$-wave $\tilde{J}$ shows qualitatively similar behaviour as the corresponding $p$-wave case (thick line). For more details on the $d$-wave calculation, see \cref{sec:Jfactor_dwave} and \cref{app:Jfactor_dwave}.

It should be noted that, for a more realistic scenario, the anisotropy parameter should be taken as a radially varying function (see \cite{osipkov1979spherical, merritt1985spherical, gerhard1991new}). We do not consider these models here, given that the three extreme scenarios for the anisotropy parameter we pick should bracket reality. In our fiducial analysis we choose $\beta = 0$, however, and show in \cref{sec:systematics_J} that this is a conservative choice. The corresponding $J$ factor calculation then simply consists of substituting \cref{eq:Dispersion beta=0} into \cref{eq:Jfactor pwave}, noting that in this case $\sigma_{\theta} = \sigma_r$, and integrating the final expression for all our halos. We do this numerically using the \clumpy package at a \healpix resolution of $\nside=2048$, which is then subsequently degraded to the coarser resolution ($\nside=256$) at which we perform the inference. This allows a more faithful calculation of the $J$ factor than performing the calculation at $\nside=256$, since the dependence of $J$ on $\rho$ is non-linear.

\subsubsection{\texorpdfstring{$J$}{J} factor for \texorpdfstring{$d$}{d}-wave annihilation}
\label{sec:Jfactor_dwave}

For $d$-wave annihilation the corresponding expression for the $J$ factor takes the form (see also \citep{boucher2022j})
\begin{equation}
    \label{eq:Jfactor dwave}
    J^{(\ell=2)}_{\mu,p} 
    = 
    \int_{\mu,p} 
    \dd\Omega \dd \mu 
    \rho^2 (\bm{r}) 
    \left(
    2
    \left< 
    \frac{\bm{v}^4}{c^4} \right>(\bm{r})
    +
    \frac{10}{3}
    \left<
    \frac{\bm{v}^2}{c^2}
    \right>^2(\bm{r})
    \right).
\end{equation}

We see that the second term appearing can be computed from a power of the velocity dispersion (as in $p$-wave; see \cref{eq:Jfactor pwave}), but the first term requires the evaluation of the fourth moment of the velocity distribution. In general, this would require for solving the Boltzmann hierarchy up to the fourth moment. It will not be possible to do so, however, since this system of equations is not closed for a general self-gravitating stellar system. In order to solve the Jeans system of differential equations, one must impose additional constraints. This can be done by requiring the stellar system has certain symmetries. One of the possibilities is to assume an ergodic distribution function ($\beta=0$, see Section 4 - Box 4.3 of \citep{binney2011galactic}). This then allows to write down the following equation, relating $\left<v_r^2\right>$ and $\left<v_r^4\right>$
\begin{equation}
    \label{eq:4thvel_moment}
    \frac{\dd }{\dd r}
    \left(\rho \left<v_r^4\right>
    \right)
    \stackrel{\beta=0}{=}
    -3 \rho \left< v_r^2 \right>
    \frac{\dd \Phi}{\dd r},
\end{equation}
where by $v_r$ we have denoted the projection of the velocity to the radial direction (fixing the coordinate system to the halo centre of mass). In addition to calculating $\left<v_r^4\right>$, we need to also calculate the other projected moments as well as the cross-terms in order to obtain the total fourth moment of the velocity, as is necessary for evaluating the \cref{eq:Jfactor dwave}. This is straightforward, due to the ergodicity of the distribution function, as shown in \cref{app:Jfactor_dwave}. The final expression for the fourth velocity moment is then given by
\begin{equation}
    \label{eq:4thvel_momen_final}
    \left<\bm{v}^4\right>
    \stackrel{\beta=0}{=}
    5 \left<v_r^4\right>.
\end{equation}
This, and the solution for $\left<\bm{v}^2\right>$ from \cref{eq:Dispersion beta=0}, are then enough to evaluate the total $J$ factor for $d$-wave annihilation. See \cref{app:Jfactor_dwave} for more details.

Our analytical expressions for the velocity moments needed to evaluate $p$- and $d$-wave $J$ factors have been obtained under the assumption of ergodicity of the phase-space distribution function but without any assumption on its shape, e.g. the extent to which it is Maxwellian. One could instead extract the moments from the velocity distribution function empirically derived from simulations (see e.g. \cite{maohalo2halo_2013}). However, as can be seen from fig. 2 of that paper, this empirical velocity distribution function lies in between the analytical models we consider here, and thus our model should bracket reality. Similarly, \citep{christyDM_velocitydist_2023} demonstrates that the analytical prediction for the velocity distribution within the scale radius of the halo matches well the (dark matter-only) numerical prediction. In particular, \citet{christyDM_velocitydist_2023} argue that the analytical prediction is especially useful for probing the high-end of the velocity distribution, which is of particular importance for $p$- and $d$-wave annihilation channels but severely impacted by the finite resolution of the simulations. Our use of the Jeans equation makes us much less sensitive to resolution effects.

\subsection{Gamma ray data}

We compare the dark matter annihilation templates to the gamma ray observations from \textit{Fermi} Large Area Telescope between mission weeks 9 and 634. We select photons in the upper quartile of angular resolution (PS3) and event class SOURCEVETO in the energy range $500{\rm \, MeV} - 50 {\rm \, GeV}$. The maximum zenith angle is chosen to be $90^\circ$. These are binned spatially onto \healpix \citep{Zonca_2019,Gorski_2005} maps with resolution $\nside = 256$ and into 9 logarithmically spaced energy bins. The processing of the photon files is performed using the Fermi Tools\footnote{\url{https://fermi.gsfc.nasa.gov/ssc/data/analysis/software/}} and \fermipy \citep{FermiPy}. We also use this software to convolve our templates with the \textit{Fermi} point spread function and to obtain the exposure maps. To reduce the impact of galactic emission and the \ac{GCE}, we mask the region with galactic latitude $|\lambda|<30^\circ$.

\begin{figure*}
    \centering
    \includegraphics[width=\textwidth]{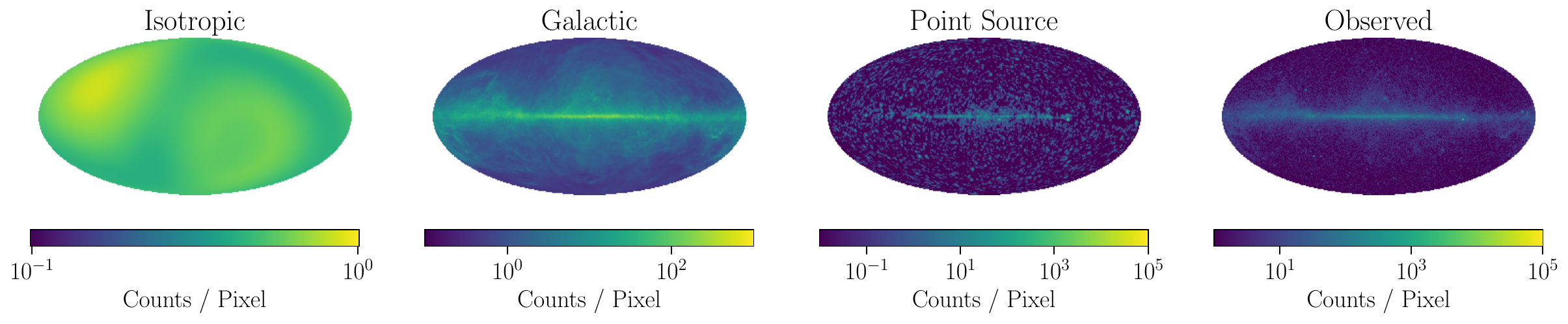}
    \caption{Astrophysical templates used to model the non-dark-matter contribution to the observed gamma ray flux, which is given in the final column. These templates are for the energy range $0.8-1.4{\rm \, GeV}$ and are shown after convolution with the point spread function and exposure maps, hence the `isotropic' contribution appears anisotropic.}
    \label{fig:templates_observed}
\end{figure*}

Although some of the photons detected by \textit{Fermi} could have been produced by dark matter annihilation, there are other sources which will also contribute, and thus it is important to incorporate these in the analysis. As in our previous work \citep{FLAT_2022}, we consider three contributions: point sources (psc), an isotropic background (iso) and the Milky Way (gal). Each contribution is modelled with a fixed spatial template for energy bin $i$, $\{T_{i}^t \left(\hat{r}\right)  :  t \in \{ {\rm iso, \, gal, \, psc} \} \}$, where we assign a different normalisation, $A_i^t$, for each energy bin and template. We jointly infer these parameters with the amplitude of the $J$ factor template. The point source template is comprised of all extended and point sources in the Large Area Telescope 12-year Source Catalog (4FGL-DR3)\footnote{\url{https://heasarc.gsfc.nasa.gov/W3Browse/fermi/fermilpsc.html}} and the galactic component is modelled using the templates described in \citep{Acero_2016}. The scalings $\{A_i^t\}$ are defined such that they all equal unity if the \textit{Fermi} data is perfectly described by these templates, however jointly inferring these parameters will allow us to capture any imperfections in the spectral modelling. We thus require an initial guess for the isotropic component's spectral shape, for which we use the \textit{Fermi} Isotropic Spectral Template\footnote{\url{https://fermi.gsfc.nasa.gov/ssc/data/access/lat/BackgroundModels.html}}. All templates are convolved with the \textit{Fermi} point spread function using the Fermi Tools.
We plot these templates after this convolution and after accounting for the varying exposure alongside the observed data in \cref{fig:templates_observed} for energies $0.8-1.4{\rm \, GeV}$.

\subsection{Likelihood model}
\label{subsec:likelihood_model}

As in \cite{FLAT_2022}, we constrain the dark matter annihilation cross-section in two steps. First, we do not assume a functional form for the spectrum, but infer the normalisation of each template in each energy bin for each \csiborg{} simulation separately. We then fit this inferred spectrum to a functional form in the second step, which allows us to constrain $a_\ell$.

For pixel $p$ and \csiborg{} simulation $j$, the mean number of counts in energy bin $i$ is predicted to be 
\begin{equation}
    \lambda_{ipj} = \mathcal{F}_{ip} \times \left(A_i^{\rm J} J_{ipj} 
    + \sum_t A_i^t T_{ip}^t \right),
\end{equation}
where the $\mathcal{F}_{ip}$ is the \textit{Fermi} exposure, and we have defined
\begin{equation}
    J_{ipj} \equiv \frac{J_{p j}}{J_0} \Delta E_i, \quad 
    T_{ip}^t \equiv \int_p T_{i}^t \left(\hat{r}\right) \dd \Omega,
\end{equation}
where $J_0 = 10^{13}  {\rm \, GeV^2\, cm^{-5}}$ sets the units. A depiction of the $J$-factor templates is available in \cref{fig:jd_ensemble}. Next, we assume that the photon counts are Poisson distributed, such that the likelihood for observing $n_{ip}$ counts in pixel $p$ and energy bin $i$ given the mean $\lambda_{ipj}$ is
\begin{equation}
    \label{eq:Poisson likelihood}
    \mathcal{L} \left( n_{ip} | \lambda_{ipj} \right) = \frac{\lambda_{ipj}^{n_{ip}} \exp\left(- \lambda_{ipj}\right)}{n_{ip} ! }.
\end{equation}
Given this likelihood, we then apply Bayes' identity
\begin{multline}
    \label{eq:Bayes theorem}
        \mathcal{P} \left(A_i^{\rm J}, \{A_i^t\}, j| \mathcal{D}_i \right) \\
        = \frac{\mathcal{L} \left(\mathcal{D}_i | A_i^{\rm J}, \{A_i^t\}, j \right) P \left( A_i^{\rm J} \right)   P\left(\{A_i^t\}\right) P \left(j \right) }{\mathcal{Z}\left(\mathcal{D}_i\right)},
\end{multline}
to obtain the posterior distribution for the model parameters, $\mathcal{P}$, where $\mathcal{Z}$ is the evidence. For our priors, $P$, we enforce the improper, uniform priors $a_\ell, A_i^{\rm J} \geq 0$, and a uniform prior on our non-dark matter templates, $A_i^{t}$, between 0.5 and 1.5, where $A_i^{t} \in \{A_i^\text{gal}, A_i^\text{iso}, A_i^\text{psc}\}$. We run the No U-Turn Sampler implemented in \numpyro \citep{phan2019composable,bingham2019pyro} for 5000 steps after an initial 1000 warmup steps. We find this typically leads to at least $\mathcal{O}\left(2000\right)$ effective samples in all parameters, and Gelman-Rubin statistics consistent with unity within $10^{-2}$.

We then marginalise over all other parameters and average over the $N_\text{sim}=101$ \csiborg{} realisations, assigning them equal weights, to obtain the one-dimensional posterior for the annihilation flux in a given energy bin
\begin{equation}
    \label{eq:spectrum likelihood}
        \mathcal{P} \left(A_i^{\rm J}| \mathcal{D}_i \right) \\
        = \frac{1}{N_{\rm sim}} \sum_j \int \dd \{A_i^t\}
        \mathcal{P} \left(A_i^{\rm J}, \{A_i^t\}, j| \mathcal{D}_i \right).
\end{equation}
Now, given some model $f_i(\bm{\theta})$ describing the predicted flux in each energy bin, we wish to find the model parameters, $\bm{\theta}$. Concretely, we will choose some annihilation channel and particle mass, and wish to find the corresponding $\bm{\theta}=a_\ell$. For $f_i(\bm{\theta})$, we use the pre-computed spectra calculated by \citet{Jeltema_2008} and provided by the \textit{Fermi} collaboration\footnote{\url{https://fermi.gsfc.nasa.gov/ssc/data/analysis/scitools/source_models.html}} which capture the energy spectra of intermediately produced standard model particles and their final products. 
As our model is deterministic,
\begin{equation}
    \mathcal{L} \left(A_i^{\rm J}| \bm{\theta} \right) = \delta \left( A_i^{\rm J} - f_i \left(\bm{\theta}\right) \right),
\end{equation}
one can then obtain the likelihood of the observed gamma-ray sky as 
\begin{align}
        \mathcal{L} \left(\mathcal{D}_i | \bm{\theta}\right) &=
    \int \dd A_i^{\rm J} \, \mathcal{L} \left(\mathcal{D}_i | A_i^{\rm J}\right) \mathcal{L} \left(A_i^{\rm J} | \bm{\theta} \right) \nonumber \\
    &= \int \dd A_i^{\rm J} \, \frac{\mathcal{P} \left(A_i^{\rm J} | \mathcal{D}_i \right) \mathcal{Z}\left(\mathcal{D}_i\right)}{P \left(A_i^{\rm J} \right)} \delta \left( A_i^{\rm J} - f_i \left(\bm{\theta}\right) \right).
\end{align}
Assuming each energy bin is independent, one can then use Bayes' theorem to find the posterior distribution for the model parameters to be
\begin{equation}
    \mathcal{P} \left(\bm{\theta}| \mathcal{D}_i \right) \propto P(\bm{\theta}) \prod_i \mathcal{L} \left(\mathcal{D}_i | \bm{\theta} \right),
\end{equation}
where $P(\bm{\theta})$ is the prior on the model parameters.

\section{Results}
\label{sec:Results}

\begin{figure}
    \centering
    \includegraphics[width=\columnwidth]{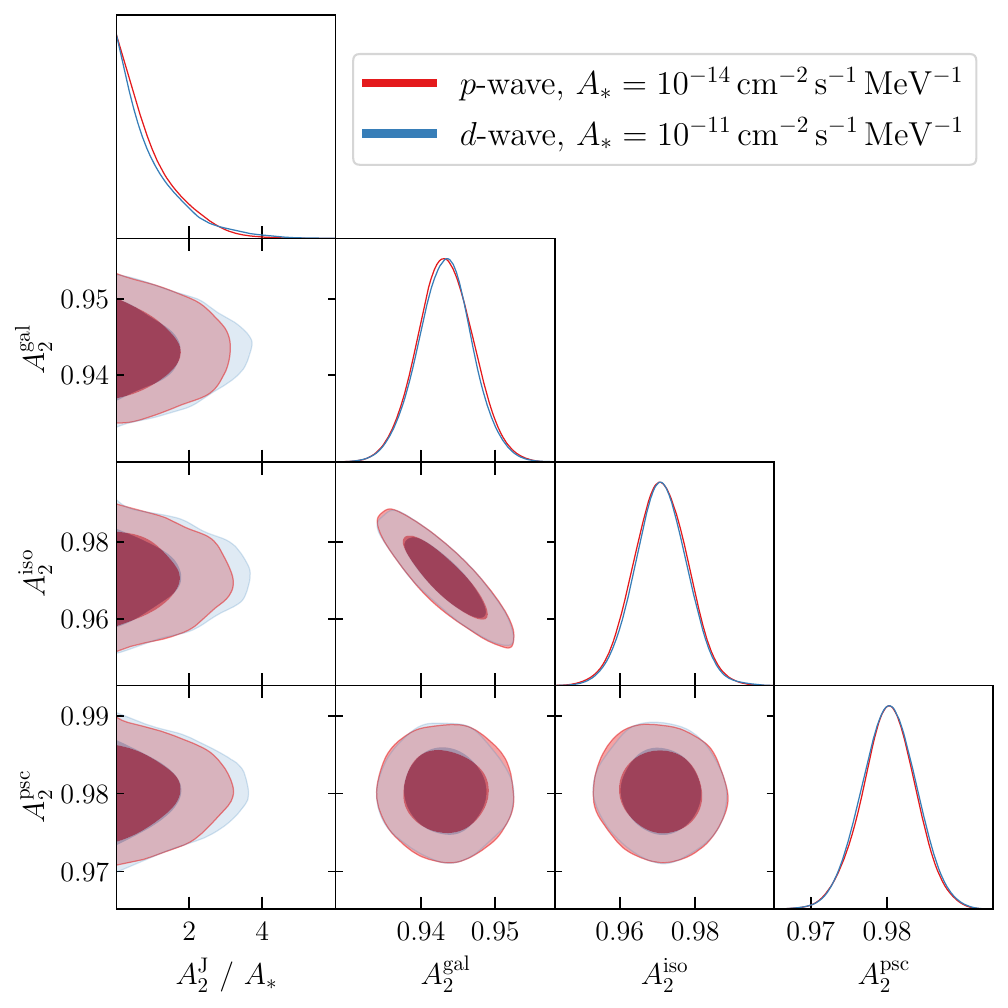}
    \caption{Posterior distributions of the parameters determining the gamma-ray flux in the energy range $0.8-1.4{\rm \, GeV}$ for \csiborg{} simulation number 9844. The contours show the $1$ and $2\sigma$ confidence intervals. We fit the amplitudes of the galactic diffuse ($A_2^{\rm gal}$), isotropic ($A_2^{\rm iso}$), and point source ($A_2^{\rm psc}$) templates, as well as a contribution from dark matter annihilation, $A_2^{\rm J}$. We show results for both $p$- and $d$-wave annihilation in this plot, although note that these amplitudes are normalised differently by a factor of 
    $10^3$.}
    \label{fig:corner}
\end{figure}

\begin{figure}
    \centering
    \includegraphics[width=\columnwidth]{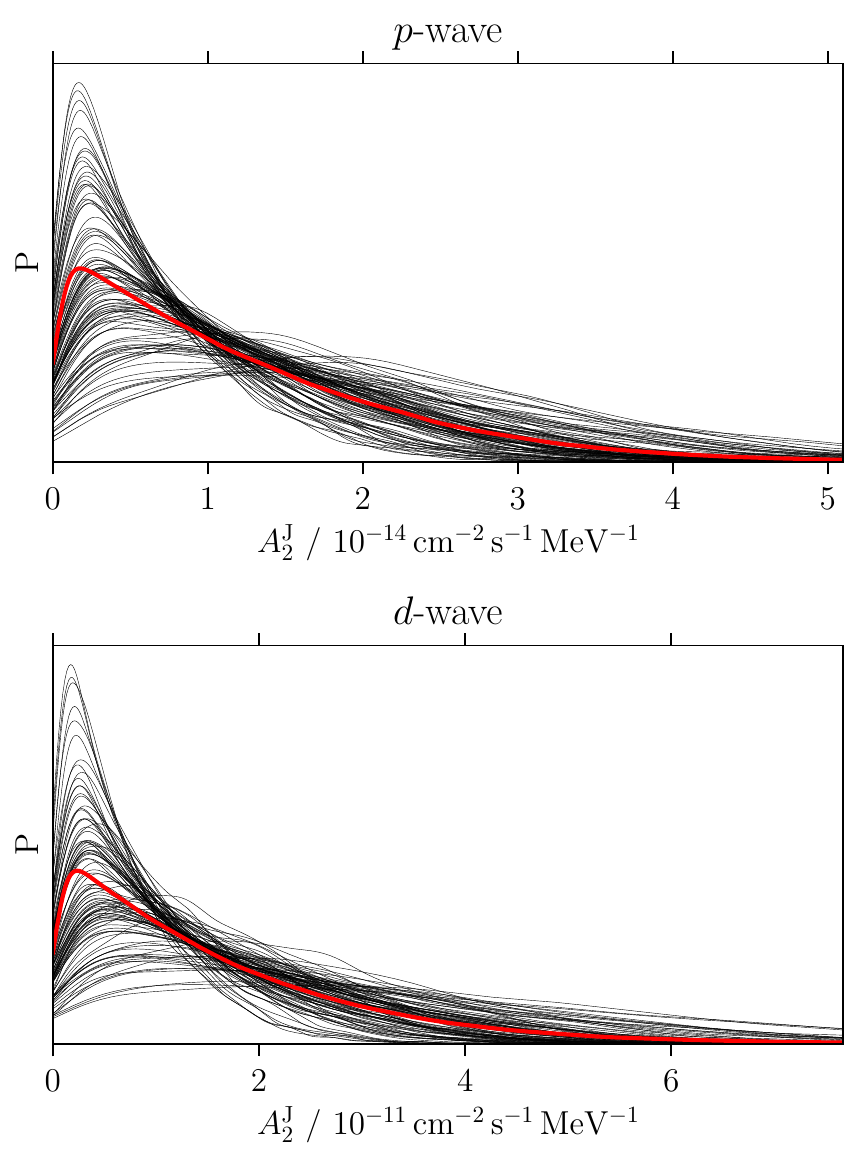}
    \caption{One dimensional posterior distributions of the coefficients describing the flux proportional to the $J$ factor for $p$-wave (upper) and $d$-wave (lower) annihilation in the energy range $0.8-1.4{\rm \, GeV}$. Each black line is the distribution for a given \csiborg{} simulation, and the red line gives the mean of these, which is the final posterior distribution used to constrain $a_\ell$.
    }
    \label{fig:AJ_posteriors}
\end{figure}

\begin{figure*}
    \centering
    \includegraphics[width=\textwidth]{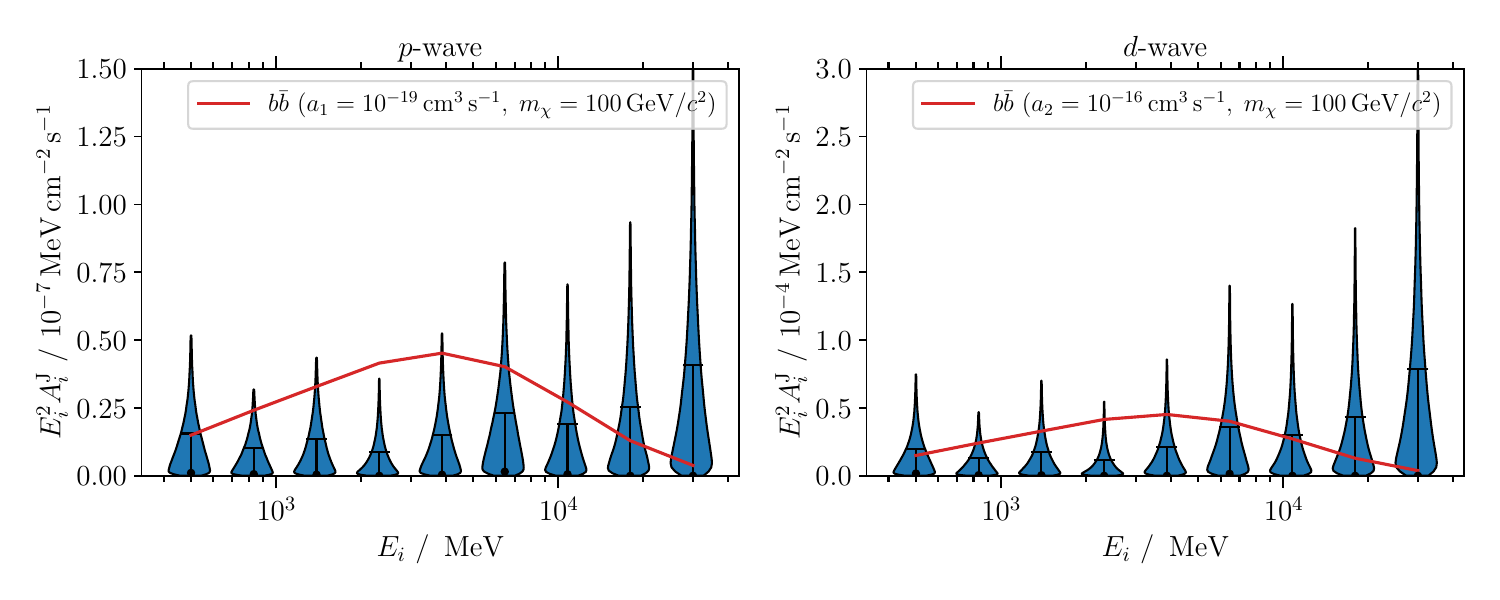}
      \caption{Posterior distributions of the amplitudes of the $J$ factor templates for each energy bin, $i$, for $p$-wave (left) and $d$-wave (right) annihilation. The black points are the maximum posterior points and the error bars indicate the $1\sigma$ confidence intervals. For reference we also plot the expected $A_i^{\rm J}$ for dark matter annihilation via the $b\bar{b}$ channel for dark matter particles of mass $m_\chi=100{\rm \, GeV}/c^2$, with $a_1 = 10^{-19} {\rm \, cm^3 s^{-1}}$ ($p$-wave) and $a_2 = 10^{-16} {\rm \, cm^3 s^{-1}}$ ($d$-wave). All fluxes are consistent with zero, indicating no detection of dark matter $p$-wave or $d$-wave annihilation.}
    \label{fig:AJ_violin}
\end{figure*}

\begin{figure}
    \centering
    \includegraphics[width=\columnwidth]{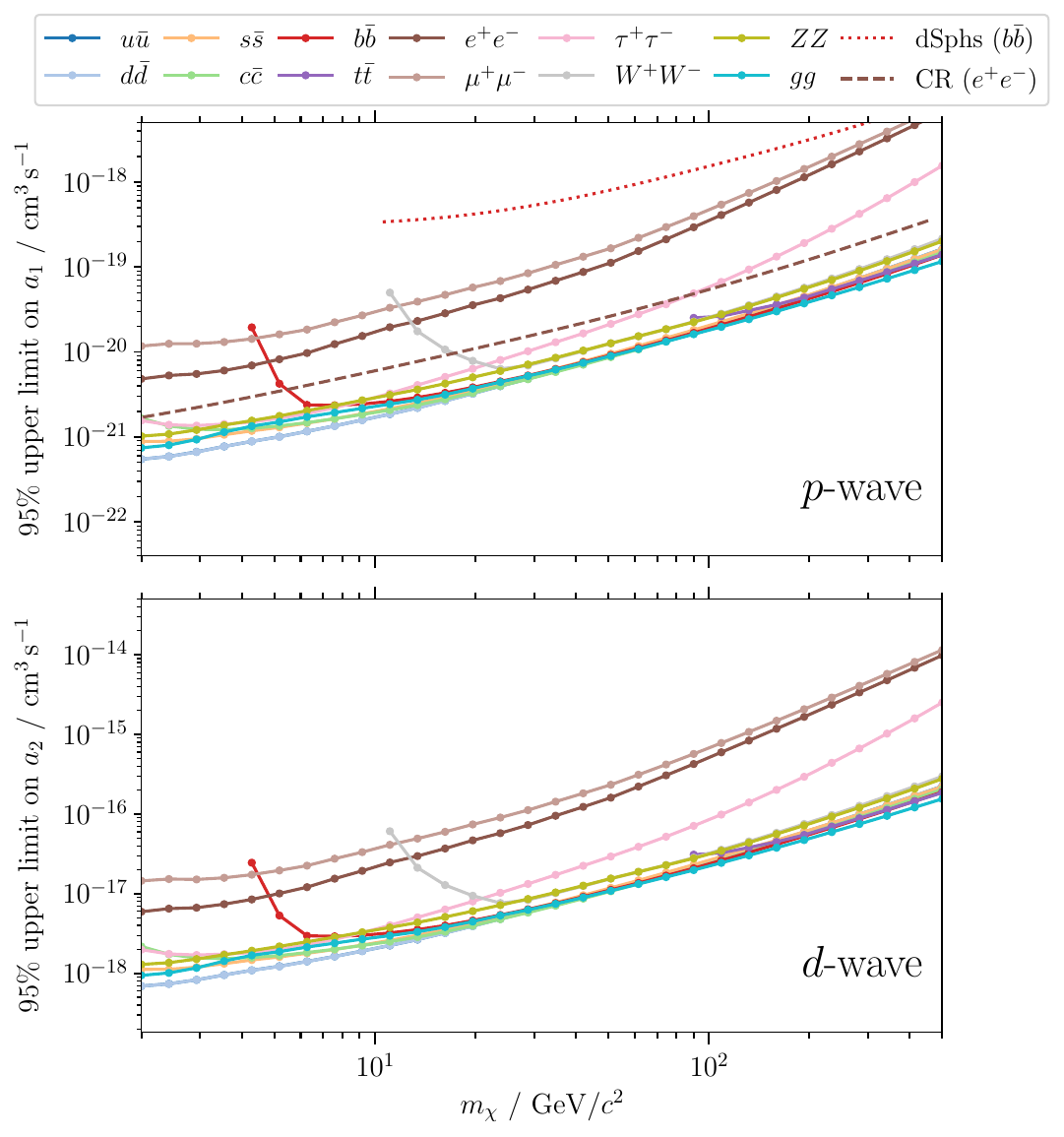}
    \caption{95\% upper limit on the dark matter $p$-wave (upper, $\ell=1$) and $d$-wave (lower, $\ell=2$) annihilation cross-section parameter, $a_\ell$, where the annihilation cross-section is $\sigma v = a_\ell (v/c)^{2\ell}$. The solid lines and points are from this work. Each constraint is derived at a fixed dark matter particle mass $m_\chi$ as indicated on the $x$-axis. The red dotted line is a constraint obtained from dwarf spheroidal (dSph) satellites of the Milky Way \citep{Zhao_2016,Boddy_2020} for the $b\bar{b}$ channel. We find that our constraints are approximately two orders of magnitude tighter than those from dSphs due to the larger velocity dispersions in massive extragalactic halos. The dSph constraint for $d$-wave is off the top of the plot. The brown dashed line is the constraint for the $e^+e^-$ channel, directly detecting the resulting electrons and positrons, assuming they are produced by the galactic halo \citep{Boudaud_2019}. This is slightly tighter than our $e^+e^-$ constraint.}
    \label{fig:constraints}
\end{figure}

\begin{figure}
    \centering
    \includegraphics[width=\columnwidth]{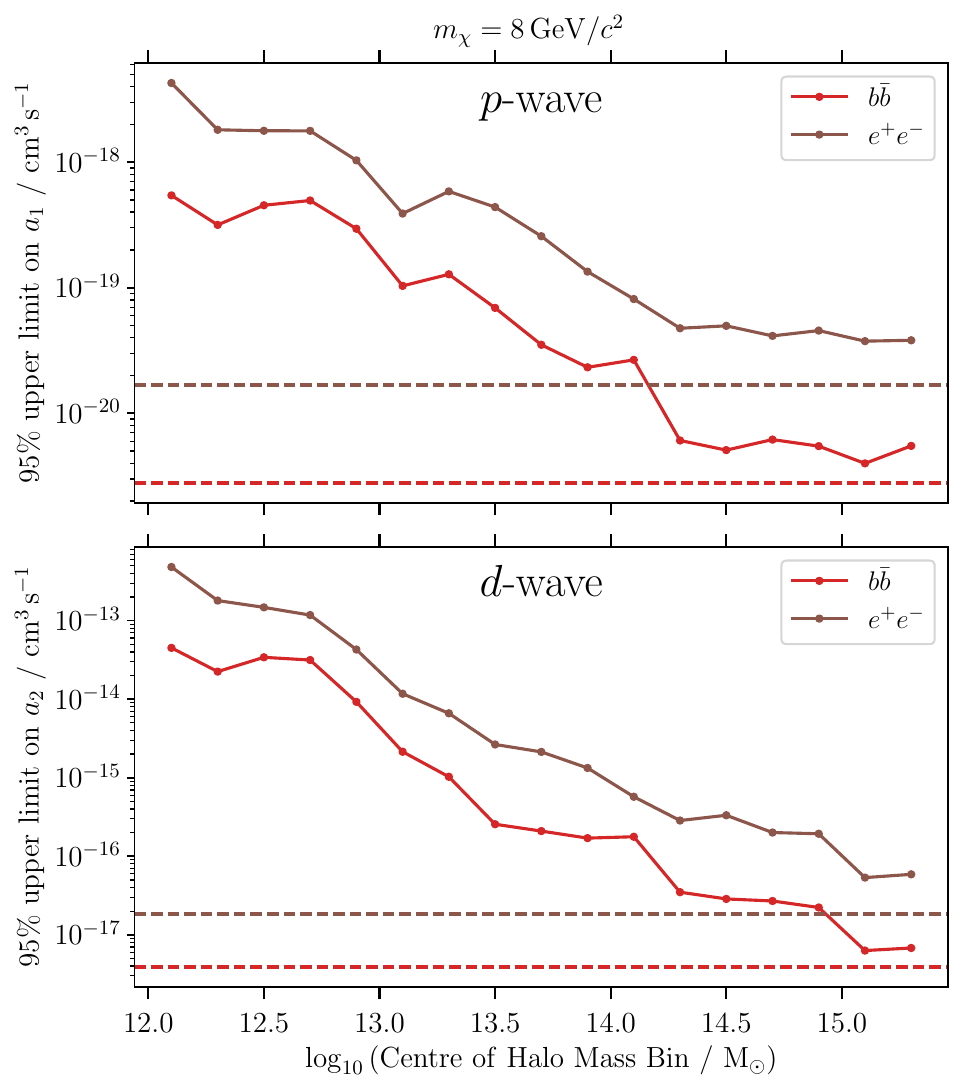}
    \caption{Constraints on $p$-wave (upper, $\ell=1$) and $d$-wave (lower, $\ell=2$) annihilation cross-section, $\sigma v = a_\ell \left(v/c\right)^{2\ell}$, from halos in a given mass range from a single \csiborg{} simulation (9844). Each mass bin has a width $\Delta \log_{10} \left( M_{\rm h} / 
    \Msun \right) = 1$, and we use a sliding window. We plot the 95\% upper limits on $a_\ell$ for the $b\bar{b}$ and $e^+e^-$ channels; the dashed horizontal lines are the constraints obtained from all halos together. We see that our constraints are driven by the most massive halos and that this is more important for $d$-wave annihilation due to the stronger dependence of annihilation rate on the relative velocity between dark matter particles.}
    \label{fig:mass_dependence_constraints}
\end{figure}

The first step of our analysis infers the amplitudes of the $J$ factor templates, $A_i^{\rm J}$, alongside the amplitudes of the isotropic, galactic and point source templates. We do this separately for each energy bin, $i$, and each \csiborg{} simulation. In \cref{fig:corner} we show a corner plot from one of these runs, for the energy range $0.8-1.4{\rm \, GeV}$ and \csiborg{} simulation number 9844. In this figure we show the results from both our $p$- and $d$-wave analyses, although we note that these are run separately. There is very little degeneracy between the normalisation of the $J$-factor templates and those giving the astrophysical contributions, and we note that the posteriors on the latter are almost independent of the type of annihilation studied. As in \citet{FLAT_2022}, we find that the astrophysical amplitudes are less than one and that there is a strong degeneracy between the galactic and isotropic normalisation. We attribute the former to the region over which the fits are performed; the templates are calibrated over $10^\circ < \left| \lambda \right| < 60^\circ$, whereas we use $\left| \lambda \right| > 30^\circ$.

Once we have run this first stage for each \csiborg{} simulation, we combine these to marginalise over the uncertainty in the density field reconstruction and the small-scale unconstrained modes of the simulation. In \cref{fig:AJ_posteriors}, we plot (in black) the posterior on $A_2^{\rm J}$ (corresponding again to the energy range $0.8-1.4{\rm \, GeV}$) using each of the \csiborg{} simulations.  The mean of these is given in red, which is the posterior distribution of $A_2^{\rm J}$ once we have marginalised over the reconstruction uncertainty.

At this stage of the inference we have eighteen such posteriors: for each energy bin we have one for $p$-wave and one for $d$-wave annihilation. These distributions are plotted in \cref{fig:AJ_violin}, and we see that these coefficients are consistent with zero at $1\sigma$ confidence for all energy bins for both $p$- and $d$-wave annihilation.
In the second step of the inference we fit this inferred spectrum to a dark matter annihilation model and hence constrain $a_\ell$. For reference, in \cref{fig:AJ_violin} we plot example spectra for annihilation via the $b\bar{b}$ channel for $m_\chi = 100 {\rm \, GeV}/c^2$ at a fiducial $a_\ell$.

In \cref{fig:constraints} we convert these posteriors to 95\% upper limits on the $p$- and $d$-wave self-annihilation cross-section ($\sigma v=a_\ell (v/c)^{2\ell}$) as a function of dark matter particle mass, $m_\chi$, for the production of any standard-model quark, charged lepton or gauge boson (except photons). We do not attempt to constrain $m_\chi$, and instead find constraints on $a_\ell$ conditioned on this variable for the range $m_\chi \in [2, 500] {\rm \, GeV}/c^2$.
In all cases we find no evidence for $p$-wave or $d$-wave dark matter annihilation, as all the fluxes in \cref{fig:AJ_violin} are consistent with zero. For dark matter annihilating to $b\bar{b}$, if $m_\chi = 10 {\rm \, GeV}/c^2$, we constrain $a_{1} < 2.4 \times 10^{-21} {\rm \, cm^3 \, s^{-1}}$ and $a_{2} < 3.0 \times 10^{-18} {\rm \, cm^3 \, s^{-1}}$ at 95\% confidence. For an annihilation which forms lighter quarks, we obtain practically identical constraints, whereas lepton production yields constraints which are approximately an order of magnitude less stringent. For large $m_\chi \gtrsim 100 {\rm \, GeV}/c^2$, our upper limits for the production of top quarks or $W$ or $Z$ bosons are similar to those for lighter quarks, but these processes cannot occur at lower $m_\chi$ so we do not obtain constraints for smaller dark matter particle masses.

Note that, in \cref{fig:constraints}, the constraints weaken as $m_\chi$ approaches $m_A$, if the annihilation product is $A\bar{A}$. Near this value, the process is strongly suppressed and hence the constraints no longer lie within the range of the plot, so we do not plot these points. 

For comparison, for $p$-wave, in \cref{fig:constraints} we also plot the constraints one obtains for the $b\bar{b}$ channel from dwarf spheroidals. As anticipated from the much smaller velocity dispersion in dwarf galaxies, our constraints are significantly tighter, by approximately two orders of magnitude for $p$-wave annihilation, and seven orders of magnitude for $d$-wave. This demonstrates the increased importance of large halo masses for velocity-dependent annihilation, whereas for $s$-wave annihilation the close proximity of these objects more than compensates for their small masses, leading to tighter constraints \citep[see e.g. Figure 7 of][]{FLAT_2022}. We perform a more detailed comparison to the literature in \Cref{sec:literature}. 

\section{Discussion}
\label{sec:Discussion}

\subsection{Understanding the results}

To determine which halos are responsible for our constraints, we produce $J$ factor maps for halos only within a given mass range and run the inference using only those halos. For computational convenience, we perform this on one representative \csiborg{} simulation only (9844). The results of this inference are plotted in \cref{fig:mass_dependence_constraints}, where we use moving mass bins of width $\Delta \log_{10} \left( M_{\rm h} / \Msun \right) = 1$ and consider dark matter particles of mass $m_\chi= 8 {\rm \, GeV}/c^2$ annihilating via the $b\bar{b}$ or $e^+e^-$ channels. We see that our constraints are dominated by the most massive objects in our simulations, i.e. halos of mass $M_{\rm h} \sim 10^{14-16} \,\Msun$. This is due to both the larger densities and velocity dispersions in massive objects, as was predicted by \citet{Baxter_2022} (see their Fig.~3). If we continued \cref{fig:mass_dependence_constraints} to larger $M_{\rm h}$ we would see that the constraints would begin to weaken, due to the lack of very massive objects in our simulation. We observe that the constraining power as a function of mass for $p$-wave annihilation plateaus at an earlier halo mass than for $d$-wave annihilation due to the weaker dependence on the relative velocity between dark matter particles, which increases with mass. Qualitatively similar results are seen for other channels and dark matter particle masses. Again, this is consistent with the expectations of \citet{Baxter_2022}. We find that the total $J$ factors we obtained, for the 128 most massive halos of one of our \csiborg{} realisation (9844), agree to within factor of $\sim 5$ when compared to the expression in Eq.~(3.3) of \citep{Baxter_2022}.
This is a consequence of more detailed modelling of the $J$-factor in this work; for example, if \citet{Baxter_2022} chose a different mass definition, given the strong dependence of the $p$- and $d$-wave annihilation channels on total mass of the object, this difference could be alleviated.

To assess which energy bins drive our constraints, for each channel and $m_\chi$ we calculate the log-likelihood at the 95\% upper limit on $a_\ell$ for each energy bin and compare to the values for $a_\ell=0$. For both $p$- and $d$-wave annihilation, we find that the first three energy bins are the most constraining for $m_\chi \lesssim 30 {\rm \, GeV}/c^2$, whereas the fourth energy bin becomes particularly important for higher masses. This is the same behaviour as for $s$-wave annihilation \citep{FLAT_2022}, which is attributed to the lower expected photon counts in the higher energy bins, and therefore weaker constraining power. As one increases $m_\chi$, one expects more photons at higher energy, and thus the relative importance of these bins increases.

\subsection{Systematic uncertainties}

\subsubsection{\texorpdfstring{$J$}{J} factor calculation}
\label{sec:systematics_J}

In \Cref{eq:spectrum likelihood} we marginalised over uncertainties in the large-scale density modes which are constrained from the \borg algorithm, as well as smaller-scale unconstrained modes which were added as white noise in the initial conditions. In our fiducial analysis we use the full suite of 101 \csiborg{} simulations, however to verify this is a sufficiently large number, we rerun our analysis one hundred times using fifty randomly selected \csiborg{} simulations (with repeats) each time. This yields an uncertainty on the 95\% upper limit of $a_\ell$ of no more than 5.9\% and 6.5\% for $p$-wave and $d$-wave annihilation, respectively, when considering all channels and particle masses separately. The median change in the constraint across these runs at a given mass and channel changes by no more than 1.2\% compared to the fiducial analysis. Thus, we conclude that this is an appropriate number of simulations.

For our $p$-wave analysis, we assumed circular orbits for dark matter particles ($\beta = 0$), leading to the radial velocity dispersion given in \cref{eq:Dispersion beta=0}. Similar expressions exist for alternative values of $\beta$, thus to determine the impact of this assumption on our constraints, we compute $J$ factor maps assuming $\beta=0.5$ or $\beta=1$ (although the latter case is unrealistic as this corresponds to purely radial orbits)
for a representative \csiborg{} simulation (9844) and rerun our analysis to obtain alternative constraints on $a_1$.

For $\beta=1$, one finds that the $J$ factor diverges logarithmically towards the centre, and thus simply computing \cref{eq:Jfactor pwave} would predict an infinite annihilation flux from each halo. In practice, if the annihilation rate becomes sufficiently high, then it will disrupt the density profile of the halo as the dark matter particles will be depleted towards the centre, presumably forming a cored profile. Thus, \cref{eq:Jfactor pwave} is only valid up to some core radius, $r_{\rm c}$. We estimate this to be the radius at which the dynamical time scale of the halo ($1/\sqrt{G \rho}$) equals the annihilation time scale ($\rho \sigma v / m_\chi$). Since we expect this to occur close to the centre of the halo, we use the leading order term for $\sigma_r^2$,
\begin{equation}
    \frac{\sigma_r^2}{V_{\rm vir}^2} \stackrel{\beta=1}{=} \frac{\pi^2 -9}{6 \left( \log \left( 1 + c \right) - \frac{c}{1+c} \right)} \frac{1}{\tilde{r}} + \mathcal{O}\left( \tilde{r}^0 \right),
\end{equation}
and assume $\rho \sim \rho_0 r_{\rm s} / r$ (see \cref{eq:NFW_profile}). This leads to a core radius of
\begin{equation}
    \label{eq:core_radius}
    r_{\rm c} \sim \left( \frac{\pi^2 -9}{6} \frac{a_1}{m_\chi} \frac{V_{\rm vir}^2}{\log \left( 1 + c \right) - \frac{c}{1+c}} \sqrt{\frac{\rho_0}{G}} \right)^{\frac{2}{3}} r_{\rm vir}.
\end{equation}
We have verified that, for the core radii that this implies, the $\mathcal{O}(s^0)$ term is negligible compared to the leading order term.

For simplicity, we compute \cref{eq:Jfactor pwave} using $r_{\rm c}$ as the lower limit of the integral and add the contribution from the cored region separately. Furthermore, we note that \cref{eq:core_radius} depends on $a_1$ and $m_\chi$, and thus a fully-consistent analysis should recompute the $J$ factor at each step in the MCMC and for each channel. However, this is computationally infeasible. Instead, we set $a_1 = 2.4 \times 10^{-21} \, {\rm cm^3 \, s^{-1}}$ and $m_\chi = 100{\rm \, GeV}/c^2$ in this formula; this is a characteristic WIMP mass with $a_1$ at approximately the 20\textsuperscript{th} percentile for the $b\bar{b}$ channel when $\beta=0.5$. This typically leads to a core radius of $r_{\rm c} \sim 0.1-1 {\rm \, pc}$. We find that the constraints on $a_1$ for $\beta = 1$ are much tighter than this, and thus our approximation yields a conservative constraint. 

For both $\beta=0.5$ and $\beta=1$, we find that the anisotropic $J$ factors lead to tighter constraints for all masses and channels, by up to a factor of $2.3$ for $\beta=0.5$ and up to two orders of magnitude for $\beta=1$. This simply reflects the general behaviour of the $J$ factor for these two different anisotropic profiles. As can be seen from \cref{fig:pwave_vs_dwave_Jtheta}, the $\beta=0.5$ and $\beta=1.0$ profiles are more strongly peaked towards the centre than the corresponding $\beta=0$ case, with $\beta=1.0$ diverging more quickly towards the centre and giving total $J$ factors between $50-100$ times larger than the corresponding $\beta=0$ case. From this and \cref{eq:fractional_flux}, it is then easy to see that the derived constraints for these respective cases will show similar trends. Choosing $\beta=0$ is therefore the most conservative of the cases considered, justifying our use of it in our fiducial analysis.

For all halos considered in this work, we have not attempted to add corrections to the halo density profiles due to baryonic effects, and have instead used profiles calibrated against dark-matter-only simulations. In the presence of baryons, adiabatic contraction during galaxy formation \citep{Blumenthal, Gnedin} can steepen the central density profile, or it can be made shallower due to the subsequent stellar feedback \citep{Pontzen_Governato, DP_CuspCore}. The different resulting density profiles would change our $J$ factor calculations and hence our constraints on $a_\ell$. Indeed, cosmological zoom hydrodynamical simulations have demonstrated that baryons do affect the $J$ factors for velocity-dependent annihilation, typically increasing $J$ from a factor of a few up to factors of several hundred \citep{Board_2021,McKeown_2022}. Assuming this increase in $J$ can be applied to our halos, then our constraints on $a_\ell$ are conservative, however we leave it to future work to implement more precise corrections for baryonic physics.

\subsubsection{Constrained simulation volume}

As already argued in Section VI.B.5 of \citep{FLAT_2022}, the choice of \healpix \texttt{nside=256} for our inference pipeline allows for circumventing the issue of the limited volume we have for our \csiborg{} simulation suite ($z \lesssim 0.05$). This is because any sources beyond the \csiborg{} volume would most likely be unresolved at this resolution and hence captured by the isotropic template. 

Our discovery that the most massive halos in the local universe dominate the $p$- and $d$-wave constraints (running the inference using the $\sim 20$ largest halos gives constraints which differ by a factor only $\mathcal{O}(1)$ compared to the full result) raises the question of whether stronger constraints could be obtained using a dedicated cluster catalogue extending to higher redshift than \csiborg. To address this we calculate the $J$ factor produced by the Master Catalogue of X-ray Clusters (MCXC; \citep{piffaretti2011mcxc}), a compilation of clusters detected through their X-ray-emitting gas.

We find that the total $J$ factor of the \csiborg{} catalogue is larger by factor of $\sim 10$ than that from MCXC. This is a result of the incompleteness of the MCXC catalogue as well as the various selection effects involved, which outweigh its larger redshift range and are absent from \csiborg. The \csiborg{} suite is therefore optimal in terms of its all-sky constraining power.

Nevertheless, since the $p$- and $d$-wave annihilation channels are strongly dependent on relative velocity of dark matter particles, and thus the total halo mass, it might still be worthwhile to explore constrained simulations with larger simulation volumes than \csiborg{} to include the contributions from more distant massive halos. Further constraining power may be extracted from the angular dependence of the $s$-, $p$- and $d$-wave annihilation signals, which we do not fully exploit due to the relatively coarse resolution of our all-sky maps. Useful future work would therefore be to use higher-resolution simulations or observations of clusters to compare their predicted and observed flux profiles in greater detail. Note that the angular resolution of the \textit{Fermi} data is approximately $\nside \approx 1024$, so significant improvements in resolution are possible. The positions and masses of local dark matter clusters may be found in the public \csiborg{} halo catalogues \citep{max_zenodo}.
    
\subsubsection{Cluster masses}\label{sec:cluster_masses}

As demonstrated in \cref{fig:mass_dependence_constraints} and indicated by \citet{Baxter_2022}, for $p$-wave and $d$-wave annihilation, the constraints on cross-section are dominated by the largest mass objects. It is therefore more important to correctly obtain cluster masses than it is when considering $s$-wave models \citep[cf. Fig.~9 of][]{FLAT_2022}. In this work we have used the masses obtained from constrained $N$-body simulations, with initial conditions inferred using the \borg algorithm. This is an alternative to traditional estimates of cluster masses, which typically use one of the virial theorem \citep{Meritt_1987}, X-ray measurements \citep{Evrard_1996}, the Sunyaev Zel’dovich effect \citep{Sunyaev_1970,Sunyaev_1980} or weak lensing \citep{Bonnet_1994,Fahlman_1994}. These measurements tend to lead to significantly different measurements of masses for the same clusters \citep[e.g. Fig.~3 of][]{Stopyra_2021}, so it is not clear whether these traditional methods offer a more appropriate route to obtaining $J$ factor maps as each method would produce very different constraints.

It has been found that the halo mass function in \csiborg{} is higher than the average $\Lambda$CDM expectation at the massive end \citep{Hutt_2022}. Although cosmic variance is high in this regime, this still represents a $\sim 2 \sigma$ effect. To determine the effect of this potential mass discrepancy, we obtain a single new posterior sample of the initial conditions with a different gravity model (using the COmoving Lagrangian Acceleration (COLA) method \citep{Tassev_2013}), different time-stepping in the forward-model, a different method for generating power spectra (\texttt{CLASS} \citep{Blas_2011} instead of Eisenstein \& Hu \citep{Eisenstein_1998}) and a more robust likelihood \citep{Porqueres_2019}, designed to remove the effects of unknown systematic uncertainties on scales larger than $1.4^\circ$. We then run a simulation with these initial conditions but otherwise identical to \csiborg. This reconstruction produces 2- and 3-point statistics of the density field and a halo mass function which is closer to that expected from a random $\Lambda$CDM realisation. However, the re-simulated halo masses are systematically smaller than those observed and some objects (e.g. the Perseus cluster) are less prominent than in our fiducial reconstruction. These modifications to the older \borg algorithm are studied in more detail in \cite{Stopyra_2023}.

Due to the variety of changes, this simulation is useful to test the robustness of our results to the reconstruction analysis. We produce $J$ factor maps for this simulation and rerun the end-to-end inference. We find this new simulation produces weaker constraints than our fiducial analysis, with changes in the constraints of up to a factor of $\sim5$ and $\sim12$ for $p$- and $d$-wave, respectively.
This is to be expected given the systematically smaller cluster masses obtained with the updated initial conditions and is much greater than the variation in constraints between \csiborg{} realisations, which is typically $\sim 40-50\%$. We note that this variation is comparable to what one would obtain using measured clustered mass because different measurements give highly variable results \citep{Stopyra_2021}. It is the most significant uncertainty in our analysis.

\subsubsection{Astrophysical templates}

In our fiducial analysis, we use the latest galactic diffuse model provided by the Fermi Collaboration (\texttt{gll\_iem\_v07}). Although we find little degeneracy between the normalisation of the astrophysical templates and the amplitude of the $J$-factor maps (and hence $a_\ell$ constraints), we nonetheless re-run our inference using an older template (\texttt{gll\_iem\_v02}) to estimate the systematic effect of the choice of galactic template. As in \citet{FLAT_2022}, we find that the constraints usually weaken slightly when using the older model, by up to 24\% for $p$-wave and up to 31\% for $d$-wave, although at low $m_\chi$ the $d$-wave constraints can be up to 7\% tighter with the older model. This variation is small compared to that caused by the cluster masses.

\subsubsection{Point sources}

\begin{figure*}
    \centering
    \includegraphics[width=\textwidth]{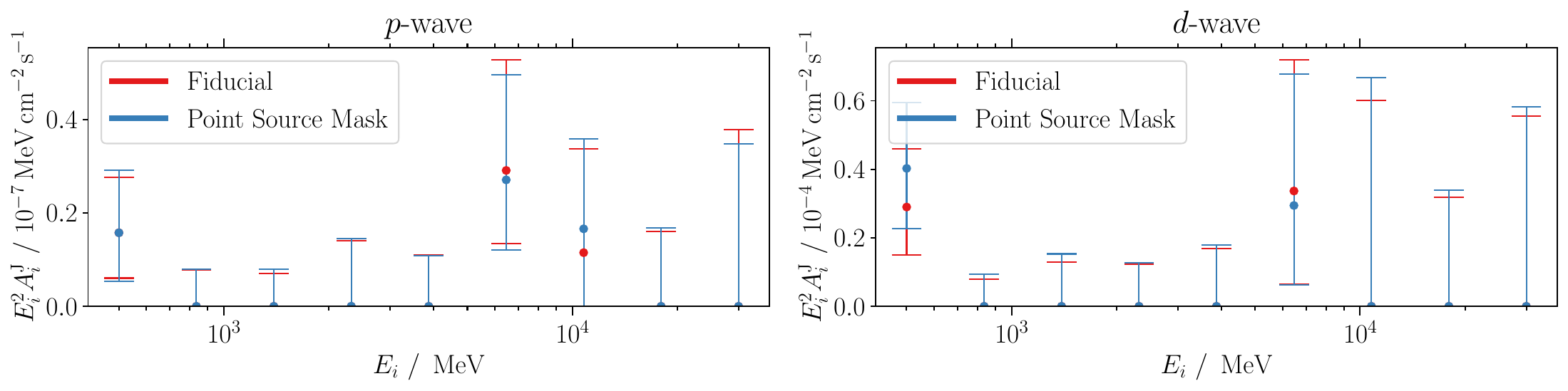}
      \caption{
      Comparison of posterior distributions on the amplitude of the $J$ factor templates as a function of energy if one includes (red) or masks (blue) point sources for one of the \csiborg{} simulations (9844). There is little change between the posteriors, indicating that the point source regions do not drive the constraints.
         }
    \label{fig:point_source_mask}
\end{figure*}

In the all-sky likelihood, point sources have a very large weight. This is an unwanted feature since we want our constraints to be driven by the large-scale structure rather than point sources. 
To test their impact on our constraints, we mask all point sources with an energy-dependent mask of radius equal to the point-spread function width (which is energy dependent) and repeat our analysis.
Although we mask the point sources, we still keep a point source template to capture any residual flux from these outside the mask.

In \cref{fig:point_source_mask} we plot the resulting constraints on $A^{\rm J}$ for the $p$-wave and $d$-wave cases using our fiducial simulation (9844). We see that there is not a significant shift in the posteriors, indicating that our constraints are not driven by point sources.
Once we convert these posteriors into constraints on the cross-section of $p$-wave annihilation, we find that our constraints weaken, but typically by only 5-10\%. 
The effect is slightly larger for $d$-wave annihilation, with changes of up to 40\%; however, in this case the constraints sometimes tighten for the masked data.
This degree of change is comparable to other systematic uncertainties, and thus we conclude that our main results are robust to whether one masks or keeps the point sources in the inference.

\subsection{Thermal relics}

In this work we have focused on dark matter particles of mass $2-500{\rm \, GeV}/c^2$ since, if the thermally-average annihilation cross-section is \hbox{$\sigv_{\rm th} \approx 3 \times 10^{-26} {\rm \, cm^3 s^{-1}}$} \citep{Steigman_2012} and these particles are thermal relics, then the current abundance of dark matter is reproduced. This cross-section is comparable to the electroweak coupling of the Standard Model, motivating the study of these dark matter candidates. In fact, the previously quoted $\sigv_{\rm th}$ is appropriate for $s$-wave annihilation only. Hence, in this section we extend the analysis of \citet{Steigman_2012} to obtain the equivalent values for $p$- and $d$-wave annihilation, and compare to the $a_\ell$ we obtain.

We assume that $\sigma v = a_\ell v^{2\ell}$ and that dark matter particles are non-relativistic and follow a Boltzmann distribution. The thermally-averaged cross-section at temperature $T$ is then
\begin{equation}
    \sigv = \frac{\int \dd^3 k_1 \dd^3 k_2 \exp \left( - \frac{E_1 + E_2}{T} \right) \sigma v}{\int \dd^3 k_1 \dd^3 k_2 \exp \left( - \frac{E_1 + E_2}{T} \right)}
    = \frac{4^\ell \Gamma\left( \ell + \frac{3}{2} \right)}{\Gamma\left(\frac{3}{2} \right)} \frac{a_\ell}{x^{\ell}},
\end{equation}
where $k_1$ and $k_2$ are the phase-space momenta and $x = m_\chi / T$. One must then solve the equation governing the number density, $n$, of dark matter
\begin{equation}
    \frac{\dd Y}{\dd x} = \frac{s \sigv}{H x} \left[ 1 + \frac{1}{3} \frac{\dd \left(\log g_{\rm s} \right)}{\dd \left(\log T \right)} \right] \left(Y_{\rm eq}^2 - Y^2 \right),
\end{equation}
where $Y = n / s$ for entropy density $s$, $g_{\rm s}$ is the number of relativistic degrees of freedom contributing to the entropy density, $H$ is the Hubble parameter, and the equilibrium value of $Y$ is
\begin{equation}
    Y_{\rm eq} = \frac{45}{2\pi^4} \left( \frac{\pi}{8} \right)^{1/2} \frac{g_\chi}{g_{\rm s}} x^{3/2} \exp\left( - x \right),
\end{equation}
where the dark matter particle has $g_\chi$ degrees of freedom. Since this is a numerically stiff differential equation, we follow \citet{Steigman_2012} and solve for $W=\log Y$ instead. We also take $g_{\rm s}(T)$ from the calculations of \citet{Laine_2006}. Using an implicit Runge-Kutta method of the Radau IIA family of order 5 \citep{Haire_1996}, we solve from $x=1$ to $x=10^3$, with an initial condition set using the approximate solution \citep{Kolb_1990}
\begin{equation}
    Y \left( x \right) \approx \frac{x^{\ell + 2}}{2 \lambda} + Y_{\rm eq} \left( x \right),
\end{equation}
which is valid long before freeze-out, where $\lambda \equiv 2.76 \times 10^{35} m_\chi \sigv_0$ for $\sigv = \sigv_0 x^{-\ell}$ and where $m_\chi$ is in ${\rm GeV}$ and $\sigv_0$ is in ${\rm cm^3 \, s^{-1}}$. We then convert our solution to a density parameter
\begin{equation}
    \label{eq:Omega_chi}
    \Omega_\chi = \frac{8 \pi G}{3 H_0^2} m_\chi s_0 Y_0,
\end{equation}
where $H_0$, $s_0$ and $Y_0$ are the values of $H$, $s$ and $Y$ evaluated today. By comparing this value to $\Omega_\chi h^2 = 0.11$ \citep{Komatsu_2011}, for $H_0 = 100 h {\rm \, km \, s^{-1} \, Mpc^{-1}}$, we repeat the analysis for varying $a_\ell$ at fixed $m_\chi$ to obtain the thermal relic cross-section which reproduces the present-day dark matter abundance.

\begin{figure}
    \centering
    \includegraphics[width=\columnwidth]{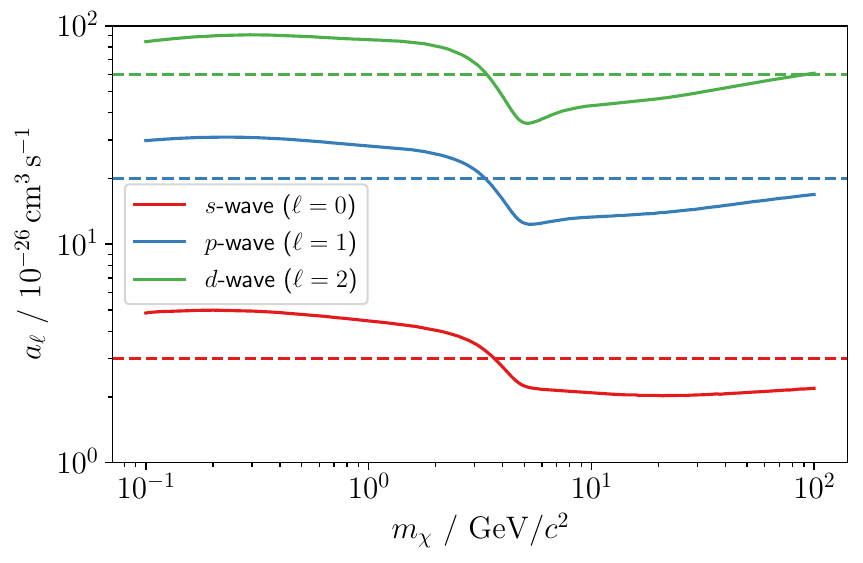}
    \caption{Annihilation cross-section coefficient ($\sigma v = a_\ell \left(v/c\right)^\ell$) as a function of dark matter particle mass, $m_\chi$, required to give a present-day dark matter abundance of $\Omega_\chi h^2 = 0.11$ for a thermal relic. We plot the results for $s$-, $p$- and $d$-wave annihilation and see that the required $a_\ell$ increases by less than an order of magnitude for a unit change of $\ell$. The dashed horizontal lines give the standard value of $a_0 = 3 \times 10^{-26} {\rm \, cm^3 s^{-1}}$ for $s$-wave, and this value multiplied by $x_{\rm f}/3$ and $x_{\rm f}^2/20$ for $p$- and $d$-wave, assuming a freeze-out value of $x$ of $x_{\rm f}=20$.}
    \label{fig:thermal_relic_prediction}
\end{figure}

This is plotted in \cref{fig:thermal_relic_prediction} as a function of $m_\chi$ for $\ell=0,1,2$. We find that the required $p$-wave coefficient is $a_1\sim (1-3) \times 10^{-25} {\rm \, cm^3 \, s^{-1}}$, whereas for $d$-wave this is $a_2\sim (4-9) \times 10^{-25} {\rm \, cm^3 \, s^{-1}}$. In both cases, this is several orders of magnitude smaller than the constraints we find in \cref{fig:constraints}, and thus we cannot rule out dark matter being a thermal relic with a velocity-dependent cross-section using large scale structure. These conclusions are consistent with those discussed in the context of the extragalactic gamma-ray background \citep{Campbell_2010,Campbell_2011}; a detection of a $p$-wave annihilation signal would suggest a non-thermal origin of the present-day dark matter abundance.

We note that approximate analytic solutions of \cref{eq:Omega_chi} exist for $p$- and $d$-wave annihilation \citep[e.g.][]{Griest_1991,Giacchino_2013}. In these cases, if freeze-out occurs when $x=x_{\rm f}$ and $x_{\rm f}$ is the same for all $\ell$, then to obtain the $p$- and $d$-wave $\sigv_{\rm th}$ from the $s$-wave analysis, one would multiply $a_{0, {\rm th}}$ by $x_{\rm f}/3 \approx 6.7$ and $x_{\rm f}^2 / 20 \approx 20$, respectively, where we have used $x_{\rm f} \approx 20$. These ratios give the dashed horizontal lines in \cref{fig:thermal_relic_prediction}, when multiplied by the fiducial value of $a_0 = 3 \times 10^{-26} {\rm \, cm^3 s^{-1}}$.

\subsection{Comparison to the literature}
\label{sec:literature}

An attempt to combine the relic abundance requirements with the GCE and lack of detection in dwarf galaxies was proposed by \citet{Choquette_2016}. They suggested that a metastable particle undergoing $s$-wave annihilation can produce the thermal abundance of dark matter particles, and these daughter particles can then undergo $p$-wave annihilation in the present epoch. They find that a dark matter particle mass $m_\chi \sim 90 {\rm \, GeV}/c^2$ with an annihilation cross-section given by $a_1 \sim 10^{-20} {\rm \, cm^3 s^{-1}}$ for the $b\bar{b}$ channel can reproduce the GCE and survive all other tests. For this mass and channel, the $a_1$ required to explain the GCE is ruled out at over $1\sigma$ confidence in our analysis, suggesting that large scale structure disfavours this interpretation. In particular, they convert constraints on $a_0$ from clusters to constraints on $a_1$ by multiplying by multiples of the square of the velocity dispersion, and they find the Fornax cluster to be the most constraining object, with constraints which are $\sim 2$ orders of magnitude weaker than those given in \cref{fig:constraints}. Since we perform a more careful analysis and obtain tight constraints on $a_1$, future work should be dedicated to repeating the analyses cited by \citet{Choquette_2016} but with the correct velocity-dependent $J$ factors.

If dark matter annihilated via the $e^+e^-$ channel, then the resulting cosmic rays from the Milky Way could be detectable in Voyager 1 and the AMS-02 data. \citet{Boudaud_2019} considered this effect, and found constraints on $p$-wave annihilation for this channel which are tighter than our constraints, by up to approximately one order of magnitude (see \cref{fig:constraints}). Their results are found to be robust to choice of profile used for the galactic halo, although are only applicable to the $e^+e^-$ channel.

This is supported by \citet{Johnson_2019}, who obtain a null detection of $p$-wave annihilation as the cause of the GCE, since they do not find a spike in emission around Sgr A*. Moreover, since $s$- and $p$-wave annihilation produce different spatial distributions of emission \citep{Boddy_2018} for the same density profile, it is argued that, to explain the GCE, $p$-wave annihilation would require a slope of the density profile of the Milky Way much steeper than normally considered ($\gamma \sim 1.5-1.7$ compared to $\gamma \sim 1.2-1.4$ for $s$-wave) \citep{Kiriu_2022}, further questioning whether velocity-dependent annihilation is the cause of this detection.

Alongside the GCE, one of the most common targets for indirect dark matter annihilation searches are dwarf galaxies in the Local Group due to their proximity and low baryon fraction. In \cref{fig:constraints} we compare our $p$-wave constraints to those obtained by these objects \citep{Zhao_2016,Zhao_2018,Boddy_2020} for the $b\bar{b}$ channel, and find that our constraints are up to two orders of magnitude tighter. Similarly, we find that our $d$-wave constraints are approximately seven orders of magnitude tighter than those previously found from studying 25 dwarf spheroidal galaxies of the Milky Way \citep{Boddy_2020}, and hence lie outside the range of \cref{fig:constraints}. This is in contrast to the results of $s$-wave annihilation, where dwarf galaxies give a constraint which is an order of magnitude tighter than that from large scale structure \citep{FLAT_2022}. Assuming that clusters and dwarf galaxies have velocity dispersions of $\mathcal{O}(100) {\rm \, km \, s^{-1}}$ and $\mathcal{O}(1-10) {\rm \, km \, s^{-1}}$, respectively, then one could estimate the $p$- and $d$-wave constraints by multiplying the $s$-wave ones by $\sigma^{2\ell}$. Given this ratio of velocity dispersions, one would anticipate that clusters can produce $p$- and $d$-wave constraints which are factors of $10^{2-3}$ and $10^{5-6}$ tighter, approximately as we find. Despite being less constraining for velocity-dependent annihilation, if one did detect such a signal, then dwarf galaxies could still have an important role, since each object would have the same spectrum, but the different density and velocity profiles would result in different spatial distributions of the signal, which one could use to determine $\ell$ \citep{Baxter_2021}.

If dark matter annihilated, then the energy injected into the intergalactic medium would modify the Universe's post-recombination ionisation history, and hence the observed cosmic microwave background (CMB). The large velocity dispersion in halos means that this is a predominantly low-redshift effect and could be detectable in the lowest multipoles of the CMB. Combining results from HST with BAO measurements, a joint analysis of data from Planck, WMAP9, ACT and SPT found no evidence of $p$-wave dark matter annihilation, ruling out cross-sections corresponding to $a_1 \sim 10^{-16} - 10^{-12} {\rm \, cm^3 \, s^{-1}}$ for the production of leptons at masses $m_\chi \sim 1 - 10^3 {\rm \, GeV}/c^2$ at 95\% confidence \citep{Diamanti_2014}. Our constraints are approximately four to five orders of magnitude more stringent than these results. One can similarly constrain dark matter annihilation via this process using the Lyman-$\alpha$ forest, but these constraints are also $\sim4$ orders of magnitude weaker than those found in this work \citep{Liu_2021}.

\section{Conclusions}
\label{sec:Conclusions}

Annihilation of dark matter particles with velocity-dependent cross-sections can result in fluxes of gamma rays which are higher from extragalactic halos than from dwarf galaxies in the Local Group. In this work, we use the \csiborg{} suite of constrained $N$-body simulations of the nearby Universe (within $155\Mpch$) to forward-model the expected emission from all halos down to $M_\text{vir}=4.4\times10^{11} M_\odot$ and therefore produce full-sky, field-level, energy-dependent intensity predictions for $p$-wave and $d$-wave dark matter annihilation. We combine these with templates describing emission by baryonic processes in the Milky Way, an isotropic background and point sources, and compare to observations from the \textit{Fermi} Large Area Telescope. Marginalising over the amplitudes of these templates and uncertainties in the reconstructed dark matter density field, we find no evidence for velocity-dependent dark matter annihilation to any standard model particle.

We place upper limits on the annihilation cross-section as a function of dark matter particles mass, $m_\chi$, in the range $m_\chi = 2-500{\rm \, GeV}/c^2$ and for different annihilation channels. For the $b\bar{b}$ channel, defining $\sigma v = a_\ell \left(v/c\right)^{2\ell}$ where $\ell = 1, 2$ for $p$ and $d$-wave, respectively, we constrain $a_{1} < 2.4 \times 10^{-21} {\rm \, cm^3 \, s^{-1}}$ and $a_{2} < 3.0 \times 10^{-18} {\rm \, cm^3 \, s^{-1}}$ at 95\% confidence for $m_\chi = 10 {\rm \, GeV}/c^2$.
Our constraints are weaker for the $e^+e^-$ channel, with $a_{1} < 1.5 \times 10^{-20} {\rm \, cm^3 \, s^{-1}}$ and $a_{2} < 1.9 \times 10^{-17} {\rm \, cm^3 \, s^{-1}}$ at 95\% confidence for this particle mass.
We find that these constraints are dominated by the most massive halos ($M_{\rm h} \sim 10^{14-16} \,\Msun$). Our constraints are four to five orders of magnitude more stringent than those on $p$-wave annihilation from cosmic microwave background analyses, and two (seven) orders of magnitude tighter than $p$-wave ($d$-wave) constraints from dwarf galaxies, as one would expect from the larger velocity dispersion in higher mass objects. Our bounds are however still several orders of magnitude above the value required to rule out dark matter being a thermal relic with a velocity-dependent annihilation cross-section.

In this work we have demonstrated the constraining power of the full halo field on velocity-dependent dark matter annihilation through a full-sky field-level inference, and obtained constraints which are several orders of magnitude tighter than alternative probes. Future work should be dedicated to higher-resolution inferences using the most constraining (highest mass) objects, incorporating other observations to constrain and marginalise over the density and velocity profiles, as well as investigating the case of Sommerfeld-enhanced annihilation channels as proposed in \citep{Facchinetti2022,Lacroix_2022}.

The dominant systematic uncertainty in our analysis is the density field reconstruction (in particular its ability to reproduce high halo masses accurately), and hence future work should focus on understanding and improving this modelling. Indeed there are promising developments within the \texttt{BORG} framework: not only has \cite{Stopyra_2023} largely addressed the cluster mass overprediction problem (see Sec.~\ref{sec:cluster_masses}), but a new reconstruction \texttt{Manticore} has recently been released~\cite{Manticore} and demonstrated to be the premier local-Universe reconstruction according to a diverse range of metrics (see also \cite{VFO}). This bodes well for future, even more precise constraints on dark matter properties from field-level inference in the local Universe.

\section*{Data availability}

This work required modifications to the \clumpy package in order to model velocity-dependent dark matter annihilation. These modifications will be included in a future \clumpy release. Other code and data---including our $J$ and $D$ factor maps, which may be useful for other indirect detection analyses of current or future data---will be made available upon request. 
As supplementary material on \texttt{arXiv} \citep{data_available} we provide the \texttt{Mathematica} notebooks used in deriving \cref{eq:rhov4_dwave} (see `TeX Source'). 

\acknowledgements
{
We thank C\'{e}line Combet, Sten Delos, Moritz H\"{u}tten, Jens Jasche, Eiichiro Komatsu, Guilhem Lavaux, David Maurin, Fabian Schmidt and Ewoud Wempe for useful discussions. We thank Jonathan Patterson for smoothly running the Glamdring Cluster hosted by the University of Oxford, where most of the data processing was performed.

AK acknowledges support from the Starting Grant (ERC-2015-STG 678652) ``GrInflaGal'' of the European Research Council at MPA. 
DJB is supported by the Simons Collaboration on ``Learning the Universe'' and was supported by STFC and Oriel College, Oxford.
HD is supported by a Royal Society University Research Fellowship (grant no. 211046). This project has received funding from the European Research Council (ERC) under the European Union’s Horizon 2020 research and innovation programme (grant agreement No 693024).
This work was done within the Aquila Consortium (\url{https://www.aquila-consortium.org/}).

Some of the results in this paper have been derived using the
\healpy and \healpix \citep{Zonca_2019,Gorski_2005},
\clumpy \citep{Charbonnier_2012,Bonnivard_2016,Hutten_2019}, 
\numpyro \citep{phan2019composable,bingham2019pyro},
\textsc{numpy} \citep{Numpy},
\textsc{scipy} \citep{Scipy} and
\fermipy \citep{FermiPy}
packages.
}

For the purpose of open access, the authors have applied a Creative Commons Attribution (CC BY) licence to any Author Accepted Manuscript version arising.

\bibliographystyle{apsrev4-1}
\bibliography{references}

\begin{thebibliography}{124}%
\makeatletter
\providecommand \@ifxundefined [1]{%
 \@ifx{#1\undefined}
}%
\providecommand \@ifnum [1]{%
 \ifnum #1\expandafter \@firstoftwo
 \else \expandafter \@secondoftwo
 \fi
}%
\providecommand \@ifx [1]{%
 \ifx #1\expandafter \@firstoftwo
 \else \expandafter \@secondoftwo
 \fi
}%
\providecommand \natexlab [1]{#1}%
\providecommand \enquote  [1]{``#1''}%
\providecommand \bibnamefont  [1]{#1}%
\providecommand \bibfnamefont [1]{#1}%
\providecommand \citenamefont [1]{#1}%
\providecommand \href@noop [0]{\@secondoftwo}%
\providecommand \href [0]{\begingroup \@sanitize@url \@href}%
\providecommand \@href[1]{\@@startlink{#1}\@@href}%
\providecommand \@@href[1]{\endgroup#1\@@endlink}%
\providecommand \@sanitize@url [0]{\catcode `\\12\catcode `\$12\catcode
  `\&12\catcode `\#12\catcode `\^12\catcode `\_12\catcode `\%12\relax}%
\providecommand \@@startlink[1]{}%
\providecommand \@@endlink[0]{}%
\providecommand \url  [0]{\begingroup\@sanitize@url \@url }%
\providecommand \@url [1]{\endgroup\@href {#1}{\urlprefix }}%
\providecommand \urlprefix  [0]{URL }%
\providecommand \Eprint [0]{\href }%
\providecommand \doibase [0]{http://dx.doi.org/}%
\providecommand \selectlanguage [0]{\@gobble}%
\providecommand \bibinfo  [0]{\@secondoftwo}%
\providecommand \bibfield  [0]{\@secondoftwo}%
\providecommand \translation [1]{[#1]}%
\providecommand \BibitemOpen [0]{}%
\providecommand \bibitemStop [0]{}%
\providecommand \bibitemNoStop [0]{.\EOS\space}%
\providecommand \EOS [0]{\spacefactor3000\relax}%
\providecommand \BibitemShut  [1]{\csname bibitem#1\endcsname}%
\let\auto@bib@innerbib\@empty
\bibitem [{\citenamefont {{Goodenough}}\ and\ \citenamefont
  {{Hooper}}(2009)}]{Goodenough_2009}%
  \BibitemOpen
  \bibfield  {author} {\bibinfo {author} {\bibfnamefont {L.}~\bibnamefont
  {{Goodenough}}}\ and\ \bibinfo {author} {\bibfnamefont {D.}~\bibnamefont
  {{Hooper}}},\ }\href@noop {} {\bibfield  {journal} {\bibinfo  {journal}
  {arXiv e-prints}\ ,\ \bibinfo {eid} {arXiv:0910.2998}} (\bibinfo {year}
  {2009})}\BibitemShut {NoStop}%
\bibitem [{\citenamefont {{Ajello}}\ \emph {et~al.}(2016)\citenamefont
  {{Ajello}} \emph {et~al.}}]{Ajello_2016}%
  \BibitemOpen
  \bibfield  {author} {\bibinfo {author} {\bibfnamefont {M.}~\bibnamefont
  {{Ajello}}} \emph {et~al.},\ }\href {\doibase 10.3847/0004-637X/819/1/44}
  {\bibfield  {journal} {\bibinfo  {journal} {\apj}\ }\textbf {\bibinfo
  {volume} {819}},\ \bibinfo {eid} {44} (\bibinfo {year} {2016})}\BibitemShut
  {NoStop}%
\bibitem [{\citenamefont {{Linden}}\ \emph {et~al.}(2016)\citenamefont
  {{Linden}}, \citenamefont {{Rodd}}, \citenamefont {{Safdi}},\ and\
  \citenamefont {{Slatyer}}}]{Linden_2016}%
  \BibitemOpen
  \bibfield  {author} {\bibinfo {author} {\bibfnamefont {T.}~\bibnamefont
  {{Linden}}}, \bibinfo {author} {\bibfnamefont {N.~L.}\ \bibnamefont
  {{Rodd}}}, \bibinfo {author} {\bibfnamefont {B.~R.}\ \bibnamefont {{Safdi}}},
  \ and\ \bibinfo {author} {\bibfnamefont {T.~R.}\ \bibnamefont {{Slatyer}}},\
  }\href {\doibase 10.1103/PhysRevD.94.103013} {\bibfield  {journal} {\bibinfo
  {journal} {\prd}\ }\textbf {\bibinfo {volume} {94}},\ \bibinfo {eid} {103013}
  (\bibinfo {year} {2016})}\BibitemShut {NoStop}%
\bibitem [{\citenamefont {{Hooper}}\ and\ \citenamefont
  {{Goodenough}}(2011)}]{Hooper_2011}%
  \BibitemOpen
  \bibfield  {author} {\bibinfo {author} {\bibfnamefont {D.}~\bibnamefont
  {{Hooper}}}\ and\ \bibinfo {author} {\bibfnamefont {L.}~\bibnamefont
  {{Goodenough}}},\ }\href {\doibase 10.1016/j.physletb.2011.02.029} {\bibfield
   {journal} {\bibinfo  {journal} {Physics Letters B}\ }\textbf {\bibinfo
  {volume} {697}},\ \bibinfo {pages} {412} (\bibinfo {year}
  {2011})}\BibitemShut {NoStop}%
\bibitem [{\citenamefont {{Hooper}}\ and\ \citenamefont
  {{Linden}}(2011)}]{Hooper_2011PRD}%
  \BibitemOpen
  \bibfield  {author} {\bibinfo {author} {\bibfnamefont {D.}~\bibnamefont
  {{Hooper}}}\ and\ \bibinfo {author} {\bibfnamefont {T.}~\bibnamefont
  {{Linden}}},\ }\href {\doibase 10.1103/PhysRevD.84.123005} {\bibfield
  {journal} {\bibinfo  {journal} {\prd}\ }\textbf {\bibinfo {volume} {84}},\
  \bibinfo {eid} {123005} (\bibinfo {year} {2011})}\BibitemShut {NoStop}%
\bibitem [{\citenamefont {{Hooper}}\ and\ \citenamefont
  {{Slatyer}}(2013)}]{Hooper_2013PDU}%
  \BibitemOpen
  \bibfield  {author} {\bibinfo {author} {\bibfnamefont {D.}~\bibnamefont
  {{Hooper}}}\ and\ \bibinfo {author} {\bibfnamefont {T.~R.}\ \bibnamefont
  {{Slatyer}}},\ }\href {\doibase 10.1016/j.dark.2013.06.003} {\bibfield
  {journal} {\bibinfo  {journal} {Physics of the Dark Universe}\ }\textbf
  {\bibinfo {volume} {2}},\ \bibinfo {pages} {118} (\bibinfo {year}
  {2013})}\BibitemShut {NoStop}%
\bibitem [{\citenamefont {{Zhou}}\ \emph {et~al.}(2015)\citenamefont {{Zhou}}
  \emph {et~al.}}]{Zhou_2015}%
  \BibitemOpen
  \bibfield  {author} {\bibinfo {author} {\bibfnamefont {B.}~\bibnamefont
  {{Zhou}}} \emph {et~al.},\ }\href {\doibase 10.1103/PhysRevD.91.123010}
  {\bibfield  {journal} {\bibinfo  {journal} {\prd}\ }\textbf {\bibinfo
  {volume} {91}},\ \bibinfo {eid} {123010} (\bibinfo {year}
  {2015})}\BibitemShut {NoStop}%
\bibitem [{\citenamefont {{Daylan}}\ \emph {et~al.}(2016)\citenamefont
  {{Daylan}} \emph {et~al.}}]{Daylan_2016}%
  \BibitemOpen
  \bibfield  {author} {\bibinfo {author} {\bibfnamefont {T.}~\bibnamefont
  {{Daylan}}} \emph {et~al.},\ }\href {\doibase 10.1016/j.dark.2015.12.005}
  {\bibfield  {journal} {\bibinfo  {journal} {Physics of the Dark Universe}\
  }\textbf {\bibinfo {volume} {12}},\ \bibinfo {pages} {1} (\bibinfo {year}
  {2016})}\BibitemShut {NoStop}%
\bibitem [{\citenamefont {{Cholis}}\ \emph {et~al.}(2021)\citenamefont
  {{Cholis}}, \citenamefont {{Zhong}}, \citenamefont {{McDermott}},\ and\
  \citenamefont {{Surdutovich}}}]{Cholis_2021}%
  \BibitemOpen
  \bibfield  {author} {\bibinfo {author} {\bibfnamefont {I.}~\bibnamefont
  {{Cholis}}}, \bibinfo {author} {\bibfnamefont {Y.-M.}\ \bibnamefont
  {{Zhong}}}, \bibinfo {author} {\bibfnamefont {S.~D.}\ \bibnamefont
  {{McDermott}}}, \ and\ \bibinfo {author} {\bibfnamefont {J.~P.}\ \bibnamefont
  {{Surdutovich}}},\ }\href@noop {} {\bibfield  {journal} {\bibinfo  {journal}
  {arXiv e-prints}\ ,\ \bibinfo {eid} {arXiv:2112.09706}} (\bibinfo {year}
  {2021})}\BibitemShut {NoStop}%
\bibitem [{\citenamefont {{Grand}}\ and\ \citenamefont
  {{White}}(2022)}]{Grand_2022}%
  \BibitemOpen
  \bibfield  {author} {\bibinfo {author} {\bibfnamefont {R.~J.~J.}\
  \bibnamefont {{Grand}}}\ and\ \bibinfo {author} {\bibfnamefont {S.~D.~M.}\
  \bibnamefont {{White}}},\ }\href {\doibase 10.1093/mnrasl/slac011} {\bibfield
   {journal} {\bibinfo  {journal} {\mnras}\ }\textbf {\bibinfo {volume}
  {511}},\ \bibinfo {pages} {L55} (\bibinfo {year} {2022})}\BibitemShut
  {NoStop}%
\bibitem [{\citenamefont {{Abazajian}}(2011)}]{Abazajian_2011}%
  \BibitemOpen
  \bibfield  {author} {\bibinfo {author} {\bibfnamefont {K.~N.}\ \bibnamefont
  {{Abazajian}}},\ }\href {\doibase 10.1088/1475-7516/2011/03/010} {\bibfield
  {journal} {\bibinfo  {journal} {\jcap}\ }\textbf {\bibinfo {volume} {2011}},\
  \bibinfo {eid} {010} (\bibinfo {year} {2011})}\BibitemShut {NoStop}%
\bibitem [{\citenamefont {{O'Leary}}\ \emph {et~al.}(2015)\citenamefont
  {{O'Leary}}, \citenamefont {{Kistler}}, \citenamefont {{Kerr}},\ and\
  \citenamefont {{Dexter}}}]{OLeary_2015}%
  \BibitemOpen
  \bibfield  {author} {\bibinfo {author} {\bibfnamefont {R.~M.}\ \bibnamefont
  {{O'Leary}}}, \bibinfo {author} {\bibfnamefont {M.~D.}\ \bibnamefont
  {{Kistler}}}, \bibinfo {author} {\bibfnamefont {M.}~\bibnamefont {{Kerr}}}, \
  and\ \bibinfo {author} {\bibfnamefont {J.}~\bibnamefont {{Dexter}}},\
  }\href@noop {} {\bibfield  {journal} {\bibinfo  {journal} {arXiv e-prints}\
  ,\ \bibinfo {eid} {arXiv:1504.02477}} (\bibinfo {year} {2015})}\BibitemShut
  {NoStop}%
\bibitem [{\citenamefont {{Petrovi{\'c}}}\ \emph {et~al.}(2015)\citenamefont
  {{Petrovi{\'c}}}, \citenamefont {{Serpico}},\ and\ \citenamefont
  {{Zaharijas}}}]{Petrovic_2015}%
  \BibitemOpen
  \bibfield  {author} {\bibinfo {author} {\bibfnamefont {J.}~\bibnamefont
  {{Petrovi{\'c}}}}, \bibinfo {author} {\bibfnamefont {P.~D.}\ \bibnamefont
  {{Serpico}}}, \ and\ \bibinfo {author} {\bibfnamefont {G.}~\bibnamefont
  {{Zaharijas}}},\ }\href {\doibase 10.1088/1475-7516/2015/02/023} {\bibfield
  {journal} {\bibinfo  {journal} {\jcap}\ }\textbf {\bibinfo {volume} {2015}},\
  \bibinfo {eid} {023} (\bibinfo {year} {2015})}\BibitemShut {NoStop}%
\bibitem [{\citenamefont {{Lee}}\ \emph {et~al.}(2016)\citenamefont {{Lee}},
  \citenamefont {{Lisanti}}, \citenamefont {{Safdi}}, \citenamefont
  {{Slatyer}},\ and\ \citenamefont {{Xue}}}]{Lee_2016}%
  \BibitemOpen
  \bibfield  {author} {\bibinfo {author} {\bibfnamefont {S.~K.}\ \bibnamefont
  {{Lee}}}, \bibinfo {author} {\bibfnamefont {M.}~\bibnamefont {{Lisanti}}},
  \bibinfo {author} {\bibfnamefont {B.~R.}\ \bibnamefont {{Safdi}}}, \bibinfo
  {author} {\bibfnamefont {T.~R.}\ \bibnamefont {{Slatyer}}}, \ and\ \bibinfo
  {author} {\bibfnamefont {W.}~\bibnamefont {{Xue}}},\ }\href {\doibase
  10.1103/PhysRevLett.116.051103} {\bibfield  {journal} {\bibinfo  {journal}
  {\prl}\ }\textbf {\bibinfo {volume} {116}},\ \bibinfo {eid} {051103}
  (\bibinfo {year} {2016})}\BibitemShut {NoStop}%
\bibitem [{\citenamefont {{Buschmann}}\ \emph {et~al.}(2020)\citenamefont
  {{Buschmann}}, \citenamefont {{Rodd}}, \citenamefont {{Safdi}}, \citenamefont
  {{Chang}}, \citenamefont {{Mishra-Sharma}}, \citenamefont {{Lisanti}},\ and\
  \citenamefont {{Macias}}}]{Buschmann_2020}%
  \BibitemOpen
  \bibfield  {author} {\bibinfo {author} {\bibfnamefont {M.}~\bibnamefont
  {{Buschmann}}}, \bibinfo {author} {\bibfnamefont {N.~L.}\ \bibnamefont
  {{Rodd}}}, \bibinfo {author} {\bibfnamefont {B.~R.}\ \bibnamefont {{Safdi}}},
  \bibinfo {author} {\bibfnamefont {L.~J.}\ \bibnamefont {{Chang}}}, \bibinfo
  {author} {\bibfnamefont {S.}~\bibnamefont {{Mishra-Sharma}}}, \bibinfo
  {author} {\bibfnamefont {M.}~\bibnamefont {{Lisanti}}}, \ and\ \bibinfo
  {author} {\bibfnamefont {O.}~\bibnamefont {{Macias}}},\ }\href {\doibase
  10.1103/PhysRevD.102.023023} {\bibfield  {journal} {\bibinfo  {journal}
  {\prd}\ }\textbf {\bibinfo {volume} {102}},\ \bibinfo {eid} {023023}
  (\bibinfo {year} {2020})}\BibitemShut {NoStop}%
\bibitem [{\citenamefont {{Gautam}}\ \emph {et~al.}(2021)\citenamefont
  {{Gautam}} \emph {et~al.}}]{Gautam_2021}%
  \BibitemOpen
  \bibfield  {author} {\bibinfo {author} {\bibfnamefont {A.}~\bibnamefont
  {{Gautam}}} \emph {et~al.},\ }\href@noop {} {\bibfield  {journal} {\bibinfo
  {journal} {arXiv e-prints}\ ,\ \bibinfo {eid} {arXiv:2106.00222}} (\bibinfo
  {year} {2021})}\BibitemShut {NoStop}%
\bibitem [{\citenamefont {{Hooper}}\ \emph {et~al.}(2013)\citenamefont
  {{Hooper}}, \citenamefont {{Cholis}}, \citenamefont {{Linden}}, \citenamefont
  {{Siegal-Gaskins}},\ and\ \citenamefont {{Slatyer}}}]{Hooper_2013}%
  \BibitemOpen
  \bibfield  {author} {\bibinfo {author} {\bibfnamefont {D.}~\bibnamefont
  {{Hooper}}}, \bibinfo {author} {\bibfnamefont {I.}~\bibnamefont {{Cholis}}},
  \bibinfo {author} {\bibfnamefont {T.}~\bibnamefont {{Linden}}}, \bibinfo
  {author} {\bibfnamefont {J.~M.}\ \bibnamefont {{Siegal-Gaskins}}}, \ and\
  \bibinfo {author} {\bibfnamefont {T.~R.}\ \bibnamefont {{Slatyer}}},\ }\href
  {\doibase 10.1103/PhysRevD.88.083009} {\bibfield  {journal} {\bibinfo
  {journal} {\prd}\ }\textbf {\bibinfo {volume} {88}},\ \bibinfo {eid} {083009}
  (\bibinfo {year} {2013})}\BibitemShut {NoStop}%
\bibitem [{\citenamefont {{Cholis}}\ \emph
  {et~al.}(2015{\natexlab{a}})\citenamefont {{Cholis}}, \citenamefont
  {{Hooper}},\ and\ \citenamefont {{Linden}}}]{Cholis_2015_pulsars}%
  \BibitemOpen
  \bibfield  {author} {\bibinfo {author} {\bibfnamefont {I.}~\bibnamefont
  {{Cholis}}}, \bibinfo {author} {\bibfnamefont {D.}~\bibnamefont {{Hooper}}},
  \ and\ \bibinfo {author} {\bibfnamefont {T.}~\bibnamefont {{Linden}}},\
  }\href {\doibase 10.1088/1475-7516/2015/06/043} {\bibfield  {journal}
  {\bibinfo  {journal} {\jcap}\ }\textbf {\bibinfo {volume} {2015}},\ \bibinfo
  {eid} {043} (\bibinfo {year} {2015}{\natexlab{a}})}\BibitemShut {NoStop}%
\bibitem [{\citenamefont {{Leane}}\ and\ \citenamefont
  {{Slatyer}}(2019)}]{Leane_2019}%
  \BibitemOpen
  \bibfield  {author} {\bibinfo {author} {\bibfnamefont {R.~K.}\ \bibnamefont
  {{Leane}}}\ and\ \bibinfo {author} {\bibfnamefont {T.~R.}\ \bibnamefont
  {{Slatyer}}},\ }\href@noop {} {\bibfield  {journal} {\bibinfo  {journal}
  {arXiv e-prints}\ ,\ \bibinfo {eid} {arXiv:1904.08430}} (\bibinfo {year}
  {2019})}\BibitemShut {NoStop}%
\bibitem [{\citenamefont {{Abazajian}}\ and\ \citenamefont
  {{Kaplinghat}}(2012)}]{Abazajian_2012}%
  \BibitemOpen
  \bibfield  {author} {\bibinfo {author} {\bibfnamefont {K.~N.}\ \bibnamefont
  {{Abazajian}}}\ and\ \bibinfo {author} {\bibfnamefont {M.}~\bibnamefont
  {{Kaplinghat}}},\ }\href {\doibase 10.1103/PhysRevD.86.083511} {\bibfield
  {journal} {\bibinfo  {journal} {\prd}\ }\textbf {\bibinfo {volume} {86}},\
  \bibinfo {eid} {083511} (\bibinfo {year} {2012})}\BibitemShut {NoStop}%
\bibitem [{\citenamefont {{Gordon}}\ and\ \citenamefont
  {{Mac{\'\i}as}}(2013)}]{Gordon_2013}%
  \BibitemOpen
  \bibfield  {author} {\bibinfo {author} {\bibfnamefont {C.}~\bibnamefont
  {{Gordon}}}\ and\ \bibinfo {author} {\bibfnamefont {O.}~\bibnamefont
  {{Mac{\'\i}as}}},\ }\href {\doibase 10.1103/PhysRevD.88.083521} {\bibfield
  {journal} {\bibinfo  {journal} {\prd}\ }\textbf {\bibinfo {volume} {88}},\
  \bibinfo {eid} {083521} (\bibinfo {year} {2013})}\BibitemShut {NoStop}%
\bibitem [{\citenamefont {{Abazajian}}\ \emph {et~al.}(2014)\citenamefont
  {{Abazajian}}, \citenamefont {{Canac}}, \citenamefont {{Horiuchi}},\ and\
  \citenamefont {{Kaplinghat}}}]{Abazajian_2014}%
  \BibitemOpen
  \bibfield  {author} {\bibinfo {author} {\bibfnamefont {K.~N.}\ \bibnamefont
  {{Abazajian}}}, \bibinfo {author} {\bibfnamefont {N.}~\bibnamefont
  {{Canac}}}, \bibinfo {author} {\bibfnamefont {S.}~\bibnamefont {{Horiuchi}}},
  \ and\ \bibinfo {author} {\bibfnamefont {M.}~\bibnamefont {{Kaplinghat}}},\
  }\href {\doibase 10.1103/PhysRevD.90.023526} {\bibfield  {journal} {\bibinfo
  {journal} {\prd}\ }\textbf {\bibinfo {volume} {90}},\ \bibinfo {eid} {023526}
  (\bibinfo {year} {2014})}\BibitemShut {NoStop}%
\bibitem [{\citenamefont {{Calore}}\ \emph {et~al.}(2015)\citenamefont
  {{Calore}}, \citenamefont {{Cholis}},\ and\ \citenamefont
  {{Weniger}}}]{Calore_2015}%
  \BibitemOpen
  \bibfield  {author} {\bibinfo {author} {\bibfnamefont {F.}~\bibnamefont
  {{Calore}}}, \bibinfo {author} {\bibfnamefont {I.}~\bibnamefont {{Cholis}}},
  \ and\ \bibinfo {author} {\bibfnamefont {C.}~\bibnamefont {{Weniger}}},\
  }\href {\doibase 10.1088/1475-7516/2015/03/038} {\bibfield  {journal}
  {\bibinfo  {journal} {\jcap}\ }\textbf {\bibinfo {volume} {2015}},\ \bibinfo
  {eid} {038} (\bibinfo {year} {2015})}\BibitemShut {NoStop}%
\bibitem [{\citenamefont {{Horiuchi}}\ \emph {et~al.}(2016)\citenamefont
  {{Horiuchi}}, \citenamefont {{Kaplinghat}},\ and\ \citenamefont
  {{Kwa}}}]{Horiuchi_2016}%
  \BibitemOpen
  \bibfield  {author} {\bibinfo {author} {\bibfnamefont {S.}~\bibnamefont
  {{Horiuchi}}}, \bibinfo {author} {\bibfnamefont {M.}~\bibnamefont
  {{Kaplinghat}}}, \ and\ \bibinfo {author} {\bibfnamefont {A.}~\bibnamefont
  {{Kwa}}},\ }\href {\doibase 10.1088/1475-7516/2016/11/053} {\bibfield
  {journal} {\bibinfo  {journal} {\jcap}\ }\textbf {\bibinfo {volume} {2016}},\
  \bibinfo {eid} {053} (\bibinfo {year} {2016})}\BibitemShut {NoStop}%
\bibitem [{\citenamefont {{Petrovi{\'c}}}\ \emph {et~al.}(2014)\citenamefont
  {{Petrovi{\'c}}}, \citenamefont {{Serpico}},\ and\ \citenamefont
  {{Zaharija{\v{s}}}}}]{Petrovic_2014}%
  \BibitemOpen
  \bibfield  {author} {\bibinfo {author} {\bibfnamefont {J.}~\bibnamefont
  {{Petrovi{\'c}}}}, \bibinfo {author} {\bibfnamefont {P.~D.}\ \bibnamefont
  {{Serpico}}}, \ and\ \bibinfo {author} {\bibfnamefont {G.}~\bibnamefont
  {{Zaharija{\v{s}}}}},\ }\href {\doibase 10.1088/1475-7516/2014/10/052}
  {\bibfield  {journal} {\bibinfo  {journal} {\jcap}\ }\textbf {\bibinfo
  {volume} {2014}},\ \bibinfo {eid} {052} (\bibinfo {year} {2014})}\BibitemShut
  {NoStop}%
\bibitem [{\citenamefont {{Carlson}}\ and\ \citenamefont
  {{Profumo}}(2014)}]{Carlson_2014}%
  \BibitemOpen
  \bibfield  {author} {\bibinfo {author} {\bibfnamefont {E.}~\bibnamefont
  {{Carlson}}}\ and\ \bibinfo {author} {\bibfnamefont {S.}~\bibnamefont
  {{Profumo}}},\ }\href {\doibase 10.1103/PhysRevD.90.023015} {\bibfield
  {journal} {\bibinfo  {journal} {\prd}\ }\textbf {\bibinfo {volume} {90}},\
  \bibinfo {eid} {023015} (\bibinfo {year} {2014})}\BibitemShut {NoStop}%
\bibitem [{\citenamefont {{Cholis}}\ \emph
  {et~al.}(2015{\natexlab{b}})\citenamefont {{Cholis}}, \citenamefont
  {{Evoli}}, \citenamefont {{Calore}}, \citenamefont {{Linden}}, \citenamefont
  {{Weniger}},\ and\ \citenamefont {{Hooper}}}]{Cholis_2015_rays}%
  \BibitemOpen
  \bibfield  {author} {\bibinfo {author} {\bibfnamefont {I.}~\bibnamefont
  {{Cholis}}}, \bibinfo {author} {\bibfnamefont {C.}~\bibnamefont {{Evoli}}},
  \bibinfo {author} {\bibfnamefont {F.}~\bibnamefont {{Calore}}}, \bibinfo
  {author} {\bibfnamefont {T.}~\bibnamefont {{Linden}}}, \bibinfo {author}
  {\bibfnamefont {C.}~\bibnamefont {{Weniger}}}, \ and\ \bibinfo {author}
  {\bibfnamefont {D.}~\bibnamefont {{Hooper}}},\ }\href {\doibase
  10.1088/1475-7516/2015/12/005} {\bibfield  {journal} {\bibinfo  {journal}
  {\jcap}\ }\textbf {\bibinfo {volume} {2015}},\ \bibinfo {eid} {005} (\bibinfo
  {year} {2015}{\natexlab{b}})}\BibitemShut {NoStop}%
\bibitem [{\citenamefont {{Gaggero}}\ \emph {et~al.}(2015)\citenamefont
  {{Gaggero}}, \citenamefont {{Taoso}}, \citenamefont {{Urbano}}, \citenamefont
  {{Valli}},\ and\ \citenamefont {{Ullio}}}]{Gaggero_2015}%
  \BibitemOpen
  \bibfield  {author} {\bibinfo {author} {\bibfnamefont {D.}~\bibnamefont
  {{Gaggero}}}, \bibinfo {author} {\bibfnamefont {M.}~\bibnamefont {{Taoso}}},
  \bibinfo {author} {\bibfnamefont {A.}~\bibnamefont {{Urbano}}}, \bibinfo
  {author} {\bibfnamefont {M.}~\bibnamefont {{Valli}}}, \ and\ \bibinfo
  {author} {\bibfnamefont {P.}~\bibnamefont {{Ullio}}},\ }\href {\doibase
  10.1088/1475-7516/2015/12/056} {\bibfield  {journal} {\bibinfo  {journal}
  {\jcap}\ }\textbf {\bibinfo {volume} {2015}},\ \bibinfo {eid} {056} (\bibinfo
  {year} {2015})}\BibitemShut {NoStop}%
\bibitem [{\citenamefont {{Macias}}\ \emph {et~al.}(2018)\citenamefont
  {{Macias}}, \citenamefont {{Gordon}}, \citenamefont {{Crocker}},
  \citenamefont {{Coleman}}, \citenamefont {{Paterson}}, \citenamefont
  {{Horiuchi}},\ and\ \citenamefont {{Pohl}}}]{Macias_2018}%
  \BibitemOpen
  \bibfield  {author} {\bibinfo {author} {\bibfnamefont {O.}~\bibnamefont
  {{Macias}}}, \bibinfo {author} {\bibfnamefont {C.}~\bibnamefont {{Gordon}}},
  \bibinfo {author} {\bibfnamefont {R.~M.}\ \bibnamefont {{Crocker}}}, \bibinfo
  {author} {\bibfnamefont {B.}~\bibnamefont {{Coleman}}}, \bibinfo {author}
  {\bibfnamefont {D.}~\bibnamefont {{Paterson}}}, \bibinfo {author}
  {\bibfnamefont {S.}~\bibnamefont {{Horiuchi}}}, \ and\ \bibinfo {author}
  {\bibfnamefont {M.}~\bibnamefont {{Pohl}}},\ }\href {\doibase
  10.1038/s41550-018-0414-3} {\bibfield  {journal} {\bibinfo  {journal} {Nature
  Astronomy}\ }\textbf {\bibinfo {volume} {2}},\ \bibinfo {pages} {387}
  (\bibinfo {year} {2018})}\BibitemShut {NoStop}%
\bibitem [{\citenamefont {{Bartels}}\ \emph {et~al.}(2018)\citenamefont
  {{Bartels}}, \citenamefont {{Storm}}, \citenamefont {{Weniger}},\ and\
  \citenamefont {{Calore}}}]{Bartels_2018}%
  \BibitemOpen
  \bibfield  {author} {\bibinfo {author} {\bibfnamefont {R.}~\bibnamefont
  {{Bartels}}}, \bibinfo {author} {\bibfnamefont {E.}~\bibnamefont {{Storm}}},
  \bibinfo {author} {\bibfnamefont {C.}~\bibnamefont {{Weniger}}}, \ and\
  \bibinfo {author} {\bibfnamefont {F.}~\bibnamefont {{Calore}}},\ }\href
  {\doibase 10.1038/s41550-018-0531-z} {\bibfield  {journal} {\bibinfo
  {journal} {Nature Astronomy}\ }\textbf {\bibinfo {volume} {2}},\ \bibinfo
  {pages} {819} (\bibinfo {year} {2018})}\BibitemShut {NoStop}%
\bibitem [{\citenamefont {{Ackermann}}\ \emph {et~al.}(2015)\citenamefont
  {{Ackermann}} \emph {et~al.}}]{dSph_0}%
  \BibitemOpen
  \bibfield  {author} {\bibinfo {author} {\bibfnamefont {M.}~\bibnamefont
  {{Ackermann}}} \emph {et~al.},\ }\href {\doibase
  10.1103/PhysRevLett.115.231301} {\bibfield  {journal} {\bibinfo  {journal}
  {\prl}\ }\textbf {\bibinfo {volume} {115}},\ \bibinfo {eid} {231301}
  (\bibinfo {year} {2015})}\BibitemShut {NoStop}%
\bibitem [{\citenamefont {{Albert}}\ \emph {et~al.}(2017)\citenamefont
  {{Albert}} \emph {et~al.}}]{dSph}%
  \BibitemOpen
  \bibfield  {author} {\bibinfo {author} {\bibfnamefont {A.}~\bibnamefont
  {{Albert}}} \emph {et~al.},\ }\href {\doibase 10.3847/1538-4357/834/2/110}
  {\bibfield  {journal} {\bibinfo  {journal} {\apj}\ }\textbf {\bibinfo
  {volume} {834}},\ \bibinfo {eid} {110} (\bibinfo {year} {2017})}\BibitemShut
  {NoStop}%
\bibitem [{\citenamefont {{Gammaldi}}\ \emph {et~al.}(2021)\citenamefont
  {{Gammaldi}} \emph {et~al.}}]{dIrr}%
  \BibitemOpen
  \bibfield  {author} {\bibinfo {author} {\bibfnamefont {V.}~\bibnamefont
  {{Gammaldi}}} \emph {et~al.},\ }\href {\doibase 10.1103/PhysRevD.104.083026}
  {\bibfield  {journal} {\bibinfo  {journal} {\prd}\ }\textbf {\bibinfo
  {volume} {104}},\ \bibinfo {eid} {083026} (\bibinfo {year}
  {2021})}\BibitemShut {NoStop}%
\bibitem [{\citenamefont {{Kim}}\ and\ \citenamefont {{Lee}}(2007)}]{Kim_2007}%
  \BibitemOpen
  \bibfield  {author} {\bibinfo {author} {\bibfnamefont {Y.~G.}\ \bibnamefont
  {{Kim}}}\ and\ \bibinfo {author} {\bibfnamefont {K.~Y.}\ \bibnamefont
  {{Lee}}},\ }\href {\doibase 10.1103/PhysRevD.75.115012} {\bibfield  {journal}
  {\bibinfo  {journal} {\prd}\ }\textbf {\bibinfo {volume} {75}},\ \bibinfo
  {eid} {115012} (\bibinfo {year} {2007})}\BibitemShut {NoStop}%
\bibitem [{\citenamefont {{Lee}}\ \emph {et~al.}(2008)\citenamefont {{Lee}},
  \citenamefont {{Kim}},\ and\ \citenamefont {{Shin}}}]{Lee_2008}%
  \BibitemOpen
  \bibfield  {author} {\bibinfo {author} {\bibfnamefont {K.~Y.}\ \bibnamefont
  {{Lee}}}, \bibinfo {author} {\bibfnamefont {Y.~G.}\ \bibnamefont {{Kim}}}, \
  and\ \bibinfo {author} {\bibfnamefont {S.}~\bibnamefont {{Shin}}},\ }\href
  {\doibase 10.1088/1126-6708/2008/05/100} {\bibfield  {journal} {\bibinfo
  {journal} {Journal of High Energy Physics}\ }\textbf {\bibinfo {volume}
  {2008}},\ \bibinfo {eid} {100} (\bibinfo {year} {2008})}\BibitemShut
  {NoStop}%
\bibitem [{\citenamefont {{Kumar}}\ and\ \citenamefont
  {{Marfatia}}(2013)}]{Kumar_2013}%
  \BibitemOpen
  \bibfield  {author} {\bibinfo {author} {\bibfnamefont {J.}~\bibnamefont
  {{Kumar}}}\ and\ \bibinfo {author} {\bibfnamefont {D.}~\bibnamefont
  {{Marfatia}}},\ }\href {\doibase 10.1103/PhysRevD.88.014035} {\bibfield
  {journal} {\bibinfo  {journal} {\prd}\ }\textbf {\bibinfo {volume} {88}},\
  \bibinfo {eid} {014035} (\bibinfo {year} {2013})}\BibitemShut {NoStop}%
\bibitem [{\citenamefont {{Giacchino}}\ \emph {et~al.}(2013)\citenamefont
  {{Giacchino}}, \citenamefont {{Lopez-Honorez}},\ and\ \citenamefont
  {{Tytgat}}}]{Giacchino_2013}%
  \BibitemOpen
  \bibfield  {author} {\bibinfo {author} {\bibfnamefont {F.}~\bibnamefont
  {{Giacchino}}}, \bibinfo {author} {\bibfnamefont {L.}~\bibnamefont
  {{Lopez-Honorez}}}, \ and\ \bibinfo {author} {\bibfnamefont {M.~H.~G.}\
  \bibnamefont {{Tytgat}}},\ }\href {\doibase 10.1088/1475-7516/2013/10/025}
  {\bibfield  {journal} {\bibinfo  {journal} {\jcap}\ }\textbf {\bibinfo
  {volume} {2013}},\ \bibinfo {eid} {025} (\bibinfo {year} {2013})}\BibitemShut
  {NoStop}%
\bibitem [{\citenamefont {{Toma}}(2013)}]{Toma_2013}%
  \BibitemOpen
  \bibfield  {author} {\bibinfo {author} {\bibfnamefont {T.}~\bibnamefont
  {{Toma}}},\ }\href {\doibase 10.1103/PhysRevLett.111.091301} {\bibfield
  {journal} {\bibinfo  {journal} {\prl}\ }\textbf {\bibinfo {volume} {111}},\
  \bibinfo {eid} {091301} (\bibinfo {year} {2013})}\BibitemShut {NoStop}%
\bibitem [{\citenamefont {{Baxter}}\ \emph {et~al.}(2022)\citenamefont
  {{Baxter}}, \citenamefont {{Kumar}}, \citenamefont {{Paul}},\ and\
  \citenamefont {{Runburg}}}]{Baxter_2022}%
  \BibitemOpen
  \bibfield  {author} {\bibinfo {author} {\bibfnamefont {E.~J.}\ \bibnamefont
  {{Baxter}}}, \bibinfo {author} {\bibfnamefont {J.}~\bibnamefont {{Kumar}}},
  \bibinfo {author} {\bibfnamefont {A.~D.}\ \bibnamefont {{Paul}}}, \ and\
  \bibinfo {author} {\bibfnamefont {J.}~\bibnamefont {{Runburg}}},\ }\href
  {\doibase 10.1088/1475-7516/2022/09/026} {\bibfield  {journal} {\bibinfo
  {journal} {\jcap}\ }\textbf {\bibinfo {volume} {2022}},\ \bibinfo {eid} {026}
  (\bibinfo {year} {2022})}\BibitemShut {NoStop}%
\bibitem [{\citenamefont {{Bartlett}}\ \emph {et~al.}(2022)\citenamefont
  {{Bartlett}}, \citenamefont {{Kosti{\'c}}}, \citenamefont {{Desmond}},
  \citenamefont {{Jasche}},\ and\ \citenamefont {{Lavaux}}}]{FLAT_2022}%
  \BibitemOpen
  \bibfield  {author} {\bibinfo {author} {\bibfnamefont {D.~J.}\ \bibnamefont
  {{Bartlett}}}, \bibinfo {author} {\bibfnamefont {A.}~\bibnamefont
  {{Kosti{\'c}}}}, \bibinfo {author} {\bibfnamefont {H.}~\bibnamefont
  {{Desmond}}}, \bibinfo {author} {\bibfnamefont {J.}~\bibnamefont {{Jasche}}},
  \ and\ \bibinfo {author} {\bibfnamefont {G.}~\bibnamefont {{Lavaux}}},\
  }\href {\doibase 10.1103/PhysRevD.106.103526} {\bibfield  {journal} {\bibinfo
   {journal} {\prd}\ }\textbf {\bibinfo {volume} {106}},\ \bibinfo {eid}
  {103526} (\bibinfo {year} {2022})}\BibitemShut {NoStop}%
\bibitem [{\citenamefont {{Bartlett}}\ \emph {et~al.}(2021)\citenamefont
  {{Bartlett}}, \citenamefont {{Desmond}},\ and\ \citenamefont
  {{Ferreira}}}]{Bartlett_2021_VS}%
  \BibitemOpen
  \bibfield  {author} {\bibinfo {author} {\bibfnamefont {D.~J.}\ \bibnamefont
  {{Bartlett}}}, \bibinfo {author} {\bibfnamefont {H.}~\bibnamefont
  {{Desmond}}}, \ and\ \bibinfo {author} {\bibfnamefont {P.~G.}\ \bibnamefont
  {{Ferreira}}},\ }\href {\doibase 10.1103/PhysRevD.103.023523} {\bibfield
  {journal} {\bibinfo  {journal} {\prd}\ }\textbf {\bibinfo {volume} {103}},\
  \bibinfo {eid} {023523} (\bibinfo {year} {2021})}\BibitemShut {NoStop}%
\bibitem [{\citenamefont {{Desmond}}\ \emph {et~al.}(2022)\citenamefont
  {{Desmond}}, \citenamefont {{Hutt}}, \citenamefont {{Devriendt}},\ and\
  \citenamefont {{Slyz}}}]{antihalos}%
  \BibitemOpen
  \bibfield  {author} {\bibinfo {author} {\bibfnamefont {H.}~\bibnamefont
  {{Desmond}}}, \bibinfo {author} {\bibfnamefont {M.~L.}\ \bibnamefont
  {{Hutt}}}, \bibinfo {author} {\bibfnamefont {J.}~\bibnamefont {{Devriendt}}},
  \ and\ \bibinfo {author} {\bibfnamefont {A.}~\bibnamefont {{Slyz}}},\ }\href
  {\doibase 10.1093/mnrasl/slac008} {\bibfield  {journal} {\bibinfo  {journal}
  {\mnras}\ }\textbf {\bibinfo {volume} {511}},\ \bibinfo {pages} {L45}
  (\bibinfo {year} {2022})}\BibitemShut {NoStop}%
\bibitem [{\citenamefont {{Hutt}}\ \emph {et~al.}(2022)\citenamefont {{Hutt}},
  \citenamefont {{Desmond}}, \citenamefont {{Devriendt}},\ and\ \citenamefont
  {{Slyz}}}]{Hutt_2022}%
  \BibitemOpen
  \bibfield  {author} {\bibinfo {author} {\bibfnamefont {M.~L.}\ \bibnamefont
  {{Hutt}}}, \bibinfo {author} {\bibfnamefont {H.}~\bibnamefont {{Desmond}}},
  \bibinfo {author} {\bibfnamefont {J.}~\bibnamefont {{Devriendt}}}, \ and\
  \bibinfo {author} {\bibfnamefont {A.}~\bibnamefont {{Slyz}}},\ }\href
  {\doibase 10.1093/mnras/stac2407} {\bibfield  {journal} {\bibinfo  {journal}
  {\mnras}\ }\textbf {\bibinfo {volume} {516}},\ \bibinfo {pages} {3592}
  (\bibinfo {year} {2022})}\BibitemShut {NoStop}%
\bibitem [{\citenamefont {{Lavaux}}\ and\ \citenamefont
  {{Jasche}}(2016)}]{Lavaux_2016}%
  \BibitemOpen
  \bibfield  {author} {\bibinfo {author} {\bibfnamefont {G.}~\bibnamefont
  {{Lavaux}}}\ and\ \bibinfo {author} {\bibfnamefont {J.}~\bibnamefont
  {{Jasche}}},\ }\href {\doibase 10.1093/mnras/stv2499} {\bibfield  {journal}
  {\bibinfo  {journal} {\mnras}\ }\textbf {\bibinfo {volume} {455}},\ \bibinfo
  {pages} {3169} (\bibinfo {year} {2016})}\BibitemShut {NoStop}%
\bibitem [{\citenamefont {Jasche}\ and\ \citenamefont
  {Lavaux}(2019)}]{Jasche_Lavaux}%
  \BibitemOpen
  \bibfield  {author} {\bibinfo {author} {\bibfnamefont {J.}~\bibnamefont
  {Jasche}}\ and\ \bibinfo {author} {\bibfnamefont {G.}~\bibnamefont
  {Lavaux}},\ }\href {\doibase 10.1051/0004-6361/201833710} {\bibfield
  {journal} {\bibinfo  {journal} {Astron. Astrophys.}\ }\textbf {\bibinfo
  {volume} {625}},\ \bibinfo {pages} {A64} (\bibinfo {year}
  {2019})}\BibitemShut {NoStop}%
\bibitem [{\citenamefont {{Jasche}}\ and\ \citenamefont
  {{Wandelt}}(2012)}]{BORG_1}%
  \BibitemOpen
  \bibfield  {author} {\bibinfo {author} {\bibfnamefont {J.}~\bibnamefont
  {{Jasche}}}\ and\ \bibinfo {author} {\bibfnamefont {B.~D.}\ \bibnamefont
  {{Wandelt}}},\ }\href {\doibase 10.1111/j.1365-2966.2012.21423.x} {\bibfield
  {journal} {\bibinfo  {journal} {\mnras}\ }\textbf {\bibinfo {volume} {425}},\
  \bibinfo {pages} {1042} (\bibinfo {year} {2012})}\BibitemShut {NoStop}%
\bibitem [{\citenamefont {{Jasche}}\ and\ \citenamefont
  {{Wandelt}}(2013)}]{BORG_2}%
  \BibitemOpen
  \bibfield  {author} {\bibinfo {author} {\bibfnamefont {J.}~\bibnamefont
  {{Jasche}}}\ and\ \bibinfo {author} {\bibfnamefont {B.~D.}\ \bibnamefont
  {{Wandelt}}},\ }\href {\doibase 10.1093/mnras/stt449} {\bibfield  {journal}
  {\bibinfo  {journal} {\mnras}\ }\textbf {\bibinfo {volume} {432}},\ \bibinfo
  {pages} {894} (\bibinfo {year} {2013})}\BibitemShut {NoStop}%
\bibitem [{\citenamefont {{Jasche}}\ \emph {et~al.}(2010)\citenamefont
  {{Jasche}}, \citenamefont {{Kitaura}}, \citenamefont {{Wandelt}},\ and\
  \citenamefont {{En{\ss}lin}}}]{BORG_3}%
  \BibitemOpen
  \bibfield  {author} {\bibinfo {author} {\bibfnamefont {J.}~\bibnamefont
  {{Jasche}}}, \bibinfo {author} {\bibfnamefont {F.~S.}\ \bibnamefont
  {{Kitaura}}}, \bibinfo {author} {\bibfnamefont {B.~D.}\ \bibnamefont
  {{Wandelt}}}, \ and\ \bibinfo {author} {\bibfnamefont {T.~A.}\ \bibnamefont
  {{En{\ss}lin}}},\ }\href {\doibase 10.1111/j.1365-2966.2010.16610.x}
  {\bibfield  {journal} {\bibinfo  {journal} {\mnras}\ }\textbf {\bibinfo
  {volume} {406}},\ \bibinfo {pages} {60} (\bibinfo {year} {2010})}\BibitemShut
  {NoStop}%
\bibitem [{\citenamefont {{Jasche}}\ \emph {et~al.}(2015)\citenamefont
  {{Jasche}}, \citenamefont {{Leclercq}},\ and\ \citenamefont
  {{Wandelt}}}]{BORG_4}%
  \BibitemOpen
  \bibfield  {author} {\bibinfo {author} {\bibfnamefont {J.}~\bibnamefont
  {{Jasche}}}, \bibinfo {author} {\bibfnamefont {F.}~\bibnamefont
  {{Leclercq}}}, \ and\ \bibinfo {author} {\bibfnamefont {B.~D.}\ \bibnamefont
  {{Wandelt}}},\ }\href {\doibase 10.1088/1475-7516/2015/01/036} {\bibfield
  {journal} {\bibinfo  {journal} {\jcap}\ }\textbf {\bibinfo {volume} {2015}},\
  \bibinfo {eid} {036} (\bibinfo {year} {2015})}\BibitemShut {NoStop}%
\bibitem [{\citenamefont {{Teyssier}}(2002)}]{ramses}%
  \BibitemOpen
  \bibfield  {author} {\bibinfo {author} {\bibfnamefont {R.}~\bibnamefont
  {{Teyssier}}},\ }\href {\doibase 10.1051/0004-6361:20011817} {\bibfield
  {journal} {\bibinfo  {journal} {\aap}\ }\textbf {\bibinfo {volume} {385}},\
  \bibinfo {pages} {337} (\bibinfo {year} {2002})}\BibitemShut {NoStop}%
\bibitem [{\citenamefont {{Bleuler}}\ \emph {et~al.}(2015)\citenamefont
  {{Bleuler}}, \citenamefont {{Teyssier}}, \citenamefont {{Carassou}},\ and\
  \citenamefont {{Martizzi}}}]{PHEW}%
  \BibitemOpen
  \bibfield  {author} {\bibinfo {author} {\bibfnamefont {A.}~\bibnamefont
  {{Bleuler}}}, \bibinfo {author} {\bibfnamefont {R.}~\bibnamefont
  {{Teyssier}}}, \bibinfo {author} {\bibfnamefont {S.}~\bibnamefont
  {{Carassou}}}, \ and\ \bibinfo {author} {\bibfnamefont {D.}~\bibnamefont
  {{Martizzi}}},\ }\href {\doibase 10.1186/s40668-015-0009-7} {\bibfield
  {journal} {\bibinfo  {journal} {Computational Astrophysics and Cosmology}\
  }\textbf {\bibinfo {volume} {2}},\ \bibinfo {eid} {5} (\bibinfo {year}
  {2015})}\BibitemShut {NoStop}%
\bibitem [{\citenamefont {Hutt}\ \emph {et~al.}(2022)\citenamefont {Hutt},
  \citenamefont {Desmond}, \citenamefont {Devriendt},\ and\ \citenamefont
  {Slyz}}]{max_zenodo}%
  \BibitemOpen
  \bibfield  {author} {\bibinfo {author} {\bibfnamefont {M.~L.}\ \bibnamefont
  {Hutt}}, \bibinfo {author} {\bibfnamefont {H.}~\bibnamefont {Desmond}},
  \bibinfo {author} {\bibfnamefont {J.}~\bibnamefont {Devriendt}}, \ and\
  \bibinfo {author} {\bibfnamefont {A.}~\bibnamefont {Slyz}},\ }\href {\doibase
  10.5281/zenodo.5851241} {\enquote {\bibinfo {title} {{The effect of local
  universe constraints on halo abundance and clustering}},}\ } (\bibinfo {year}
  {2022})\BibitemShut {NoStop}%
\bibitem [{\citenamefont {Boucher}\ \emph {et~al.}(2022)\citenamefont
  {Boucher}, \citenamefont {Kumar}, \citenamefont {Le},\ and\ \citenamefont
  {Runburg}}]{boucher2022j}%
  \BibitemOpen
  \bibfield  {author} {\bibinfo {author} {\bibfnamefont {B.}~\bibnamefont
  {Boucher}}, \bibinfo {author} {\bibfnamefont {J.}~\bibnamefont {Kumar}},
  \bibinfo {author} {\bibfnamefont {V.~B.}\ \bibnamefont {Le}}, \ and\ \bibinfo
  {author} {\bibfnamefont {J.}~\bibnamefont {Runburg}},\ }\href@noop {}
  {\bibfield  {journal} {\bibinfo  {journal} {Physical Review D}\ }\textbf
  {\bibinfo {volume} {106}},\ \bibinfo {pages} {023025} (\bibinfo {year}
  {2022})}\BibitemShut {NoStop}%
\bibitem [{\citenamefont {{Charbonnier}}\ \emph {et~al.}(2012)\citenamefont
  {{Charbonnier}}, \citenamefont {{Combet}},\ and\ \citenamefont
  {{Maurin}}}]{Charbonnier_2012}%
  \BibitemOpen
  \bibfield  {author} {\bibinfo {author} {\bibfnamefont {A.}~\bibnamefont
  {{Charbonnier}}}, \bibinfo {author} {\bibfnamefont {C.}~\bibnamefont
  {{Combet}}}, \ and\ \bibinfo {author} {\bibfnamefont {D.}~\bibnamefont
  {{Maurin}}},\ }\href {\doibase 10.1016/j.cpc.2011.10.017} {\bibfield
  {journal} {\bibinfo  {journal} {Computer Physics Communications}\ }\textbf
  {\bibinfo {volume} {183}},\ \bibinfo {pages} {656} (\bibinfo {year}
  {2012})}\BibitemShut {NoStop}%
\bibitem [{\citenamefont {{Bonnivard}}\ \emph {et~al.}(2016)\citenamefont
  {{Bonnivard}}, \citenamefont {{H{\"u}tten}}, \citenamefont {{Nezri}},
  \citenamefont {{Charbonnier}}, \citenamefont {{Combet}},\ and\ \citenamefont
  {{Maurin}}}]{Bonnivard_2016}%
  \BibitemOpen
  \bibfield  {author} {\bibinfo {author} {\bibfnamefont {V.}~\bibnamefont
  {{Bonnivard}}}, \bibinfo {author} {\bibfnamefont {M.}~\bibnamefont
  {{H{\"u}tten}}}, \bibinfo {author} {\bibfnamefont {E.}~\bibnamefont
  {{Nezri}}}, \bibinfo {author} {\bibfnamefont {A.}~\bibnamefont
  {{Charbonnier}}}, \bibinfo {author} {\bibfnamefont {C.}~\bibnamefont
  {{Combet}}}, \ and\ \bibinfo {author} {\bibfnamefont {D.}~\bibnamefont
  {{Maurin}}},\ }\href {\doibase 10.1016/j.cpc.2015.11.012} {\bibfield
  {journal} {\bibinfo  {journal} {Computer Physics Communications}\ }\textbf
  {\bibinfo {volume} {200}},\ \bibinfo {pages} {336} (\bibinfo {year}
  {2016})}\BibitemShut {NoStop}%
\bibitem [{\citenamefont {{H{\"u}tten}}\ \emph {et~al.}(2019)\citenamefont
  {{H{\"u}tten}}, \citenamefont {{Combet}},\ and\ \citenamefont
  {{Maurin}}}]{Hutten_2019}%
  \BibitemOpen
  \bibfield  {author} {\bibinfo {author} {\bibfnamefont {M.}~\bibnamefont
  {{H{\"u}tten}}}, \bibinfo {author} {\bibfnamefont {C.}~\bibnamefont
  {{Combet}}}, \ and\ \bibinfo {author} {\bibfnamefont {D.}~\bibnamefont
  {{Maurin}}},\ }\href {\doibase 10.1016/j.cpc.2018.10.001} {\bibfield
  {journal} {\bibinfo  {journal} {Computer Physics Communications}\ }\textbf
  {\bibinfo {volume} {235}},\ \bibinfo {pages} {336} (\bibinfo {year}
  {2019})}\BibitemShut {NoStop}%
\bibitem [{\citenamefont {{Navarro}}\ \emph {et~al.}(1997)\citenamefont
  {{Navarro}}, \citenamefont {{Frenk}},\ and\ \citenamefont
  {{White}}}]{NFW_1997}%
  \BibitemOpen
  \bibfield  {author} {\bibinfo {author} {\bibfnamefont {J.~F.}\ \bibnamefont
  {{Navarro}}}, \bibinfo {author} {\bibfnamefont {C.~S.}\ \bibnamefont
  {{Frenk}}}, \ and\ \bibinfo {author} {\bibfnamefont {S.~D.~M.}\ \bibnamefont
  {{White}}},\ }\href {\doibase 10.1086/304888} {\bibfield  {journal} {\bibinfo
   {journal} {\apj}\ }\textbf {\bibinfo {volume} {490}},\ \bibinfo {pages}
  {493} (\bibinfo {year} {1997})}\BibitemShut {NoStop}%
\bibitem [{\citenamefont {{Prada}}\ \emph {et~al.}(2012)\citenamefont
  {{Prada}}, \citenamefont {{Klypin}}, \citenamefont {{Cuesta}}, \citenamefont
  {{Betancort-Rijo}},\ and\ \citenamefont {{Primack}}}]{Prada12}%
  \BibitemOpen
  \bibfield  {author} {\bibinfo {author} {\bibfnamefont {F.}~\bibnamefont
  {{Prada}}}, \bibinfo {author} {\bibfnamefont {A.~A.}\ \bibnamefont
  {{Klypin}}}, \bibinfo {author} {\bibfnamefont {A.~J.}\ \bibnamefont
  {{Cuesta}}}, \bibinfo {author} {\bibfnamefont {J.~E.}\ \bibnamefont
  {{Betancort-Rijo}}}, \ and\ \bibinfo {author} {\bibfnamefont
  {J.}~\bibnamefont {{Primack}}},\ }\href {\doibase
  10.1111/j.1365-2966.2012.21007.x} {\bibfield  {journal} {\bibinfo  {journal}
  {\mnras}\ }\textbf {\bibinfo {volume} {423}},\ \bibinfo {pages} {3018}
  (\bibinfo {year} {2012})},\ \Eprint {http://arxiv.org/abs/1104.5130}
  {arXiv:1104.5130 [astro-ph.CO]} \BibitemShut {NoStop}%
\bibitem [{\citenamefont {{S{\'a}nchez-Conde}}\ and\ \citenamefont
  {{Prada}}(2014)}]{Sanchez_Conde_2014}%
  \BibitemOpen
  \bibfield  {author} {\bibinfo {author} {\bibfnamefont {M.~A.}\ \bibnamefont
  {{S{\'a}nchez-Conde}}}\ and\ \bibinfo {author} {\bibfnamefont
  {F.}~\bibnamefont {{Prada}}},\ }\href {\doibase 10.1093/mnras/stu1014}
  {\bibfield  {journal} {\bibinfo  {journal} {\mnras}\ }\textbf {\bibinfo
  {volume} {442}},\ \bibinfo {pages} {2271} (\bibinfo {year}
  {2014})}\BibitemShut {NoStop}%
\bibitem [{\citenamefont {{Dutton}}\ and\ \citenamefont
  {{Macci{\`o}}}(2014)}]{Dutton_2014}%
  \BibitemOpen
  \bibfield  {author} {\bibinfo {author} {\bibfnamefont {A.~A.}\ \bibnamefont
  {{Dutton}}}\ and\ \bibinfo {author} {\bibfnamefont {A.~V.}\ \bibnamefont
  {{Macci{\`o}}}},\ }\href {\doibase 10.1093/mnras/stu742} {\bibfield
  {journal} {\bibinfo  {journal} {\mnras}\ }\textbf {\bibinfo {volume} {441}},\
  \bibinfo {pages} {3359} (\bibinfo {year} {2014})}\BibitemShut {NoStop}%
\bibitem [{\citenamefont {{Diemer}}\ and\ \citenamefont
  {{Kravtsov}}(2015)}]{diemerkravtsov_2015}%
  \BibitemOpen
  \bibfield  {author} {\bibinfo {author} {\bibfnamefont {B.}~\bibnamefont
  {{Diemer}}}\ and\ \bibinfo {author} {\bibfnamefont {A.~V.}\ \bibnamefont
  {{Kravtsov}}},\ }\href {\doibase 10.1088/0004-637X/799/1/108} {\bibfield
  {journal} {\bibinfo  {journal} {\apj}\ }\textbf {\bibinfo {volume} {799}},\
  \bibinfo {eid} {108} (\bibinfo {year} {2015})},\ \Eprint
  {http://arxiv.org/abs/1407.4730} {arXiv:1407.4730 [astro-ph.CO]} \BibitemShut
  {NoStop}%
\bibitem [{\citenamefont {{Piccirillo}}\ \emph {et~al.}(2022)\citenamefont
  {{Piccirillo}} \emph {et~al.}}]{Piccirillo_2022}%
  \BibitemOpen
  \bibfield  {author} {\bibinfo {author} {\bibfnamefont {E.}~\bibnamefont
  {{Piccirillo}}} \emph {et~al.},\ }\href {\doibase
  10.1088/1475-7516/2022/08/058} {\bibfield  {journal} {\bibinfo  {journal}
  {\jcap}\ }\textbf {\bibinfo {volume} {2022}},\ \bibinfo {eid} {058} (\bibinfo
  {year} {2022})}\BibitemShut {NoStop}%
\bibitem [{\citenamefont {{Facchinetti}}\ \emph {et~al.}(2022)\citenamefont
  {{Facchinetti}}, \citenamefont {{Stref}}, \citenamefont {{Lacroix}},
  \citenamefont {{Lavalle}}, \citenamefont {{P{\'e}rez-Romero}}, \citenamefont
  {{Maurin}},\ and\ \citenamefont {{S{\'a}nchez-Conde}}}]{Facchinetti2022}%
  \BibitemOpen
  \bibfield  {author} {\bibinfo {author} {\bibfnamefont {G.}~\bibnamefont
  {{Facchinetti}}}, \bibinfo {author} {\bibfnamefont {M.}~\bibnamefont
  {{Stref}}}, \bibinfo {author} {\bibfnamefont {T.}~\bibnamefont {{Lacroix}}},
  \bibinfo {author} {\bibfnamefont {J.}~\bibnamefont {{Lavalle}}}, \bibinfo
  {author} {\bibfnamefont {J.}~\bibnamefont {{P{\'e}rez-Romero}}}, \bibinfo
  {author} {\bibfnamefont {D.}~\bibnamefont {{Maurin}}}, \ and\ \bibinfo
  {author} {\bibfnamefont {M.~A.}\ \bibnamefont {{S{\'a}nchez-Conde}}},\
  }\href@noop {} {\bibfield  {journal} {\bibinfo  {journal} {arXiv e-prints}\
  ,\ \bibinfo {eid} {arXiv:2203.16491}} (\bibinfo {year} {2022})},\ \Eprint
  {http://arxiv.org/abs/2203.16491} {arXiv:2203.16491 [astro-ph.CO]}
  \BibitemShut {NoStop}%
\bibitem [{\citenamefont {{Lacroix}}\ \emph {et~al.}(2022)\citenamefont
  {{Lacroix}} \emph {et~al.}}]{Lacroix_2022}%
  \BibitemOpen
  \bibfield  {author} {\bibinfo {author} {\bibfnamefont {T.}~\bibnamefont
  {{Lacroix}}} \emph {et~al.},\ }\href {\doibase 10.1088/1475-7516/2022/10/021}
  {\bibfield  {journal} {\bibinfo  {journal} {\jcap}\ }\textbf {\bibinfo
  {volume} {2022}},\ \bibinfo {eid} {021} (\bibinfo {year} {2022})}\BibitemShut
  {NoStop}%
\bibitem [{\citenamefont {{{\L}okas}}\ and\ \citenamefont
  {{Mamon}}(2001)}]{Lokas_2001}%
  \BibitemOpen
  \bibfield  {author} {\bibinfo {author} {\bibfnamefont {E.~L.}\ \bibnamefont
  {{{\L}okas}}}\ and\ \bibinfo {author} {\bibfnamefont {G.~A.}\ \bibnamefont
  {{Mamon}}},\ }\href {\doibase 10.1046/j.1365-8711.2001.04007.x} {\bibfield
  {journal} {\bibinfo  {journal} {\mnras}\ }\textbf {\bibinfo {volume} {321}},\
  \bibinfo {pages} {155} (\bibinfo {year} {2001})}\BibitemShut {NoStop}%
\bibitem [{\citenamefont {{Boddy}}\ \emph {et~al.}(2018)\citenamefont
  {{Boddy}}, \citenamefont {{Kumar}},\ and\ \citenamefont
  {{Strigari}}}]{Boddy_2018}%
  \BibitemOpen
  \bibfield  {author} {\bibinfo {author} {\bibfnamefont {K.~K.}\ \bibnamefont
  {{Boddy}}}, \bibinfo {author} {\bibfnamefont {J.}~\bibnamefont {{Kumar}}}, \
  and\ \bibinfo {author} {\bibfnamefont {L.~E.}\ \bibnamefont {{Strigari}}},\
  }\href {\doibase 10.1103/PhysRevD.98.063012} {\bibfield  {journal} {\bibinfo
  {journal} {\prd}\ }\textbf {\bibinfo {volume} {98}},\ \bibinfo {eid} {063012}
  (\bibinfo {year} {2018})}\BibitemShut {NoStop}%
\bibitem [{\citenamefont {Osipkov}(1979)}]{osipkov1979spherical}%
  \BibitemOpen
  \bibfield  {author} {\bibinfo {author} {\bibfnamefont {L.}~\bibnamefont
  {Osipkov}},\ }\href@noop {} {\bibfield  {journal} {\bibinfo  {journal} {Pisma
  v Astronomicheskii Zhurnal}\ }\textbf {\bibinfo {volume} {5}},\ \bibinfo
  {pages} {77} (\bibinfo {year} {1979})}\BibitemShut {NoStop}%
\bibitem [{\citenamefont {Merritt}(1985)}]{merritt1985spherical}%
  \BibitemOpen
  \bibfield  {author} {\bibinfo {author} {\bibfnamefont {D.}~\bibnamefont
  {Merritt}},\ }\href@noop {} {\bibfield  {journal} {\bibinfo  {journal} {The
  astronomical journal}\ }\textbf {\bibinfo {volume} {90}},\ \bibinfo {pages}
  {1027} (\bibinfo {year} {1985})}\BibitemShut {NoStop}%
\bibitem [{\citenamefont {Gerhard}(1991)}]{gerhard1991new}%
  \BibitemOpen
  \bibfield  {author} {\bibinfo {author} {\bibfnamefont {O.~E.}\ \bibnamefont
  {Gerhard}},\ }\href@noop {} {\bibfield  {journal} {\bibinfo  {journal}
  {Monthly Notices of the Royal Astronomical Society}\ }\textbf {\bibinfo
  {volume} {250}},\ \bibinfo {pages} {812} (\bibinfo {year}
  {1991})}\BibitemShut {NoStop}%
\bibitem [{\citenamefont {Binney}\ and\ \citenamefont
  {Tremaine}(2011)}]{binney2011galactic}%
  \BibitemOpen
  \bibfield  {author} {\bibinfo {author} {\bibfnamefont {J.}~\bibnamefont
  {Binney}}\ and\ \bibinfo {author} {\bibfnamefont {S.}~\bibnamefont
  {Tremaine}},\ }\href@noop {} {\emph {\bibinfo {title} {Galactic dynamics}}},\
  Vol.~\bibinfo {volume} {13}\ (\bibinfo  {publisher} {Princeton university
  press},\ \bibinfo {year} {2011})\BibitemShut {NoStop}%
\bibitem [{\citenamefont {{Mao}}\ \emph {et~al.}(2013)\citenamefont {{Mao}},
  \citenamefont {{Strigari}}, \citenamefont {{Wechsler}}, \citenamefont
  {{Wu}},\ and\ \citenamefont {{Hahn}}}]{maohalo2halo_2013}%
  \BibitemOpen
  \bibfield  {author} {\bibinfo {author} {\bibfnamefont {Y.-Y.}\ \bibnamefont
  {{Mao}}}, \bibinfo {author} {\bibfnamefont {L.~E.}\ \bibnamefont
  {{Strigari}}}, \bibinfo {author} {\bibfnamefont {R.~H.}\ \bibnamefont
  {{Wechsler}}}, \bibinfo {author} {\bibfnamefont {H.-Y.}\ \bibnamefont
  {{Wu}}}, \ and\ \bibinfo {author} {\bibfnamefont {O.}~\bibnamefont
  {{Hahn}}},\ }\href {\doibase 10.1088/0004-637X/764/1/35} {\bibfield
  {journal} {\bibinfo  {journal} {\apj}\ }\textbf {\bibinfo {volume} {764}},\
  \bibinfo {eid} {35} (\bibinfo {year} {2013})},\ \Eprint
  {http://arxiv.org/abs/1210.2721} {arXiv:1210.2721 [astro-ph.CO]} \BibitemShut
  {NoStop}%
\bibitem [{\citenamefont {{Christy}}\ \emph {et~al.}(2024)\citenamefont
  {{Christy}}, \citenamefont {{Kumar}},\ and\ \citenamefont
  {{Strigari}}}]{christyDM_velocitydist_2023}%
  \BibitemOpen
  \bibfield  {author} {\bibinfo {author} {\bibfnamefont {K.}~\bibnamefont
  {{Christy}}}, \bibinfo {author} {\bibfnamefont {J.}~\bibnamefont {{Kumar}}},
  \ and\ \bibinfo {author} {\bibfnamefont {L.~E.}\ \bibnamefont {{Strigari}}},\
  }\href {\doibase 10.1103/PhysRevD.109.063016} {\bibfield  {journal} {\bibinfo
   {journal} {\prd}\ }\textbf {\bibinfo {volume} {109}},\ \bibinfo {eid}
  {063016} (\bibinfo {year} {2024})},\ \Eprint
  {http://arxiv.org/abs/2309.01979} {arXiv:2309.01979 [astro-ph.CO]}
  \BibitemShut {NoStop}%
\bibitem [{\citenamefont {Zonca}\ \emph {et~al.}(2019)\citenamefont {Zonca},
  \citenamefont {Singer}, \citenamefont {Lenz}, \citenamefont {Reinecke},
  \citenamefont {Rosset}, \citenamefont {Hivon},\ and\ \citenamefont
  {Gorski}}]{Zonca_2019}%
  \BibitemOpen
  \bibfield  {author} {\bibinfo {author} {\bibfnamefont {A.}~\bibnamefont
  {Zonca}}, \bibinfo {author} {\bibfnamefont {L.}~\bibnamefont {Singer}},
  \bibinfo {author} {\bibfnamefont {D.}~\bibnamefont {Lenz}}, \bibinfo {author}
  {\bibfnamefont {M.}~\bibnamefont {Reinecke}}, \bibinfo {author}
  {\bibfnamefont {C.}~\bibnamefont {Rosset}}, \bibinfo {author} {\bibfnamefont
  {E.}~\bibnamefont {Hivon}}, \ and\ \bibinfo {author} {\bibfnamefont
  {K.}~\bibnamefont {Gorski}},\ }\href {\doibase 10.21105/joss.01298}
  {\bibfield  {journal} {\bibinfo  {journal} {Journal of Open Source Software}\
  }\textbf {\bibinfo {volume} {4}},\ \bibinfo {pages} {1298} (\bibinfo {year}
  {2019})}\BibitemShut {NoStop}%
\bibitem [{\citenamefont {{G{\'o}rski}}\ \emph {et~al.}(2005)\citenamefont
  {{G{\'o}rski}}, \citenamefont {{Hivon}}, \citenamefont {{Banday}},
  \citenamefont {{Wandelt}}, \citenamefont {{Hansen}}, \citenamefont
  {{Reinecke}},\ and\ \citenamefont {{Bartelmann}}}]{Gorski_2005}%
  \BibitemOpen
  \bibfield  {author} {\bibinfo {author} {\bibfnamefont {K.~M.}\ \bibnamefont
  {{G{\'o}rski}}}, \bibinfo {author} {\bibfnamefont {E.}~\bibnamefont
  {{Hivon}}}, \bibinfo {author} {\bibfnamefont {A.~J.}\ \bibnamefont
  {{Banday}}}, \bibinfo {author} {\bibfnamefont {B.~D.}\ \bibnamefont
  {{Wandelt}}}, \bibinfo {author} {\bibfnamefont {F.~K.}\ \bibnamefont
  {{Hansen}}}, \bibinfo {author} {\bibfnamefont {M.}~\bibnamefont
  {{Reinecke}}}, \ and\ \bibinfo {author} {\bibfnamefont {M.}~\bibnamefont
  {{Bartelmann}}},\ }\href {\doibase 10.1086/427976} {\bibfield  {journal}
  {\bibinfo  {journal} {\apj}\ }\textbf {\bibinfo {volume} {622}},\ \bibinfo
  {pages} {759} (\bibinfo {year} {2005})}\BibitemShut {NoStop}%
\bibitem [{\citenamefont {{Wood}}\ \emph {et~al.}(2017)\citenamefont {{Wood}},
  \citenamefont {{Caputo}}, \citenamefont {{Charles}}, \citenamefont {{Di
  Mauro}}, \citenamefont {{Magill}}, \citenamefont {{Perkins}},\ and\
  \citenamefont {{Fermi-LAT Collaboration}}}]{FermiPy}%
  \BibitemOpen
  \bibfield  {author} {\bibinfo {author} {\bibfnamefont {M.}~\bibnamefont
  {{Wood}}}, \bibinfo {author} {\bibfnamefont {R.}~\bibnamefont {{Caputo}}},
  \bibinfo {author} {\bibfnamefont {E.}~\bibnamefont {{Charles}}}, \bibinfo
  {author} {\bibfnamefont {M.}~\bibnamefont {{Di Mauro}}}, \bibinfo {author}
  {\bibfnamefont {J.}~\bibnamefont {{Magill}}}, \bibinfo {author}
  {\bibfnamefont {J.~S.}\ \bibnamefont {{Perkins}}}, \ and\ \bibinfo {author}
  {\bibnamefont {{Fermi-LAT Collaboration}}},\ }in\ \href@noop {} {\emph
  {\bibinfo {booktitle} {35th International Cosmic Ray Conference
  (ICRC2017)}}},\ \bibinfo {series} {International Cosmic Ray Conference},
  Vol.\ \bibinfo {volume} {301}\ (\bibinfo {year} {2017})\ p.\ \bibinfo {pages}
  {824}\BibitemShut {NoStop}%
\bibitem [{\citenamefont {{Acero}}\ \emph {et~al.}(2016)\citenamefont {{Acero}}
  \emph {et~al.}}]{Acero_2016}%
  \BibitemOpen
  \bibfield  {author} {\bibinfo {author} {\bibfnamefont {F.}~\bibnamefont
  {{Acero}}} \emph {et~al.},\ }\href {\doibase 10.3847/0067-0049/223/2/26}
  {\bibfield  {journal} {\bibinfo  {journal} {\apjs}\ }\textbf {\bibinfo
  {volume} {223}},\ \bibinfo {eid} {26} (\bibinfo {year} {2016})}\BibitemShut
  {NoStop}%
\bibitem [{\citenamefont {Phan}\ \emph {et~al.}(2019)\citenamefont {Phan},
  \citenamefont {Pradhan},\ and\ \citenamefont
  {Jankowiak}}]{phan2019composable}%
  \BibitemOpen
  \bibfield  {author} {\bibinfo {author} {\bibfnamefont {D.}~\bibnamefont
  {Phan}}, \bibinfo {author} {\bibfnamefont {N.}~\bibnamefont {Pradhan}}, \
  and\ \bibinfo {author} {\bibfnamefont {M.}~\bibnamefont {Jankowiak}},\
  }\href@noop {} {\bibfield  {journal} {\bibinfo  {journal} {arXiv preprint
  arXiv:1912.11554}\ } (\bibinfo {year} {2019})}\BibitemShut {NoStop}%
\bibitem [{\citenamefont {Bingham}\ \emph {et~al.}(2019)\citenamefont {Bingham}
  \emph {et~al.}}]{bingham2019pyro}%
  \BibitemOpen
  \bibfield  {author} {\bibinfo {author} {\bibfnamefont {E.}~\bibnamefont
  {Bingham}} \emph {et~al.},\ }\href {http://jmlr.org/papers/v20/18-403.html}
  {\bibfield  {journal} {\bibinfo  {journal} {J. Mach. Learn. Res.}\ }\textbf
  {\bibinfo {volume} {20}},\ \bibinfo {pages} {28:1} (\bibinfo {year}
  {2019})}\BibitemShut {NoStop}%
\bibitem [{\citenamefont {{Jeltema}}\ and\ \citenamefont
  {{Profumo}}(2008)}]{Jeltema_2008}%
  \BibitemOpen
  \bibfield  {author} {\bibinfo {author} {\bibfnamefont {T.~E.}\ \bibnamefont
  {{Jeltema}}}\ and\ \bibinfo {author} {\bibfnamefont {S.}~\bibnamefont
  {{Profumo}}},\ }\href {\doibase 10.1088/1475-7516/2008/11/003} {\bibfield
  {journal} {\bibinfo  {journal} {\jcap}\ }\textbf {\bibinfo {volume} {2008}},\
  \bibinfo {eid} {003} (\bibinfo {year} {2008})}\BibitemShut {NoStop}%
\bibitem [{\citenamefont {{Zhao}}\ \emph {et~al.}(2016)\citenamefont {{Zhao}},
  \citenamefont {{Bi}}, \citenamefont {{Jia}}, \citenamefont {{Yin}},\ and\
  \citenamefont {{Zhu}}}]{Zhao_2016}%
  \BibitemOpen
  \bibfield  {author} {\bibinfo {author} {\bibfnamefont {Y.}~\bibnamefont
  {{Zhao}}}, \bibinfo {author} {\bibfnamefont {X.-J.}\ \bibnamefont {{Bi}}},
  \bibinfo {author} {\bibfnamefont {H.-Y.}\ \bibnamefont {{Jia}}}, \bibinfo
  {author} {\bibfnamefont {P.-F.}\ \bibnamefont {{Yin}}}, \ and\ \bibinfo
  {author} {\bibfnamefont {F.-R.}\ \bibnamefont {{Zhu}}},\ }\href {\doibase
  10.1103/PhysRevD.93.083513} {\bibfield  {journal} {\bibinfo  {journal}
  {\prd}\ }\textbf {\bibinfo {volume} {93}},\ \bibinfo {eid} {083513} (\bibinfo
  {year} {2016})}\BibitemShut {NoStop}%
\bibitem [{\citenamefont {{Boddy}}\ \emph {et~al.}(2020)\citenamefont
  {{Boddy}}, \citenamefont {{Kumar}}, \citenamefont {{Pace}}, \citenamefont
  {{Runburg}},\ and\ \citenamefont {{Strigari}}}]{Boddy_2020}%
  \BibitemOpen
  \bibfield  {author} {\bibinfo {author} {\bibfnamefont {K.~K.}\ \bibnamefont
  {{Boddy}}}, \bibinfo {author} {\bibfnamefont {J.}~\bibnamefont {{Kumar}}},
  \bibinfo {author} {\bibfnamefont {A.~B.}\ \bibnamefont {{Pace}}}, \bibinfo
  {author} {\bibfnamefont {J.}~\bibnamefont {{Runburg}}}, \ and\ \bibinfo
  {author} {\bibfnamefont {L.~E.}\ \bibnamefont {{Strigari}}},\ }\href
  {\doibase 10.1103/PhysRevD.102.023029} {\bibfield  {journal} {\bibinfo
  {journal} {\prd}\ }\textbf {\bibinfo {volume} {102}},\ \bibinfo {eid}
  {023029} (\bibinfo {year} {2020})}\BibitemShut {NoStop}%
\bibitem [{\citenamefont {{Boudaud}}\ \emph {et~al.}(2019)\citenamefont
  {{Boudaud}}, \citenamefont {{Lacroix}}, \citenamefont {{Stref}},\ and\
  \citenamefont {{Lavalle}}}]{Boudaud_2019}%
  \BibitemOpen
  \bibfield  {author} {\bibinfo {author} {\bibfnamefont {M.}~\bibnamefont
  {{Boudaud}}}, \bibinfo {author} {\bibfnamefont {T.}~\bibnamefont
  {{Lacroix}}}, \bibinfo {author} {\bibfnamefont {M.}~\bibnamefont {{Stref}}},
  \ and\ \bibinfo {author} {\bibfnamefont {J.}~\bibnamefont {{Lavalle}}},\
  }\href {\doibase 10.1103/PhysRevD.99.061302} {\bibfield  {journal} {\bibinfo
  {journal} {\prd}\ }\textbf {\bibinfo {volume} {99}},\ \bibinfo {eid} {061302}
  (\bibinfo {year} {2019})},\ \Eprint {http://arxiv.org/abs/1810.01680}
  {arXiv:1810.01680 [astro-ph.HE]} \BibitemShut {NoStop}%
\bibitem [{\citenamefont {{Blumenthal}}\ \emph {et~al.}(1986)\citenamefont
  {{Blumenthal}}, \citenamefont {{Faber}}, \citenamefont {{Flores}},\ and\
  \citenamefont {{Primack}}}]{Blumenthal}%
  \BibitemOpen
  \bibfield  {author} {\bibinfo {author} {\bibfnamefont {G.~R.}\ \bibnamefont
  {{Blumenthal}}}, \bibinfo {author} {\bibfnamefont {S.~M.}\ \bibnamefont
  {{Faber}}}, \bibinfo {author} {\bibfnamefont {R.}~\bibnamefont {{Flores}}}, \
  and\ \bibinfo {author} {\bibfnamefont {J.~R.}\ \bibnamefont {{Primack}}},\
  }\href {\doibase 10.1086/163867} {\bibfield  {journal} {\bibinfo  {journal}
  {\apj}\ }\textbf {\bibinfo {volume} {301}},\ \bibinfo {pages} {27} (\bibinfo
  {year} {1986})}\BibitemShut {NoStop}%
\bibitem [{\citenamefont {{Gnedin}}\ \emph {et~al.}(2004)\citenamefont
  {{Gnedin}}, \citenamefont {{Kravtsov}}, \citenamefont {{Klypin}},\ and\
  \citenamefont {{Nagai}}}]{Gnedin}%
  \BibitemOpen
  \bibfield  {author} {\bibinfo {author} {\bibfnamefont {O.~Y.}\ \bibnamefont
  {{Gnedin}}}, \bibinfo {author} {\bibfnamefont {A.~V.}\ \bibnamefont
  {{Kravtsov}}}, \bibinfo {author} {\bibfnamefont {A.~A.}\ \bibnamefont
  {{Klypin}}}, \ and\ \bibinfo {author} {\bibfnamefont {D.}~\bibnamefont
  {{Nagai}}},\ }\href {\doibase 10.1086/424914} {\bibfield  {journal} {\bibinfo
   {journal} {\apj}\ }\textbf {\bibinfo {volume} {616}},\ \bibinfo {pages} {16}
  (\bibinfo {year} {2004})}\BibitemShut {NoStop}%
\bibitem [{\citenamefont {{Pontzen}}\ and\ \citenamefont
  {{Governato}}(2012)}]{Pontzen_Governato}%
  \BibitemOpen
  \bibfield  {author} {\bibinfo {author} {\bibfnamefont {A.}~\bibnamefont
  {{Pontzen}}}\ and\ \bibinfo {author} {\bibfnamefont {F.}~\bibnamefont
  {{Governato}}},\ }\href {\doibase 10.1111/j.1365-2966.2012.20571.x}
  {\bibfield  {journal} {\bibinfo  {journal} {Mon. Not. Roy. Astron. Soc.}\
  }\textbf {\bibinfo {volume} {421}},\ \bibinfo {pages} {3464} (\bibinfo {year}
  {2012})}\BibitemShut {NoStop}%
\bibitem [{\citenamefont {Del~Popolo}\ and\ \citenamefont
  {Pace}(2016)}]{DP_CuspCore}%
  \BibitemOpen
  \bibfield  {author} {\bibinfo {author} {\bibfnamefont {A.}~\bibnamefont
  {Del~Popolo}}\ and\ \bibinfo {author} {\bibfnamefont {F.}~\bibnamefont
  {Pace}},\ }\href@noop {} {\bibfield  {journal} {\bibinfo  {journal}
  {Astrophys. Space Sci.}\ }\textbf {\bibinfo {volume} {361}},\ \bibinfo
  {pages} {162} (\bibinfo {year} {2016})},\ \bibinfo {note} {[Erratum:
  Astrophys.Space Sci. 361, 225 (2016)]},\ \Eprint
  {http://arxiv.org/abs/1502.01947} {1502.01947} \BibitemShut {NoStop}%
\bibitem [{\citenamefont {{Board}}\ \emph {et~al.}(2021)\citenamefont {{Board}}
  \emph {et~al.}}]{Board_2021}%
  \BibitemOpen
  \bibfield  {author} {\bibinfo {author} {\bibfnamefont {E.}~\bibnamefont
  {{Board}}} \emph {et~al.},\ }\href {\doibase 10.1088/1475-7516/2021/04/070}
  {\bibfield  {journal} {\bibinfo  {journal} {\jcap}\ }\textbf {\bibinfo
  {volume} {2021}},\ \bibinfo {eid} {070} (\bibinfo {year} {2021})}\BibitemShut
  {NoStop}%
\bibitem [{\citenamefont {{McKeown}}\ \emph {et~al.}(2022)\citenamefont
  {{McKeown}} \emph {et~al.}}]{McKeown_2022}%
  \BibitemOpen
  \bibfield  {author} {\bibinfo {author} {\bibfnamefont {D.}~\bibnamefont
  {{McKeown}}} \emph {et~al.},\ }\href {\doibase 10.1093/mnras/stac966}
  {\bibfield  {journal} {\bibinfo  {journal} {\mnras}\ }\textbf {\bibinfo
  {volume} {513}},\ \bibinfo {pages} {55} (\bibinfo {year} {2022})}\BibitemShut
  {NoStop}%
\bibitem [{\citenamefont {Piffaretti}\ \emph {et~al.}(2011)\citenamefont
  {Piffaretti}, \citenamefont {Arnaud}, \citenamefont {Pratt}, \citenamefont
  {Pointecouteau},\ and\ \citenamefont {Melin}}]{piffaretti2011mcxc}%
  \BibitemOpen
  \bibfield  {author} {\bibinfo {author} {\bibfnamefont {R.}~\bibnamefont
  {Piffaretti}}, \bibinfo {author} {\bibfnamefont {M.}~\bibnamefont {Arnaud}},
  \bibinfo {author} {\bibfnamefont {G.}~\bibnamefont {Pratt}}, \bibinfo
  {author} {\bibfnamefont {E.}~\bibnamefont {Pointecouteau}}, \ and\ \bibinfo
  {author} {\bibfnamefont {J.-B.}\ \bibnamefont {Melin}},\ }\href@noop {}
  {\bibfield  {journal} {\bibinfo  {journal} {Astronomy \& Astrophysics}\
  }\textbf {\bibinfo {volume} {534}},\ \bibinfo {pages} {A109} (\bibinfo {year}
  {2011})}\BibitemShut {NoStop}%
\bibitem [{\citenamefont {{Merritt}}(1987)}]{Meritt_1987}%
  \BibitemOpen
  \bibfield  {author} {\bibinfo {author} {\bibfnamefont {D.}~\bibnamefont
  {{Merritt}}},\ }\href {\doibase 10.1086/164953} {\bibfield  {journal}
  {\bibinfo  {journal} {\apj}\ }\textbf {\bibinfo {volume} {313}},\ \bibinfo
  {pages} {121} (\bibinfo {year} {1987})}\BibitemShut {NoStop}%
\bibitem [{\citenamefont {{Evrard}}\ \emph {et~al.}(1996)\citenamefont
  {{Evrard}}, \citenamefont {{Metzler}},\ and\ \citenamefont
  {{Navarro}}}]{Evrard_1996}%
  \BibitemOpen
  \bibfield  {author} {\bibinfo {author} {\bibfnamefont {A.~E.}\ \bibnamefont
  {{Evrard}}}, \bibinfo {author} {\bibfnamefont {C.~A.}\ \bibnamefont
  {{Metzler}}}, \ and\ \bibinfo {author} {\bibfnamefont {J.~F.}\ \bibnamefont
  {{Navarro}}},\ }\href {\doibase 10.1086/177798} {\bibfield  {journal}
  {\bibinfo  {journal} {\apj}\ }\textbf {\bibinfo {volume} {469}},\ \bibinfo
  {pages} {494} (\bibinfo {year} {1996})}\BibitemShut {NoStop}%
\bibitem [{\citenamefont {{Sunyaev}}\ and\ \citenamefont
  {{Zeldovich}}(1970)}]{Sunyaev_1970}%
  \BibitemOpen
  \bibfield  {author} {\bibinfo {author} {\bibfnamefont {R.~A.}\ \bibnamefont
  {{Sunyaev}}}\ and\ \bibinfo {author} {\bibfnamefont {Y.~B.}\ \bibnamefont
  {{Zeldovich}}},\ }\href@noop {} {\bibfield  {journal} {\bibinfo  {journal}
  {Comments on Astrophysics and Space Physics}\ }\textbf {\bibinfo {volume}
  {2}},\ \bibinfo {pages} {66} (\bibinfo {year} {1970})}\BibitemShut {NoStop}%
\bibitem [{\citenamefont {{Sunyaev}}\ and\ \citenamefont
  {{Zeldovich}}(1980)}]{Sunyaev_1980}%
  \BibitemOpen
  \bibfield  {author} {\bibinfo {author} {\bibfnamefont {R.~A.}\ \bibnamefont
  {{Sunyaev}}}\ and\ \bibinfo {author} {\bibfnamefont {I.~B.}\ \bibnamefont
  {{Zeldovich}}},\ }\href {\doibase 10.1146/annurev.aa.18.090180.002541}
  {\bibfield  {journal} {\bibinfo  {journal} {\araa}\ }\textbf {\bibinfo
  {volume} {18}},\ \bibinfo {pages} {537} (\bibinfo {year} {1980})}\BibitemShut
  {NoStop}%
\bibitem [{\citenamefont {{Bonnet}}\ \emph {et~al.}(1994)\citenamefont
  {{Bonnet}}, \citenamefont {{Mellier}},\ and\ \citenamefont
  {{Fort}}}]{Bonnet_1994}%
  \BibitemOpen
  \bibfield  {author} {\bibinfo {author} {\bibfnamefont {H.}~\bibnamefont
  {{Bonnet}}}, \bibinfo {author} {\bibfnamefont {Y.}~\bibnamefont {{Mellier}}},
  \ and\ \bibinfo {author} {\bibfnamefont {B.}~\bibnamefont {{Fort}}},\ }\href
  {\doibase 10.1086/187370} {\bibfield  {journal} {\bibinfo  {journal} {\apjl}\
  }\textbf {\bibinfo {volume} {427}},\ \bibinfo {pages} {L83} (\bibinfo {year}
  {1994})}\BibitemShut {NoStop}%
\bibitem [{\citenamefont {{Fahlman}}\ \emph {et~al.}(1994)\citenamefont
  {{Fahlman}}, \citenamefont {{Kaiser}}, \citenamefont {{Squires}},\ and\
  \citenamefont {{Woods}}}]{Fahlman_1994}%
  \BibitemOpen
  \bibfield  {author} {\bibinfo {author} {\bibfnamefont {G.}~\bibnamefont
  {{Fahlman}}}, \bibinfo {author} {\bibfnamefont {N.}~\bibnamefont {{Kaiser}}},
  \bibinfo {author} {\bibfnamefont {G.}~\bibnamefont {{Squires}}}, \ and\
  \bibinfo {author} {\bibfnamefont {D.}~\bibnamefont {{Woods}}},\ }\href
  {\doibase 10.1086/174974} {\bibfield  {journal} {\bibinfo  {journal} {\apj}\
  }\textbf {\bibinfo {volume} {437}},\ \bibinfo {pages} {56} (\bibinfo {year}
  {1994})}\BibitemShut {NoStop}%
\bibitem [{\citenamefont {{Stopyra}}\ \emph {et~al.}(2021)\citenamefont
  {{Stopyra}}, \citenamefont {{Peiris}}, \citenamefont {{Pontzen}},
  \citenamefont {{Jasche}},\ and\ \citenamefont {{Natarajan}}}]{Stopyra_2021}%
  \BibitemOpen
  \bibfield  {author} {\bibinfo {author} {\bibfnamefont {S.}~\bibnamefont
  {{Stopyra}}}, \bibinfo {author} {\bibfnamefont {H.~V.}\ \bibnamefont
  {{Peiris}}}, \bibinfo {author} {\bibfnamefont {A.}~\bibnamefont {{Pontzen}}},
  \bibinfo {author} {\bibfnamefont {J.}~\bibnamefont {{Jasche}}}, \ and\
  \bibinfo {author} {\bibfnamefont {P.}~\bibnamefont {{Natarajan}}},\ }\href
  {\doibase 10.1093/mnras/stab2456} {\bibfield  {journal} {\bibinfo  {journal}
  {\mnras}\ }\textbf {\bibinfo {volume} {507}},\ \bibinfo {pages} {5425}
  (\bibinfo {year} {2021})}\BibitemShut {NoStop}%
\bibitem [{\citenamefont {{Tassev}}\ \emph {et~al.}(2013)\citenamefont
  {{Tassev}}, \citenamefont {{Zaldarriaga}},\ and\ \citenamefont
  {{Eisenstein}}}]{Tassev_2013}%
  \BibitemOpen
  \bibfield  {author} {\bibinfo {author} {\bibfnamefont {S.}~\bibnamefont
  {{Tassev}}}, \bibinfo {author} {\bibfnamefont {M.}~\bibnamefont
  {{Zaldarriaga}}}, \ and\ \bibinfo {author} {\bibfnamefont {D.~J.}\
  \bibnamefont {{Eisenstein}}},\ }\href {\doibase
  10.1088/1475-7516/2013/06/036} {\bibfield  {journal} {\bibinfo  {journal}
  {\jcap}\ }\textbf {\bibinfo {volume} {2013}},\ \bibinfo {eid} {036} (\bibinfo
  {year} {2013})}\BibitemShut {NoStop}%
\bibitem [{\citenamefont {{Blas}}\ \emph {et~al.}(2011)\citenamefont {{Blas}},
  \citenamefont {{Lesgourgues}},\ and\ \citenamefont {{Tram}}}]{Blas_2011}%
  \BibitemOpen
  \bibfield  {author} {\bibinfo {author} {\bibfnamefont {D.}~\bibnamefont
  {{Blas}}}, \bibinfo {author} {\bibfnamefont {J.}~\bibnamefont
  {{Lesgourgues}}}, \ and\ \bibinfo {author} {\bibfnamefont {T.}~\bibnamefont
  {{Tram}}},\ }\href {\doibase 10.1088/1475-7516/2011/07/034} {\bibfield
  {journal} {\bibinfo  {journal} {Journal of Cosmology and Astro-Particle
  Physics}\ }\textbf {\bibinfo {volume} {2011}},\ \bibinfo {eid} {034}
  (\bibinfo {year} {2011})}\BibitemShut {NoStop}%
\bibitem [{\citenamefont {{Eisenstein}}\ and\ \citenamefont
  {{Hu}}(1998)}]{Eisenstein_1998}%
  \BibitemOpen
  \bibfield  {author} {\bibinfo {author} {\bibfnamefont {D.~J.}\ \bibnamefont
  {{Eisenstein}}}\ and\ \bibinfo {author} {\bibfnamefont {W.}~\bibnamefont
  {{Hu}}},\ }\href {\doibase 10.1086/305424} {\bibfield  {journal} {\bibinfo
  {journal} {\apj}\ }\textbf {\bibinfo {volume} {496}},\ \bibinfo {pages} {605}
  (\bibinfo {year} {1998})}\BibitemShut {NoStop}%
\bibitem [{\citenamefont {{Porqueres}}\ \emph {et~al.}(2019)\citenamefont
  {{Porqueres}}, \citenamefont {{Kodi Ramanah}}, \citenamefont {{Jasche}},\
  and\ \citenamefont {{Lavaux}}}]{Porqueres_2019}%
  \BibitemOpen
  \bibfield  {author} {\bibinfo {author} {\bibfnamefont {N.}~\bibnamefont
  {{Porqueres}}}, \bibinfo {author} {\bibfnamefont {D.}~\bibnamefont {{Kodi
  Ramanah}}}, \bibinfo {author} {\bibfnamefont {J.}~\bibnamefont {{Jasche}}}, \
  and\ \bibinfo {author} {\bibfnamefont {G.}~\bibnamefont {{Lavaux}}},\ }\href
  {\doibase 10.1051/0004-6361/201834844} {\bibfield  {journal} {\bibinfo
  {journal} {\aap}\ }\textbf {\bibinfo {volume} {624}},\ \bibinfo {eid} {A115}
  (\bibinfo {year} {2019})},\ \Eprint {http://arxiv.org/abs/1812.05113}
  {arXiv:1812.05113 [astro-ph.CO]} \BibitemShut {NoStop}%
\bibitem [{\citenamefont {Stopyra}\ \emph {et~al.}(2023)\citenamefont
  {Stopyra}, \citenamefont {Peiris}, \citenamefont {Pontzen}, \citenamefont
  {Jasche},\ and\ \citenamefont {Lavaux}}]{Stopyra_2023}%
  \BibitemOpen
  \bibfield  {author} {\bibinfo {author} {\bibfnamefont {S.}~\bibnamefont
  {Stopyra}}, \bibinfo {author} {\bibfnamefont {H.~V.}\ \bibnamefont {Peiris}},
  \bibinfo {author} {\bibfnamefont {A.}~\bibnamefont {Pontzen}}, \bibinfo
  {author} {\bibfnamefont {J.}~\bibnamefont {Jasche}}, \ and\ \bibinfo {author}
  {\bibfnamefont {G.}~\bibnamefont {Lavaux}},\ }\href@noop {} {\enquote
  {\bibinfo {title} {Towards accurate field-level inference of massive cosmic
  structures},}\ } (\bibinfo {year} {2023}),\ \Eprint
  {http://arxiv.org/abs/2304.09193} {arXiv:2304.09193 [astro-ph.CO]}
  \BibitemShut {NoStop}%
\bibitem [{\citenamefont {{Steigman}}\ \emph {et~al.}(2012)\citenamefont
  {{Steigman}}, \citenamefont {{Dasgupta}},\ and\ \citenamefont
  {{Beacom}}}]{Steigman_2012}%
  \BibitemOpen
  \bibfield  {author} {\bibinfo {author} {\bibfnamefont {G.}~\bibnamefont
  {{Steigman}}}, \bibinfo {author} {\bibfnamefont {B.}~\bibnamefont
  {{Dasgupta}}}, \ and\ \bibinfo {author} {\bibfnamefont {J.~F.}\ \bibnamefont
  {{Beacom}}},\ }\href {\doibase 10.1103/PhysRevD.86.023506} {\bibfield
  {journal} {\bibinfo  {journal} {\prd}\ }\textbf {\bibinfo {volume} {86}},\
  \bibinfo {eid} {023506} (\bibinfo {year} {2012})}\BibitemShut {NoStop}%
\bibitem [{\citenamefont {{Laine}}\ and\ \citenamefont
  {{Schr{\"o}der}}(2006)}]{Laine_2006}%
  \BibitemOpen
  \bibfield  {author} {\bibinfo {author} {\bibfnamefont {M.}~\bibnamefont
  {{Laine}}}\ and\ \bibinfo {author} {\bibfnamefont {Y.}~\bibnamefont
  {{Schr{\"o}der}}},\ }\href {\doibase 10.1103/PhysRevD.73.085009} {\bibfield
  {journal} {\bibinfo  {journal} {\prd}\ }\textbf {\bibinfo {volume} {73}},\
  \bibinfo {eid} {085009} (\bibinfo {year} {2006})}\BibitemShut {NoStop}%
\bibitem [{\citenamefont {Hairer}\ and\ \citenamefont
  {Wanner}(1996)}]{Haire_1996}%
  \BibitemOpen
  \bibfield  {author} {\bibinfo {author} {\bibfnamefont {E.}~\bibnamefont
  {Hairer}}\ and\ \bibinfo {author} {\bibfnamefont {G.}~\bibnamefont
  {Wanner}},\ }\href {\doibase 10.1007/978-3-662-09947-6} {\emph {\bibinfo
  {title} {Solving Ordinary Differential Equations II. Stiff and
  Differential-Algebraic Problems}}},\ Vol.~\bibinfo {volume} {14}\ (\bibinfo
  {year} {1996})\BibitemShut {NoStop}%
\bibitem [{\citenamefont {Kolb}\ and\ \citenamefont
  {Turner}(1990)}]{Kolb_1990}%
  \BibitemOpen
  \bibfield  {author} {\bibinfo {author} {\bibfnamefont {E.~W.}\ \bibnamefont
  {Kolb}}\ and\ \bibinfo {author} {\bibfnamefont {M.~S.}\ \bibnamefont
  {Turner}},\ }\href {\doibase 10.1201/9780429492860} {\emph {\bibinfo {title}
  {{The Early Universe}}}},\ Vol.~\bibinfo {volume} {69}\ (\bibinfo {year}
  {1990})\BibitemShut {NoStop}%
\bibitem [{\citenamefont {{Komatsu}}\ \emph {et~al.}(2011)\citenamefont
  {{Komatsu}} \emph {et~al.}}]{Komatsu_2011}%
  \BibitemOpen
  \bibfield  {author} {\bibinfo {author} {\bibfnamefont {E.}~\bibnamefont
  {{Komatsu}}} \emph {et~al.},\ }\href {\doibase 10.1088/0067-0049/192/2/18}
  {\bibfield  {journal} {\bibinfo  {journal} {\apjs}\ }\textbf {\bibinfo
  {volume} {192}},\ \bibinfo {eid} {18} (\bibinfo {year} {2011})}\BibitemShut
  {NoStop}%
\bibitem [{\citenamefont {{Cafmpbell}}\ \emph {et~al.}(2010)\citenamefont
  {{Cafmpbell}}, \citenamefont {{Dutta}},\ and\ \citenamefont
  {{Komatsu}}}]{Campbell_2010}%
  \BibitemOpen
  \bibfield  {author} {\bibinfo {author} {\bibfnamefont {S.}~\bibnamefont
  {{Cafmpbell}}}, \bibinfo {author} {\bibfnamefont {B.}~\bibnamefont
  {{Dutta}}}, \ and\ \bibinfo {author} {\bibfnamefont {E.}~\bibnamefont
  {{Komatsu}}},\ }\href {\doibase 10.1103/PhysRevD.82.095007} {\bibfield
  {journal} {\bibinfo  {journal} {\prd}\ }\textbf {\bibinfo {volume} {82}},\
  \bibinfo {eid} {095007} (\bibinfo {year} {2010})}\BibitemShut {NoStop}%
\bibitem [{\citenamefont {{Campbell}}\ and\ \citenamefont
  {{Dutta}}(2011)}]{Campbell_2011}%
  \BibitemOpen
  \bibfield  {author} {\bibinfo {author} {\bibfnamefont {S.}~\bibnamefont
  {{Campbell}}}\ and\ \bibinfo {author} {\bibfnamefont {B.}~\bibnamefont
  {{Dutta}}},\ }\href {\doibase 10.1103/PhysRevD.84.075004} {\bibfield
  {journal} {\bibinfo  {journal} {\prd}\ }\textbf {\bibinfo {volume} {84}},\
  \bibinfo {eid} {075004} (\bibinfo {year} {2011})}\BibitemShut {NoStop}%
\bibitem [{\citenamefont {{Griest}}\ and\ \citenamefont
  {{Seckel}}(1991)}]{Griest_1991}%
  \BibitemOpen
  \bibfield  {author} {\bibinfo {author} {\bibfnamefont {K.}~\bibnamefont
  {{Griest}}}\ and\ \bibinfo {author} {\bibfnamefont {D.}~\bibnamefont
  {{Seckel}}},\ }\href {\doibase 10.1103/PhysRevD.43.3191} {\bibfield
  {journal} {\bibinfo  {journal} {\prd}\ }\textbf {\bibinfo {volume} {43}},\
  \bibinfo {pages} {3191} (\bibinfo {year} {1991})}\BibitemShut {NoStop}%
\bibitem [{\citenamefont {{Choquette}}\ \emph {et~al.}(2016)\citenamefont
  {{Choquette}}, \citenamefont {{Cline}},\ and\ \citenamefont
  {{Cornell}}}]{Choquette_2016}%
  \BibitemOpen
  \bibfield  {author} {\bibinfo {author} {\bibfnamefont {J.}~\bibnamefont
  {{Choquette}}}, \bibinfo {author} {\bibfnamefont {J.~M.}\ \bibnamefont
  {{Cline}}}, \ and\ \bibinfo {author} {\bibfnamefont {J.~M.}\ \bibnamefont
  {{Cornell}}},\ }\href {\doibase 10.1103/PhysRevD.94.015018} {\bibfield
  {journal} {\bibinfo  {journal} {\prd}\ }\textbf {\bibinfo {volume} {94}},\
  \bibinfo {eid} {015018} (\bibinfo {year} {2016})}\BibitemShut {NoStop}%
\bibitem [{\citenamefont {{Johnson}}\ \emph {et~al.}(2019)\citenamefont
  {{Johnson}}, \citenamefont {{Caputo}}, \citenamefont {{Karwin}},
  \citenamefont {{Murgia}}, \citenamefont {{Ritz}}, \citenamefont {{Shelton}},\
  and\ \citenamefont {{Fermi-LAT Collaboration}}}]{Johnson_2019}%
  \BibitemOpen
  \bibfield  {author} {\bibinfo {author} {\bibfnamefont {C.}~\bibnamefont
  {{Johnson}}}, \bibinfo {author} {\bibfnamefont {R.}~\bibnamefont {{Caputo}}},
  \bibinfo {author} {\bibfnamefont {C.}~\bibnamefont {{Karwin}}}, \bibinfo
  {author} {\bibfnamefont {S.}~\bibnamefont {{Murgia}}}, \bibinfo {author}
  {\bibfnamefont {S.}~\bibnamefont {{Ritz}}}, \bibinfo {author} {\bibfnamefont
  {J.}~\bibnamefont {{Shelton}}}, \ and\ \bibinfo {author} {\bibnamefont
  {{Fermi-LAT Collaboration}}},\ }\href {\doibase 10.1103/PhysRevD.99.103007}
  {\bibfield  {journal} {\bibinfo  {journal} {\prd}\ }\textbf {\bibinfo
  {volume} {99}},\ \bibinfo {eid} {103007} (\bibinfo {year}
  {2019})}\BibitemShut {NoStop}%
\bibitem [{\citenamefont {{Kiriu}}\ \emph {et~al.}(2022)\citenamefont
  {{Kiriu}}, \citenamefont {{Kumar}},\ and\ \citenamefont
  {{Runburg}}}]{Kiriu_2022}%
  \BibitemOpen
  \bibfield  {author} {\bibinfo {author} {\bibfnamefont {K.}~\bibnamefont
  {{Kiriu}}}, \bibinfo {author} {\bibfnamefont {J.}~\bibnamefont {{Kumar}}}, \
  and\ \bibinfo {author} {\bibfnamefont {J.}~\bibnamefont {{Runburg}}},\
  }\href@noop {} {\bibfield  {journal} {\bibinfo  {journal} {arXiv e-prints}\
  ,\ \bibinfo {eid} {arXiv:2208.14002}} (\bibinfo {year} {2022})}\BibitemShut
  {NoStop}%
\bibitem [{\citenamefont {{Zhao}}\ \emph {et~al.}(2018)\citenamefont {{Zhao}},
  \citenamefont {{Bi}}, \citenamefont {{Yin}},\ and\ \citenamefont
  {{Zhang}}}]{Zhao_2018}%
  \BibitemOpen
  \bibfield  {author} {\bibinfo {author} {\bibfnamefont {Y.}~\bibnamefont
  {{Zhao}}}, \bibinfo {author} {\bibfnamefont {X.-J.}\ \bibnamefont {{Bi}}},
  \bibinfo {author} {\bibfnamefont {P.-F.}\ \bibnamefont {{Yin}}}, \ and\
  \bibinfo {author} {\bibfnamefont {X.}~\bibnamefont {{Zhang}}},\ }\href
  {\doibase 10.1103/PhysRevD.97.063013} {\bibfield  {journal} {\bibinfo
  {journal} {\prd}\ }\textbf {\bibinfo {volume} {97}},\ \bibinfo {eid} {063013}
  (\bibinfo {year} {2018})}\BibitemShut {NoStop}%
\bibitem [{\citenamefont {{Baxter}}\ \emph {et~al.}(2021)\citenamefont
  {{Baxter}}, \citenamefont {{Kumar}}, \citenamefont {{Pace}},\ and\
  \citenamefont {{Runburg}}}]{Baxter_2021}%
  \BibitemOpen
  \bibfield  {author} {\bibinfo {author} {\bibfnamefont {E.~J.}\ \bibnamefont
  {{Baxter}}}, \bibinfo {author} {\bibfnamefont {J.}~\bibnamefont {{Kumar}}},
  \bibinfo {author} {\bibfnamefont {A.~B.}\ \bibnamefont {{Pace}}}, \ and\
  \bibinfo {author} {\bibfnamefont {J.}~\bibnamefont {{Runburg}}},\ }\href
  {\doibase 10.1088/1475-7516/2021/07/030} {\bibfield  {journal} {\bibinfo
  {journal} {\jcap}\ }\textbf {\bibinfo {volume} {2021}},\ \bibinfo {eid} {030}
  (\bibinfo {year} {2021})}\BibitemShut {NoStop}%
\bibitem [{\citenamefont {{Diamanti}}\ \emph {et~al.}(2014)\citenamefont
  {{Diamanti}}, \citenamefont {{Lopez-Honorez}}, \citenamefont {{Mena}},
  \citenamefont {{Palomares-Ruiz}},\ and\ \citenamefont
  {{Vincent}}}]{Diamanti_2014}%
  \BibitemOpen
  \bibfield  {author} {\bibinfo {author} {\bibfnamefont {R.}~\bibnamefont
  {{Diamanti}}}, \bibinfo {author} {\bibfnamefont {L.}~\bibnamefont
  {{Lopez-Honorez}}}, \bibinfo {author} {\bibfnamefont {O.}~\bibnamefont
  {{Mena}}}, \bibinfo {author} {\bibfnamefont {S.}~\bibnamefont
  {{Palomares-Ruiz}}}, \ and\ \bibinfo {author} {\bibfnamefont {A.~C.}\
  \bibnamefont {{Vincent}}},\ }\href {\doibase 10.1088/1475-7516/2014/02/017}
  {\bibfield  {journal} {\bibinfo  {journal} {\jcap}\ }\textbf {\bibinfo
  {volume} {2014}},\ \bibinfo {eid} {017} (\bibinfo {year} {2014})}\BibitemShut
  {NoStop}%
\bibitem [{\citenamefont {{Liu}}\ \emph {et~al.}(2021)\citenamefont {{Liu}},
  \citenamefont {{Qin}}, \citenamefont {{Ridgway}},\ and\ \citenamefont
  {{Slatyer}}}]{Liu_2021}%
  \BibitemOpen
  \bibfield  {author} {\bibinfo {author} {\bibfnamefont {H.}~\bibnamefont
  {{Liu}}}, \bibinfo {author} {\bibfnamefont {W.}~\bibnamefont {{Qin}}},
  \bibinfo {author} {\bibfnamefont {G.~W.}\ \bibnamefont {{Ridgway}}}, \ and\
  \bibinfo {author} {\bibfnamefont {T.~R.}\ \bibnamefont {{Slatyer}}},\ }\href
  {\doibase 10.1103/PhysRevD.104.043514} {\bibfield  {journal} {\bibinfo
  {journal} {\prd}\ }\textbf {\bibinfo {volume} {104}},\ \bibinfo {eid}
  {043514} (\bibinfo {year} {2021})},\ \Eprint
  {http://arxiv.org/abs/2008.01084} {arXiv:2008.01084 [astro-ph.CO]}
  \BibitemShut {NoStop}%
\bibitem [{\citenamefont {{McAlpine}}\ \emph {et~al.}(2025)\citenamefont
  {{McAlpine}}, \citenamefont {{Jasche}}, \citenamefont {{Ata}}, \citenamefont
  {{Lavaux}}, \citenamefont {{Stiskalek}}, \citenamefont {{Frenk}},\ and\
  \citenamefont {{Jenkins}}}]{Manticore}%
  \BibitemOpen
  \bibfield  {author} {\bibinfo {author} {\bibfnamefont {S.}~\bibnamefont
  {{McAlpine}}}, \bibinfo {author} {\bibfnamefont {J.}~\bibnamefont
  {{Jasche}}}, \bibinfo {author} {\bibfnamefont {M.}~\bibnamefont {{Ata}}},
  \bibinfo {author} {\bibfnamefont {G.}~\bibnamefont {{Lavaux}}}, \bibinfo
  {author} {\bibfnamefont {R.}~\bibnamefont {{Stiskalek}}}, \bibinfo {author}
  {\bibfnamefont {C.~S.}\ \bibnamefont {{Frenk}}}, \ and\ \bibinfo {author}
  {\bibfnamefont {A.}~\bibnamefont {{Jenkins}}},\ }\href {\doibase
  10.1093/mnras/staf767} {\bibfield  {journal} {\bibinfo  {journal} {\mnras}\
  }\textbf {\bibinfo {volume} {540}},\ \bibinfo {pages} {716} (\bibinfo {year}
  {2025})},\ \Eprint {http://arxiv.org/abs/2505.10682} {arXiv:2505.10682
  [astro-ph.CO]} \BibitemShut {NoStop}%
\bibitem [{\citenamefont {{Stiskalek}}\ \emph {et~al.}(2026)\citenamefont
  {{Stiskalek}}, \citenamefont {{Desmond}}, \citenamefont {{Devriendt}},
  \citenamefont {{Slyz}}, \citenamefont {{Lavaux}}, \citenamefont {{Hudson}},
  \citenamefont {{Bartlett}},\ and\ \citenamefont {{Courtois}}}]{VFO}%
  \BibitemOpen
  \bibfield  {author} {\bibinfo {author} {\bibfnamefont {R.}~\bibnamefont
  {{Stiskalek}}}, \bibinfo {author} {\bibfnamefont {H.}~\bibnamefont
  {{Desmond}}}, \bibinfo {author} {\bibfnamefont {J.}~\bibnamefont
  {{Devriendt}}}, \bibinfo {author} {\bibfnamefont {A.}~\bibnamefont {{Slyz}}},
  \bibinfo {author} {\bibfnamefont {G.}~\bibnamefont {{Lavaux}}}, \bibinfo
  {author} {\bibfnamefont {M.~J.}\ \bibnamefont {{Hudson}}}, \bibinfo {author}
  {\bibfnamefont {D.~J.}\ \bibnamefont {{Bartlett}}}, \ and\ \bibinfo {author}
  {\bibfnamefont {H.~M.}\ \bibnamefont {{Courtois}}},\ }\href {\doibase
  10.1093/mnras/staf1960} {\bibfield  {journal} {\bibinfo  {journal} {\mnras}\
  }\textbf {\bibinfo {volume} {545}},\ \bibinfo {eid} {staf1960} (\bibinfo
  {year} {2026})},\ \Eprint {http://arxiv.org/abs/2502.00121} {arXiv:2502.00121
  [astro-ph.CO]} \BibitemShut {NoStop}%
\bibitem [{\citenamefont {{Kosti{\'c}}}\ \emph {et~al.}(2023)\citenamefont
  {{Kosti{\'c}}}, \citenamefont {{Bartlett}},\ and\ \citenamefont
  {{Desmond}}}]{data_available}%
  \BibitemOpen
  \bibfield  {author} {\bibinfo {author} {\bibfnamefont {A.}~\bibnamefont
  {{Kosti{\'c}}}}, \bibinfo {author} {\bibfnamefont {D.~J.}\ \bibnamefont
  {{Bartlett}}}, \ and\ \bibinfo {author} {\bibfnamefont {H.}~\bibnamefont
  {{Desmond}}},\ }\href {\doibase 10.48550/arXiv.2304.10301} {\bibfield
  {journal} {\bibinfo  {journal} {arXiv e-prints}\ ,\ \bibinfo {eid}
  {arXiv:2304.10301}} (\bibinfo {year} {2023})},\ \Eprint
  {http://arxiv.org/abs/2304.10301} {arXiv:2304.10301 [astro-ph.CO]}
  \BibitemShut {NoStop}%
\bibitem [{\citenamefont {Harris}\ \emph {et~al.}(2020)\citenamefont {Harris}
  \emph {et~al.}}]{Numpy}%
  \BibitemOpen
  \bibfield  {author} {\bibinfo {author} {\bibfnamefont {C.~R.}\ \bibnamefont
  {Harris}} \emph {et~al.},\ }\href {\doibase 10.1038/s41586-020-2649-2}
  {\bibfield  {journal} {\bibinfo  {journal} {Nature}\ }\textbf {\bibinfo
  {volume} {585}},\ \bibinfo {pages} {357} (\bibinfo {year}
  {2020})}\BibitemShut {NoStop}%
\bibitem [{\citenamefont {Virtanen}\ \emph {et~al.}(2020)\citenamefont
  {Virtanen} \emph {et~al.}}]{Scipy}%
  \BibitemOpen
  \bibfield  {author} {\bibinfo {author} {\bibfnamefont {P.}~\bibnamefont
  {Virtanen}} \emph {et~al.},\ }\href {\doibase 10.1038/s41592-019-0686-2}
  {\bibfield  {journal} {\bibinfo  {journal} {Nature Methods}\ }\textbf
  {\bibinfo {volume} {17}},\ \bibinfo {pages} {261} (\bibinfo {year}
  {2020})}\BibitemShut {NoStop}%
\bibitem [{\citenamefont {{Dejonghe}}(1986)}]{dejonghe1986}%
  \BibitemOpen
  \bibfield  {author} {\bibinfo {author} {\bibfnamefont {H.}~\bibnamefont
  {{Dejonghe}}},\ }\href {\doibase 10.1016/0370-1573(86)90098-0} {\bibfield
  {journal} {\bibinfo  {journal} {\physrep}\ }\textbf {\bibinfo {volume}
  {133}},\ \bibinfo {pages} {217} (\bibinfo {year} {1986})}\BibitemShut
  {NoStop}%
\bibitem [{\citenamefont {{Roughan}}(2020)}]{roughan2020polylogarithm}%
  \BibitemOpen
  \bibfield  {author} {\bibinfo {author} {\bibfnamefont {M.}~\bibnamefont
  {{Roughan}}},\ }\href@noop {} {\bibfield  {journal} {\bibinfo  {journal}
  {arXiv e-prints}\ } (\bibinfo {year} {2020})},\ \Eprint
  {http://arxiv.org/abs/2010.09860} {arXiv:2010.09860 [math]} \BibitemShut
  {NoStop}%
\bibitem [{\citenamefont {{Voigt}}(2022)}]{voigy2022polylog}%
  \BibitemOpen
  \bibfield  {author} {\bibinfo {author} {\bibfnamefont {A.}~\bibnamefont
  {{Voigt}}},\ }\href@noop {} {\bibfield  {journal} {\bibinfo  {journal} {arXiv
  e-prints}\ } (\bibinfo {year} {2022})},\ \Eprint
  {http://arxiv.org/abs/2201.01678} {arXiv:2201.01678 [hep-ph]} \BibitemShut
  {NoStop}%
\end{thebibliography}%

\appendix

\section{Calculation of \texorpdfstring{$J$}{J} factor for \texorpdfstring{$d$}{d}-wave}
\label{app:Jfactor_dwave}

To calculate the $d$-wave $J$ factor at a given distance from the halo centre, one needs to evaluate the corresponding fourth velocity moment at that distance, $\langle \bm{v}^4 \rangle$. To do so, we assume ergodicity of the halo dark matter distribution function, which directly leads to the relation \cref{eq:4thvel_moment}. However, this equation solves for the fourth moment of the radial velocity component, while we need the total $\langle \bm{v}^4 \rangle$. One can show that in the case of an ergodic distribution function the following holds (see Problem 4.29 in \cite{binney2011galactic} and Section 1.3 from \cite{dejonghe1986})

\vspace{-0.2cm}
\begin{equation}
    \langle v_\theta^{n-j} v_\phi^{j-k} v_r^k \rangle 
    = 
    \langle v_r^n \rangle
    \frac
    {
    \Gamma(\tfrac{k+1}{2})
    \Gamma(\tfrac{n-j+1}{2}) 
    \Gamma(\tfrac{j-k+1}{2})}
    {\pi \Gamma(\tfrac{n+1}{2})}.
\end{equation}

From this, the following relations directly follow

\begin{align}
    \langle v_\theta^4 \rangle 
    = \langle v_\phi^4 \rangle 
    &= \langle v_r^4 \rangle \\
    \langle v_\theta^2 v_\phi^2 \rangle, 
    = \langle v_\theta^2 v_r^2 \rangle
    &= \langle v_\phi^2 v_r^2 \rangle
    = \frac{2}{3} \langle v_r^4 \rangle,
\end{align}

as does \cref{eq:4thvel_momen_final}. Now, the only remaining issue is finding a solution to \cref{eq:4thvel_moment}, since the second part of \cref{eq:Jfactor dwave} can be directly obtained from \cref{eq:Dispersion beta=0}. Note, that since we are assuming ergodicity, $\beta=0$. 

We assume NFW profiles (\cref{eq:NFW_profile}) for our halos, which we rewrite as 
\vspace{-0.4cm}
\begin{equation}
    \rho(s) = \frac{\rho_0}{c\tilde{r}(1 + c\tilde{r})^2},
\end{equation}

where $c$ is the concentration, and $\tilde{r}=r/r_{\mathrm{vir}}$. We can then solve \cref{eq:4thvel_moment} using \cref{eq:Dispersion beta=0}. Noting that the gravitational potential in this notation is
\vspace{-0.4cm}
\begin{equation}
    \Phi \left( \tilde{r} \right) = - 4 \pi G \rho_0 r_{\rm s}^2 \frac{\log \left( 1 + c \tilde{r}\right)}{c s},
\end{equation}

one obtains

\begin{widetext}
\begin{align*}    
    \label{eq:rhov4_dwave}
    \rho \langle v_r^4 \rangle = 
    &\frac{4}{3} 
    \pi ^2 G^2 
    \rho_0^3 r_s^4 
    \begin{aligned}[t]
        &\bigg[
            -\frac{6 (c \tilde{r} (c \tilde{r} (8 c \tilde{r}+21)+6)-1) \log ^2(c \tilde{r}+1)}{c^3 \tilde{r}^3}
            +\frac{6 \text{Li}_2
            \left(c^2 \tilde{r}^2
            \right) (10 c \tilde{r} + 9 \log (c \tilde{r}+1))}{c \tilde{r}} \\
            &-\frac{6\log (c \tilde{r}+1)}{c^2 \tilde{r}^2 (c \tilde{r}+1)} 
            \bigg(
            c \tilde{r} \left(-3 \pi ^2 (c \tilde{r}+1)+c \tilde{r} (3 c \tilde{r}+35)+25\right) 
            +3 c \tilde{r} (c \tilde{r}+1) \log (c \tilde{r})+2
            \bigg)\\
            &+288 \text{Li}_3(c \tilde{r}+1) +\text{Li}_2(c \tilde{r}) 
            \left(-\frac{108 \log (c \tilde{r}+1)}{c \tilde{r}}-120\right) +72 \text{Li}_2(c \tilde{r}+1) (3-4 \log (c \tilde{r}+1)) \\
            &-\frac{52 (\pi  c \tilde{r}+\pi )^2+174 c \tilde{r}+177}{(c \tilde{r}+1)^2}
            +\frac{6}{c \tilde{r} (c \tilde{r}+1)^2}\\ 
            &+\frac{6 (8 c \tilde{r}+9) \log ^3(c \tilde{r}+1)}{c \tilde{r}}
            +72 \log (c \tilde{r}) (3-2 \log (c \tilde{r}+1)) \log (c \tilde{r}+1) 
            +18 \log (c \tilde{r}) 
        \bigg]
    \end{aligned} \\
    &+ \mathrm{const}, \numberthis
\end{align*}
\end{widetext}

where $\Li_2$ and $\Li_3$ represent the dilogarithm and trilogarithm, respectively. We use the implementations provided in the \texttt{polylogarithm} library \cite{roughan2020polylogarithm, voigy2022polylog} to evaluate these. We impose $\lim_{\tilde{r} \to \infty} (\rho \langle v_r^4 \rangle)(\tilde{r}) = 0$, which fixes the integration constant to $0$. This is a consequence of the fact that the velocity dispersion itself goes to zero far away from the halo, for the case of ergodic phase-space distribution function which we assumed throughout for our halos. This in turn implies that the velocity distribution has a vanishing second moment and hence, by definition, the fourth moment is zero too in this limit. 

This analytical solution was implemented within the \clumpy package and tested against a numerical solution. For all our results regarding $d$-wave, we always utilise the analytical solution from \cref{eq:rhov4_dwave} and use the numerical solution only for cross-checking the results. We demonstrate the obtained $J$ factor for $d$-wave in \cref{fig:pwave_vs_dwave_Jtheta} obtained by substituting the expression from \cref{eq:rhov4_dwave} into \cref{eq:Jfactor dwave} and normalising by the total integrated $J$ factor value and dividing by the corresponding quantity for $s$-wave (labelled as $\tilde{J}^{\ell=2}/\tilde{J}^{\ell=0}$). One sees that the shape of the angular distribution of the normalised $J$ factor is qualitatively very similar to the $\beta=0$, $p$-wave case and shows the same behaviour towards the halo centre and the edge of the halo. 

Note that similar results have been obtained in \citet{Boddy_2018}, using the Eddington inversion formula. We chose to directly integrate the Jeans equations instead, since we were also interested in studying the anisotropic cases for the $p$-wave annihilation channel, and hence for consistency used the same approach for the $d$-wave $J$ factor calculation.

\end{document}